\documentclass[11pt, a4paper]{book}
\usepackage{svn-multi}
\svnid{$Id$}

\usepackage[hyperindex=true,
			bookmarks=true,
            pdftitle={}, pdfauthor={Xi Yang},
            colorlinks=false,
            pdfborder=0,
            pagebackref=false,
            citecolor=blue,
            plainpages=false,
            pdfpagelabels,
            pagebackref=true,
            hyperfootnotes=false]{hyperref}
\usepackage[all]{hypcap}
\usepackage[palatino]{anuthesis}
\usepackage{afterpage}
\usepackage{graphicx}
\usepackage{thesis}
\usepackage[square]{natbib}
\usepackage[normalem]{ulem}
\usepackage[table]{xcolor}
\usepackage{makeidx}
\usepackage{cleveref}
\usepackage[centerlast]{caption2}
\usepackage{float}
\usepackage{hyperref}
\usepackage[T1]{fontenc}
\usepackage[scaled]{beramono}
\usepackage{float}
\usepackage{multirow}
\usepackage{adjustbox}
\usepackage{tablefootnote}
\usepackage[multiple]{footmisc}

\usepackage{pdflscape}
\usepackage{enumerate}
\usepackage{mwe}

\usepackage{geometry}
\usepackage{lscape,lipsum,graphicx}
\usepackage[absolute]{textpos}
\usepackage{fancyhdr}

\fancypagestyle{lscape}{% 
\fancyhf{} % clear all header and footer fields 
\fancyfoot[LE]{%
\begin{textblock}{1}(13,10.5){\rotatebox{90}{\thepage}}\end{textblock}}
}

\usepackage{sistyle}
\SIthousandsep{,}

\renewcommand*{\backref}[1]{}
\renewcommand*{\backrefalt}[4]{
  \ifcase #1 %
  \or
    (cited on page #2)%
  \else
    (cited on pages #2)%
  \fi
}

\usepackage{bera}
\usepackage{listings}
\usepackage{xcolor}

\colorlet{punct}{red!60!black}
\definecolor{background}{HTML}{EEEEEE}
\definecolor{delim}{RGB}{20,105,176}
\colorlet{numb}{magenta!60!black}

\definecolor{eclipseStrings}{RGB}{42,0.0,255}
\definecolor{eclipseKeywords}{RGB}{127,0,85}
\colorlet{numb}{magenta!60!black}

\lstdefinelanguage{json}{
    basicstyle=\normalfont\ttfamily,
    commentstyle=\color{eclipseStrings}, % style of comment
    stringstyle=\color{eclipseKeywords}, % style of strings
    numbers=left,
    numberstyle=\scriptsize,
    stepnumber=1,
    numbersep=8pt,
    showstringspaces=false,
    breaklines=true,
    frame=lines,
    backgroundcolor=\color{background}, %only if you like
    string=[s]{"}{"},
    comment=[l]{:\ "},
    morecomment=[l]{:"},
    literate=
        *{0}{{{\color{numb}0}}}{1}
         {1}{{{\color{numb}1}}}{1}
         {2}{{{\color{numb}2}}}{1}
         {3}{{{\color{numb}3}}}{1}
         {4}{{{\color{numb}4}}}{1}
         {5}{{{\color{numb}5}}}{1}
         {6}{{{\color{numb}6}}}{1}
         {7}{{{\color{numb}7}}}{1}
         {8}{{{\color{numb}8}}}{1}
         {9}{{{\color{numb}9}}}{1}
}

\usepackage{booktabs}
\usepackage{relsize}
\usepackage{xspace}
\usepackage{subfigure}
\usepackage{listings}
\definecolor{tableheadcolor}{rgb}{0.8,0.8,1.0}
\definecolor{tablealtcolor}{rgb}{0.9,0.9,0.95}

\definecolor{todocolor}{rgb}{0.8,0.8,1.0}
\definecolor{fixcolor}{rgb}{1,0.8,0.8}
\definecolor{commentcolor}{rgb}{0.8,1.0,0.8}

\usepackage[color=todocolor, colorinlistoftodos]{todonotes}

\newcommand{\textjava}[1]{{\lstset{basicstyle=\ttfamily}\lstinline@#1@}}
\newcommand{\textjavafn}[1]{{\lstset{basicstyle=\footnotesize\ttfamily}\lstinline@#1@}}
\usepackage{setspace}
\usepackage{ifthen}

\long\def\sfootnote[#1]#2{\begingroup%
\def\thefootnote{\fnsymbol{footnote}}\footnote[#1]{#2}\endgroup}
\newcommand{\doi}[1]{\href{http://dx.doi.org/#1}{\nolinkurl{doi:#1}}}
\newcommand{\ignore}[1]{}

\title{Measuring COVID-19 Related Media Consumption on Twitter}

\author{Cai Yang}
\date{\today}

\renewcommand{\thepage}{\roman{page}}

\makeindex
\begin{document}

%\doparttoc
%%%%%%%%%%%%%%%%%%%%%%%%%%%%%%%%%%%%%%%%%%%%%%%%%%%%%%%%%%%%%%%%%%%%%%%
%% Title page
\pagestyle{empty}
\thispagestyle{empty}
%% anuthesis.sty Copyright (C) 1996, 1997 Steve Blackburn
%% Department of Computer Science, Australian National University
%%

\begin{titlepage}
  \enlargethispage{2cm}
  \begin{center}
    \makeatletter
    \Huge\textbf{\@title} \\[.4cm]
    \Huge\textbf{\thesisqualifier} \\[2.5cm]
    \huge\textbf{\@author} \\[9cm]
    \makeatother
%%   \LARGE A thesis submitted for the degree of \\
%%    Master of Philosophy at \\
%%    The Australian National University \\[2cm]
    \LARGE A thesis submitted for the degree of \\
    Bachelor of Information Technology (Honours) \\
    The Australian National University \\[2cm]
    % \thismonth
    November 2021
  \end{center}
\end{titlepage}

% %%%%%%%%%%%%%%%%%%%%%%%%%%%%%%%%%%%%%%%%%%%%%%%%%%%%%%%%%%%%%%%%%%%%%%%
% %% Here begin the preliminaries
\input{frontmatter}

%%%%%%%%%%%%%%%%%%%%%%%%%%%%%%%%%%%%%%%%%%%%%%%%%%%%%%%%%%%%%%%%%%%%%%%
%% Dedication
\cleardoublepage
\pagestyle{empty}
\vspace*{7cm}
\begin{center}
To everyone who has supported me all along the journey.
\end{center}

%%%%%%%%%%%%%%%%%%%%%%%%%%%%%%%%%%%%%%%%%%%%%%%%%%%%%%%%%%%%%%%%%%%%%%%
%% Acknowledgements
\cleardoublepage
\pagestyle{empty}
\chapter*{Acknowledgments}
\addcontentsline{toc}{chapter}{Acknowledgments}

First and foremost, I want to express my sincerest and warmest thanks to Lexing Xie and  Siqi Wu for being great supervisors and mentors. The thesis has gone through several setbacks, and there is no way for me to make it to the end without your guidance. Also, thank you for providing me with a summer research internship, which opens me the door to computational social media.

Besides my supervisors, I also want to thank Yu-Ru Lin and Xian Teng from the University of Pittsburgh for their insightful advice ever since the beginning of the project. 

Last but not least, I am also grateful for the care, encouragement, and sacrifices from my parents. 2021 has been extremely challenging for me, and I have learned that things are not always going as planned. It is your unconditional love that supports me through it.

%%%%%%%%%%%%%%%%%%%%%%%%%%%%%%%%%%%%%%%%%%%%%%%%%%%%%%%%%%%%%%%%%%%%%%%
%% Abstract
\cleardoublepage
\pagestyle{headings}
\chapter*{Abstract}
\addcontentsline{toc}{chapter}{Abstract}
\vspace{-1em}

The COVID-19 pandemic has been affecting the world dramatically ever since 2020. The minimum availability of physical interactions during the lockdown has caused more and more people to turn to online activities on social media platforms. These platforms have provided essential updates regarding the pandemic, serving as bridges for communications. 

Research on studying these communications on different platforms emerges during the meantime. Prior studies focus on areas such as topic modeling, sentiment analysis and prediction tasks such as predicting COVID-19 positive cases, misinformation spread, etc. However, online communications with media outlets remain unexplored on an international scale. We have little knowledge about the patterns of the media consumption geographically and their association with offline political preference. We believe addressing these questions could help governments and researchers better understand human behaviors during the pandemic.

In this thesis, we specifically investigate the online consumption of media outlets on Twitter through a set of quantitative analyses. We make use of several public media outlet datasets to extract media consumption from tweets collected based on COVID-19 keyword matching. We make use of a metric ``interaction'' to quantify media consumption through weighted Twitter activities. We further construct a matrix based on it which could be directly used to measure user-media consumption in different granularities.

We then conduct analyses on the United States level and global level. Both investigations have provided a high-level overview of the media consumption from users among different locations. Moreover, the study on the United States links online media consumption with offline voting results and validates the media consumption vectors. The global interaction analysis identifies user groups within and across countries based on the empirical evidence on polarization and homophily. To the best of our knowledge, this thesis presents the first-of-its-kind study on media consumption on COVID-19 across countries, it sheds light on understanding how people consume media outlets during the pandemic and provides potential insights for peer researchers.

%%%%%%%%%%%%%%%%%%%%%%%%%%%%%%%%%%%%%%%%%%%%%%%%%%%%%%%%%%%%%%%%%%%%%%%
%% Table of contents
\cleardoublepage
\pagestyle{headings}
\markboth{Contents}{Contents}
\tableofcontents
\listoffigures
\listoftables

%%%%%%%%%%%%%%%%%%%%%%%%%%%%%%%%%%%%%%%%%%%%%%%%%%%%%%%%%%%%%%%%%%%%%%
%% Here begins the main text
\mainmatter

%% Introduction
\chapter{Introduction}
\label{cha:intro}

% \section{Introduction}
% \label{sec:problemstatement}

The COVID-19 pandemic has brought seismic changes in people's communications in various aspects. Governments and health organizations across the world have implemented different social distancing policies and lockdowns to mitigate the impact of the pandemic (\cite{Nguyen2020ChangesID}). The reduced physical socialization due to such policies has forced people to shift their offline attention to online. %due to the physical lockdown and distancing policy. 

% During the massive shifts of interactions, social media platforms have played a critical role in providing the latest information about the crisis and hence attracted more attention from online users. ). 

Nonetheless, there exist very few works on studying online consumption during the pandemic. We have little knowledge about people's online consumption behaviors on social media. For instance, who are people interacting with, how do they consume information related to COVID-19, how does their online consumption relate to their offline behaviors, etc. As a result, understanding how people communicate and consume information online has become a crucial research topic.

% Addressing these questions is challenging in general for several reasons. First, the popularity on social media platforms does not reach its peak until recent years. Moreover, the rarity of pandemic-level diseases has greatly limited the timespan for analysis (\cite{Yang2021OnlineCS}). The two factors restrict the past studies on the interplay between pandemic and social media platforms. Second, exploration of online consumption is difficult to proceed without the availability of large-scale data sources.

As pointed out by \cite{Yang2021OnlineCS}, addressing these questions is challenging in general for several reasons. First, the rarity of pandemic-level diseases has greatly limited the analysis in the past. Second, the previous pandemics did not happen in a modern stage where the popularity of social media has driven numerous online communications. Such factors restrict the past studies on the interplay between pandemic and social media platforms. Last, exploration of online consumption is difficult to proceed without the availability of large-scale data sources.

In this thesis, we overcome such challenges by utilizing a collection of datasets in conjunction with tweets to study COVID-19-related media consumption on Twitter. Twitter has been one of the most popular social media platforms for providing breaking news and politics and attracted a large number of users and media outlets. Accordingly, it has drawn our attention to consider it as our research targeted platform. This thesis aims to address the following questions to help us better understand the online media consumption related to the COVID-19 pandemic: 

\begin{itemize}
    \item RQ1: What do the overall media consumption patterns look like on the United States and international scale?
    \item RQ2: Is the online media consumption correlated with the offline political preference of users?
    \item RQ3: Are there any user groups across the world and what do those groups look like?
    \item RQ4: Are there any potential user groups within each country, and how do user groups and their media consumption change across countries? 
\end{itemize}

% \section{Contributions}
% \label{sec:contri}

% The spirit of our work lies in the intersection of the current work, especially closer to some work presented in section \ref{sec:glo-med-attn}. However, the previous works mainly focus on user communications or the general media consumption before COVID-19. 

% \begin{itemize}
%     \item We are the first to study COVID-19 media consumption on Twitter with different granularity, including on both United States and countries outside the United States.
%     \item We are the first to demonstrate the association between online COVID-19-related consumption and offline political behaviors, making it possible to apply such consumption vectors onto various tasks where political partisanship is involved.
%     \item We are the first to systematically provide detailed country-level analysis on user-media interactions, including overall analysis and specific country-pairs. Our method developed can be directly adapted onto different country pairs, making it easy to generalize the study onto other countries of interest.
% \end{itemize}

The spirit of our work lies in the intersection of media consumption, political polarization, and communications related to COVID-19. We take the first step to quantify the user-media consumption on COVID-19 topics on Twitter on an international scale. In summary, our work aims to fill in the gap with the following contributions and findings:

\begin{itemize}
    \item We define a metric, ``interaction'', on quantifying Twitter users' media consumption. Meanwhile, we construct the matrix representation constructed that allows quantitative measurement on the different granularity of locations and can be generalized onto different countries (Chapter \ref{cha:interaction-intro}).
    \item We provide new evidence on the association between online COVID-19-related interaction vectors and offline political votes. The strong correlation between the online signal and offline behaviors reveals the polarization of users' online media consumption on COVID-19-related topics (Chapter \ref{cha:us-inter}).
    \item We demonstrate the existence of user groups through detailed global-level analysis. We show that there are user groups whose membership is dominated by countries outside the United States. We also identify the occurrence of various user groups within each country, differed by their ideologically different media consumption. Our method can be adapted onto different country pairs, making it easy to generalize onto other countries of interest (Chapter \ref{cha:global-inter}).
    \item As a by-product, this thesis also offers a fast and efficient pipeline for parsing Twitter user descriptions onto different countries, which can help future work on global-level studies (Chapter \ref{cha:data}).
\end{itemize}

\section{Thesis Outline}
\label{sec:outline}

The thesis is organized as follows:

\begin{itemize}
    \item Chapter~\ref{cha:background} reviews relevant literature to our work. It also highlights how this thesis has filled in the gaps of the current works.
    \item Chapter~\ref{cha:data} presents all the datasets used throughout this thesis, including how they are obtained and their statistics. It also covers how we parse user profile descriptions into locations.
    \item Chapter~\ref{cha:interaction-intro} defines the quantitative measurement ``interaction'' and the matrices of different granularities constructed from it.
    \item Chapter~\ref{cha:us-inter} offers a glance at the interactions extracted from the United States level. It covers two representations of interaction matrices. Meanwhile, it also shows the relationship between users' online media consumption vectors and political preference based on the 2020 Presidential election results.
    \item Chapter~\ref{cha:global-inter} moves the attention from the United States to a global scale. It briefly goes through interaction matrices similar to the previous chapter. Furthermore, it provides a detailed analysis of international level user consumption with media outlets.
    \item Chapter~\ref{cha:conc} concludes the thesis with a summary. It also points out the limitations of current work along with some possible future directions.
\end{itemize}

%% Chapters
\chapter{Literature Review}
% \chapter{Literature Review}
\label{cha:background}

This chapter reviews the relevant literature. Section \ref{sec:attention-consumption} presents the studies done on media attention and how people consume media. Specifically, it also offers related work on consumption across different social media platforms and global media attention. Section \ref{sec:polarization} reviews literature on studying the political polarization and partisanship on user behaviors in online social activities. Section \ref{sec:covid-communication} outlines some work regarding online communications during the pandemic. It is worth mentioning that the following topics are not mutually exclusive. Meanwhile, we briefly discuss the difference between existing work and our work under each section.

% \section{The Twitter Platform}

% Twitter has become a popular target for many social computing research in the past. 
% \cite{Cha2012TheWO}: Information propagation in online social networks like Twitter
% \cite{Kwak2010WhatIT}: what is Twitter? 
% \cite{Crilley2020UnderstandingRA}: Understanding RT’s Audiences: Exposure Not Endorsement for Twitter Followers of Russian State-Sponsored Media

%%%%%%%%%%%%%%%%%%%%%%%%%%%%%%%%%%%%%%%%%%%%%%%%%%%%%%%%%%%%%%%%%%%%%%%%%%%%%%%%%%%%%%%%%%%%%%%%%%%%%%%%%%%%%%%%%%%%%

\section{Media Attention and Consumption}
\label{sec:attention-consumption}

The following works can be considered as general media consumption and user behaviors on social media. Our work follows a similar goal and takes a step towards understanding media consumption related to the pandemic, which we believe has changed dramatically given the timespan and scope of impacts brought by COVID-19.

As one of the earliest works on studying Twitter, \cite{Wu2011WhoSW} studies the production, flow, and consumption of information. The experiments show that although celebrities are the most followed user groups on Twitter, roughly half of the attention consumed, in the form of URLs, is from elite users, which only comprises less than 0.05\% of the overall population. Meanwhile, \cite{An2011MediaLI} presents a preliminary study on user media exposure on Twitter. Their experiment has shown that there exists indirect media exposure from users to media, which significantly expands the political diversity of news that users are exposed to.

\cite{Kulshrestha2015CharacterizingID} defines the media diet of users as a topical distribution and characterizes the information consumed by different types of users on Twitter. They find that popular users tend to consume media of only a few topics, and news organizations usually produce more focused diets than their mass media diets. They also find that most users are only interested in consuming media of their interests.

\cite{Reis2017DemographicsON} presents first in-depth characterization on news spreaders in Twitter. The work investigates the demographics of those news spreaders, the content shared by them, and the audience they reach. It is shown that white users and male users are more active as news spreaders, which bias the news audience to the interests of these demographic groups.

\cite{An2017MultidimensionalAO} provides a multidimensional analysis of news reading behaviors across different demographic groups. The authors measured the difference of behaviors in four levels: actual news items, sections, topics, and sub-topics. By using topic modeling and vector representations, it is found that the differences are most noticeable at the sub-topic level.

% \cite{Falzon2017RepresentationAA}: Representation and Analysis of Twitter Activity: A Dynamic Network Perspective
% \cite{An2019PoliticalDI}: Political Discussions in Homogeneous and Cross-Cutting Communication Spaces
% \cite{Baly2020WhatWW}: What Was Written vs. Who Read It: News Media Profiling Using Text Analysis and Social Media Context
% \cite{Cointet2021UncoveringTS}: Uncovering the structure of the French media ecosystem
% \cite{Aldous2019ViewLC}: View, Like, Comment, Post: Analyzing User Engagement by Topic at 4 Levels across 5 Social Media Platforms for 53 News Organizations
% \cite{Weld2021PoliticalBA}: Political Bias and Factualness in News Sharing across more than 100,000 Online Communities Galen
% \cite{Robertson2018AuditingPA}: Auditing Partisan Audience Bias within Google Search
% \cite{Garimella2021PoliticalPI} analyzes the online web browsing 
% \cite{PrabhakarKaila2020InformationalFO}: Informational Flow on Twitter – Corona Virus Outbreak – Topic Modelling Approach

%%%%%%%%%%%%%%%%%%%%%%%%%%%%%%%%%%%%%%%%%%%%%%%%%%%%%%%%%%%%%%%%%%%%%%%%%%%%%%%%%%%%%%%%%%%%%%%%%%%%%%%%%%%%%%%%%%%%%

\subsection{Cross-community Consumption}
\label{sec:crscomm-consumption}

Results from \cite{Zannettou2017TheWC} indicate that the alt-right communities within 4chan and Reddit have brought surprising influence on Twitter. The authors provide evidence showing that such communities can often succeed in spreading news onto mainstream social media and the web. Their work emphasizes the difference of the preferred content among the three communities and the effects of one community on the other. In this thesis, we do not emphasize the effects from one user group on the other but rather focus on the change in their media consumption from an ideology perspective.

\cite{Hosseinmardi2021ExaminingTC} find out the news consumption on YouTube is mainly dominated by mainstream media and centrist sources. The authors also claim there is little evidence that YouTube recommendation algorithms drive the attention to far-right and anti-woke content. They believe the tendency in political video consumption should be attributed to more complicated factors such as a combination of user preference, platform features, and supply-and-demand dynamics of the web. However, their work addresses systematic effects at the population level while this thesis covers analysis on small user groups. 

%%%%%%%%%%%%%%%%%%%%%%%%%%%%%%%%%%%%%%%%%%%%%%%%%%%%%%%%%%%%%%%%%%%%%%%%%%%%%%%%%%%%%%%%%%%%%%%%%%%%%%%%%%%%%%%%%%%%%

\subsection{Global Media Attention}
\label{sec:glo-med-attn}

% Jisun An et al have conducted several studies on global media attention and some of her work share a similar purpose with this thesis.

\cite{Kwak2017MultiplexMA} constructs a multiplex media attention and disregard network to assess the media attention. The authors characterize the network from different levels and show that media attention is highly skewed and hierarchical. \cite{Kwak2018WhatWR} compares the attention from media organizations and the interests of the public. With data collected from Unfiltered News and web searching volume data from Google Trends, the authors build a multiplex network to study the two different attention. They find a difference between public attention and media attention in some countries, indicating that the public may be ignoring their local media outlets and consuming from other news sources.

Work from \cite{An2017WhatGM} aims to analyze the types of topics from media attention and their temporal patterns over time,  based on the empirical evidence from 196 countries with the help of Unfiltered News. \cite{An2017ConvergenceOM} assess the longitudinal tendency of similarity and dissimilarity from media attention. The authors show that there exist two complementary patterns that can explain the media attention, which is driven due to geographical proximity and other factors like historical reasons. 

Although their work shares a similar goal with ours, there are several intrinsic differences. First, most of the works are media-oriented, that is they focus on attention from media rather than the consumption on them. Although the works aim to measure media attention from a global perspective, they were conducted on a very general level and lack the details on relations between specific country groups. Second, some of these works rely on relatively complicated network structures to study media attention and public interests. They concentrate on analysis on differences between time series and shared topics. On the contrary, the metric in our work is much simpler and focuses on quantitative measurement. Last, all the work relies significantly on Unfiltered News but our work aims at measuring media consumption on Twitter.

%%%%%%%%%%%%%%%%%%%%%%%%%%%%%%%%%%%%%%%%%%%%%%%%%%%%%%%%%%%%%%%%%%%%%%%%%%%%%%%%%%%%%%%%%%%%%%%%%%%%%%%%%%%%%%%%%%%%%

\section{Political Partisanship and Polarization}
\label{sec:polarization}

\cite{Conover2011PoliticalPO} examines two networks: retweet network and user-user mention network on Twitter to investigate how social media facilitates communication between communities of different political orientations. The authors show the retweet network exhibits a highly segregated partisan structure, that is, there is minimum connectivity between left- and right-leaning users. However, the user-user mention network shows a higher rate between interactions from ideologically-opposed individuals. The authors attribute this phenomenon to the fact that politically provoked individuals tend to spread partisan content to ideologically-opposed audiences with statistical evidence.

\cite{Freelon2020FalseEO} shows that in general left-leaning and right-leaning activists use digital and legacy media for different purposes. Left-leaning activists operate through hashtag activism and offline pose while right-leaning activists are more likely to operate over legacy media and migrate onto alternative platforms. The authors also argue more research is needed to reveal the magnitude of disinformation and conspiracy from left-leaning media content.

\cite{Rao2021PoliticalPA} aims to measure political partisanship and anti-science attitudes within the online discussions on Twitter. The experiments have shown a strong correlation between polarized consumption along the science and political dimension, where politically moderate users tend to consume pro-science content while politically biased users tend to consume more anti-science content. Meanwhile, the authors also find the geographical difference in anti-science users. Moderate users tend to tweet from the western states in the United States while conservatives tend to tweet from the Southern and Northwestern states. 

All the works presented above aims to analyze the political polarization of users. It is worth mentioning that our work does not study political polarization and echo chambers explicitly but rather provides empirical evidence on their existence. Although Twitter has been involved among all the works here, most of them only use partial Twitter activities such as retweets and mentions. In contrast, our work considers almost all Twitter activities to obtain a variety of media consumption. \cite{Rao2021PoliticalPA} collects the dataset similarly to ours but with less coverage and focus on the United States only.

Meanwhile, there are also a few case studies on political polarization. \cite{Jiang2020PoliticalPD} show there exists a correlation between partisanship and the sentiment towards government on Twitter. \cite{VillaCox2021ExploringPO} explore the polarized user behaviors during South American protests on Twitter and find empirical evidence of the ``filter bubble'' phenomenon during the protest. \cite{Maulana2021PoliticalPI} represent the media landscape as a network with the shared audience and make use of community detection algorithms to study the political polarization during the Indonesian elections.

% It is obvious the analyses on these works have been limited to single countries

%%%%%%%%%%%%%%%%%%%%%%%%%%%%%%%%%%%%%%%%%%%%%%%%%%%%%%%%%%%%%%%%%%%%%%%%%%%%%%%%%%%%%%%%%%%%%%%%%%%%%%%%%%%%%%%%%%%%%

\section{Communication during COVID-19}
\label{sec:covid-communication}

During the massive shifts of interactions, social media platforms have played a critical role in providing the latest information about the crisis and hence attracted more attention from online users. Research starts to emerge on studying the interplay of COVID-19 and various social media platforms in the meantime (\cite{Yang2021OnlineCS}). However, most of the existing works done related to COVID-19 focus on misinformation detection and topics modeling. We highlight the following work that studies the communications during the pandemic.

\cite{Yang2021OnlineCS} conducts a study on how COVID-19 affects the sharing behaviors of users on Snapchat. Their experiments have shown strong evidence that the pandemic has increased private sharing propensity between users while a decrease in public sharing. They also find that the homophily between user groups of similar locations, ages, and the same gender has reduced due to social distancing. In the end, the authors also observe the association between positive COVID-19 cases in the United States and the public sharing on Snapchat.

\cite{Zhang2021UnderstandingTD} tried to characterize the trajectory of users' activities between the two communities \textit{/r/China\_flu} and \textit{/r/Coronavirus} on Reddit during COVID-19. The authors first demonstrate the connection between user activities on Reddit and some representative dates of the pandemic. They also compare the differences between new users on the two communities and show that the difference went up ever since March when Reddit has announced \textit{/r/Coronavirus} as an official community for COVID-19. Meanwhile, they also show that half of the users who began at \textit{/r/China\_flu} from January to March transferred to the other channel while users who began later at \textit{/r/China\_flu} are more loyal.

The two works above deal with the propensity of user behaviors. Meanwhile, they also demonstrate an association of online behaviors and offline phenomena such as infected cases or important dates. Part of our work has a similar purpose but rather focuses more on political behaviors.

\cite{Jiang2021SocialMP} studies the political polarization and echo chambers effect through a case study of COVID-19 on Twitter. They propose Retweet-BERT to help with estimating user polarity based on languages and user networks. The authors find that right-leaning users are more vocal and active when it comes to COVID-19 information consumption. They also demonstrate the majority of the influential users are partisan, which is considered to further contribute to polarization. The authors focus more on the roles of users in spreading information. Moreover, their work depends on a variant of BERT to predict the leanings of users from retweets while we do not use any deep learning models and cover more Twitter activities.

\chapter{Data Collection and Overview}
\label{cha:data}

Chapter \ref{cha:data} presents datasets used in the project. Firstly, section \ref{sec:tweets-collection} describes how the COVID-19-related tweets were collected. It also covers some attributes of tweets used in later work. Following this, section \ref{sec:geo-parsing} goes through one of the fundamental tasks in this thesis: parsing users' descriptions into geo-locations. Section \ref{sec:media-set} introduces the media dataset and how their relevant Twitter accounts were collected. In the end, section \ref{sec:supp-data} briefly mentions supplementary data used in this thesis.

\section{COVID-19 Tweets Collection}
\label{sec:tweets-collection}

The tweets were collected through a software \verb|Twitter-intact-stream|\footnote{\url{https://github.com/avalanchesiqi/twitter-intact-stream}}, which makes use of Twitter filtered streaming API provided by \verb|tweepy|\footnote{\url{https://github.com/tweepy/tweepy}}, by matching COVID-19 related keywords extended from \cite{info:doi/10.2196/19273}. The complete set of keywords and settings of crawlers can be found in Appendix \ref{app:keywords}. \verb|Twitter-intact-stream| splits a set of filtering keywords and languages into multiple sub-crawlers. Each sub-crawler can cover more tweets for the target stream; altogether, they could produce a dataset with higher coverage compared to crawling with a single crawler, where rate limits are likely to be hit. The table below shows the data collected from different periods along with their estimated sampling rates.

\begin{table*}[!htbp]
%\begin{adjustbox}{width=\textwidth,center}
    \centering
    \begin{tabular}{|c|c|c|c|c|} 
    \hline
    Period & Days & Tweets & Sampling Rate & Tweets/Day \\
    \hline
    2020-03-23 - 2020-04-16 & 25 & \num{525573253} & 94.9\% & 21.02 M \\
    \hline
    2020-04-19 - 2020-04-23 & 5 & \num{73989784} & 99.3\% & 14.80 M \\
    \hline
    2020-04-27 - 2020-06-27 & 62 & \num{564982759} & 99.5\% & 9.11 M \\
    \hline
    2020-07-28 - 2020-11-30 & 125 & \num{754920425} & - & 6.04 M \\
    \hline
    \end{tabular}
%\end{adjustbox}
\caption{COVID-19 Tweet dataset overview. The sampling rate for 125-day data is not available at the moment we finish this thesis.}
\label{tab:tweets-data-stats}
\end{table*}

The sampling rates are calculated based on the approach described in \cite{Wu2020VariationAS}. Twitter filtered streaming API will raise a notice when the sub-crawler has matched more tweets than what its current threshold allows to be delivered. The message contains a timestamp and a total count of undelivered tweets since the connection was established (\cite{twitterdev}). By dividing the datasets into a list of segments where (a) there is no limit notice showing up when collecting that specific dataset; (b) the collected dataset is bounded by two limit notices, we can calculate the total number of undelivered tweets and hence the estimated sampling rate of the collected tweets. Furthermore, according to \cite{Wu2020VariationAS}, using multiple sub-crawlers can collect data with a negligible difference compared to the full dataset. Hence the data used here can be considered representative of the Twitter population regarding COVID-19 related activities.

It is shown in the table that there is a declining tendency on the average number of tweets per day, which may be due to people having become fatigued about COVID-19-related topics. As suggested by a survey conducted by \cite{pewfatigue} from Pew Research Center, 71\% of Americans expressed their willingness to take breaks from coronavirus news, and 43\% said the news makes them emotionally worse.

Our studies in this thesis are \textbf{all based on the 5-day dataset} from 2020-04-19 to 2020-04-23. We leave the analysis on other datasets and exploration on the potential temporal patterns as future work.

\subsection{Tweet Overview}
\label{sec:tweet-overview}

Each tweet object contains multiple attributes and it can be rendered in JSON format. An example of tweet is shown below\footnote{Taken from \url{https://developer.twitter.com/en/docs/twitter-api/v1/data-dictionary/object-model/tweet} }. \verb|id| (integers) and \verb|id_str| (strings) at root level are the unique identifiers of the tweet. Each user is also associated with a unique identifier: \verb|id| and \verb|id_str| under \verb|user| attribute. Meanwhile each user's account, also known as \textbf{Twitter handle}, is shown in \verb|screen_name| under \verb|user| attribute.

\begin{lstlisting}[language=json,firstnumber=1]
{
 "created_at": "Wed Oct 10 20:19:24 +0000 2018",
 "id": 1050118621198921728,
 "id_str": "1050118621198921728",
 "text": "To make room for more expression, we will now count all emojis as equal-including those with gender and skin t... https://t.co/MkGjXf9aXm",
 "user": {
    "id": 6253282,
    "id_str": "6253282",
    "name": "Twitter API",
    "screen_name": "TwitterAPI",
    "location": "San Francisco, CA",
    "url": "https://developer.twitter.com",
    "description": "The Real Twitter API. tweets about API changes, service issues and our Developer Platform. Don't get an answer? It's on my website.",
    "verified": true,
    "followers_count": 6129794,
    "friends_count": 12,
    "listed_count": 12899,
    "favourites_count": 31,
    "statuses_count": 3658,
    "created_at": "Wed May 23 06:01:13 +0000 2007",
    "utc_offset": null,
    "time_zone": null,
    "geo_enabled": false,
    "lang": "en",
    "contributors_enabled": false,
    "is_translator": false,
  },
 "entities": {
    "hashtags":[],
    "urls":[],
    "user_mentions":[],
    "media":[],
    "symbols":[]
    "polls":[]
    }
}
\end{lstlisting}

There are several mutually inclusive tweet activities: mention, retweet, quote, reply and URL sharing that can be identified through the attributes of each tweet. We briefly identify some key attributes on determining the type of a tweet below.

\begin{itemize}
    \item \verb|retweeted_status|: it presents when a tweet is a retweet. A retweet is a tweet re-posting or forwarding another user's message, including the user itself.
    \item \verb|quoted_status|: it presents when a tweet is a quote. A quote can be viewed as a retweet with an additional text message. Notice that if a tweet is a retweet of a quote, both \verb|retweeted_status| and \verb|quoted_status| are present.
    \item \verb|in_reply_to_user_id| or \verb|in_reply_to_user_id_str|: it presents when a tweet is mentioning other users. Mention is a general activity that is likely to show up with other activities such as retweets and quotes.
    \item \verb|entities|: the entities provide metadata and additional contextual information about the content posted such as hashtags used in the tweet, URLs shared by the tweet, etc. The URLs contained play a significant role in the later section where we try to query the Twitter accounts for each media outlet.
\end{itemize}

\section{Geo-parsing}
\label{sec:geo-parsing}

For the later purpose of spatial analysis, we aim to associate each tweet with geo-locations, namely states for United States-level analysis and countries for global-level analysis. We refer to the task of parsing given descriptions from Twitter user profiles into actual locations as geo-parsing. The difficulty of geo-parsing arises because Twitter user profile descriptions are free-formatted, which implies they can be anything else besides locations. To tackle this problem, we exploit a two-phase parsing method of rule-based deterministic checking and geo-coding service.

The rule-based checking creates a mapping from all possible combinations of cities, states/provinces, and countries to their corresponding countries and states based on data obtained from Simplemaps\footnote{\url{https://simplemaps.com/data/world-cities}}. It serves as a simple and fast mapping to find out locations from standard geo-formatted English descriptions among free-formatted location strings. Moreover, it can provide all the possible results derived from the input string that can help reduce the conflicting locations.

Additionally, the geo-coding service aims to help with parsing the unparsable locations from the rule-based checking, for instance, due to misspelled names (\textit{Illinois} spelled as \textit{Ilinois}). In particular, we make use of \verb|local-geocode|\footnote{\url{https://github.com/mar-muel/local-geocode}}, which makes use of memory-efficient data structure called Flashtext and offline data from \verb|GeoNames|\footnote{\url{http://www.geonames.org/}}. Therefore, it is highly efficient compared to the deep learning-based method (e.g. \verb|mordecai|) and has unlimited usage compared to API-based libraries (e.g. \verb|OpenStreeMap|, \verb|Google Map API|). Furthermore, \verb|local-geocode| supports multiple languages, which can deal with non-English descriptions.

It should be noted that the geo-parsing method only parses descriptions without ambiguity. We only parse locations with ambiguity such as \textit{Cambridge} if there is corroborating evidence that narrows down the location. For instance, \textit{Cambridge} alone is considered to be inconclusive since it can be a location from \textit{the United States}, \textit{the United Kingdom}, \textit{New Zealand}, and \textit{Canada} while \textit{Cambridge, Massachusetts} can be parsed into \textit{Massachusetts, United States}.

There are \num{2032669} unique descriptions from the 5-day dataset, among which \num{911631} can be parsed into countries, including \num{278941} US locations. There are \num{174451} states parsed from such United States locations. The remaining United States locations that cannot be parsed into states are usually general descriptions such as \textit{USA}, \textit{United States}, etc. Associated with these unique descriptions, there are \num{9320202} users in total and \num{6395658} of them can be parsed. \num{1764096} are parsed into US, among which \num{1098097} are linked with states. 

\begin{figure*}[!htbp]
  \centering
  \advance\leftskip-2.27cm
  \vspace{0pt}
  \includegraphics[width=1.4\linewidth]{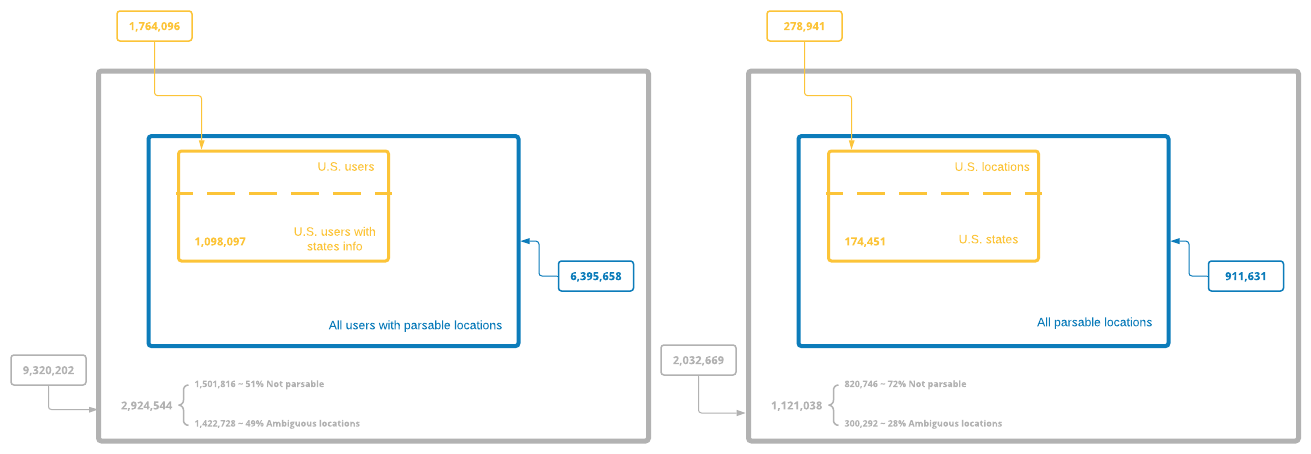}
  \caption{Inclusion illustration on users with parsable and unparsable locations.}
  \label{fig:venn}
\end{figure*}

Figure \ref{fig:venn} shows the inclusion hierarchy of the parsable and unparsable locations and the users with these locations. In addition, there are \num{2924544} unparsable locations, around 49\% of which are due to ambiguity mentioned earlier. There are \num{1121038} users associated with these unparsable locations, approximately 28\% of which have ambiguous locations.

% \begin{figure*}
%   \centering
%     \includegraphics[width=1.2\linewidth]{figs/geo-paring/geoparsing_venn.pdf}
%   \caption{Venn diagram illustration on users with parsable and unparsable locations. The two figures are used to illustrate the containing }
%   \label{fig:aus-to-us}
% \end{figure*}

\section{Media Outlets Collection}
\label{sec:media-set}

In this section, we present how media outlets and their corresponding Twitter handle(s) are collected.

\subsection{Media Bias/Fact Check}
\label{sec:mbfc}

%  where domain experts perform a careful manual analysis based on clear guidelines
 
%  It has been widely used in the research on social media.\fix{supporting arguments?}
 
\textbf{Media Bias/Fact Check}\footnote{\url{https://mediabiasfactcheck.com/}} (\textbf{MBFC}) is a comprehensive website for media bias and factuality checks. The most important attributes used in this project offered by it are the political ideologies of media, including extreme-left, left, center-left, center, center-right, right, extreme-right. To obtain these attributes, we crawled the website \footnote{Crawled on 2021-07-20} based on its 6 sub-pages: \textit{left}, \textit{leftcenter}, \textit{center}, \textit{right-center}, \textit{right}, \textit{fake-news}. Additionally, we also extract the image profile information for each media if it is present to serve as supplementary labels for those media with missing ideologies. 

\subsection{Media Attributes Overview}
\label{sec:media-features}

Each media has multiple attributes that can be obtained through MBFC such as factuality, credibility, etc. This information serves as an important aspect when analyzing clusters in chapter \ref{cha:global-inter}. The methodology for calculating ratings of bias and overall factuality is described below. A comprehensive description can be found at Media Bias/Fact Check\footnote{\url{https://mediabiasfactcheck.com/left-vs-right-bias-how-we-rate-the-bias-of-media-sources/}}.

The bias of media is determined by the number of times it falls into each category of the topics below. Left-biased or right-biased media favor more of one side while center-leaning media tend to be more balanced. It is worth mentioning the bias is determined from the American perspective and may not align with the political spectrum with all other countries. Some examples of media of different political leanings are shown in the table \ref{tab:media-example-leaning}.

\begin{itemize}
    \item General Philosophy. Left: Community over individuals (Collectivism). Right: Individuals over the community (Individualism).
    \item Abortion. Left: Mostly legal. Right: generally illegal with exceptions.
    \item Economic Policy. Left: equality on income; higher taxes on the wealthy and strong regulation on business. Right: lower taxes and less regulation over the business.
    \item Education Policy. Left: favor free and public education; Right: favor private schools. Not opposed to public education in general but critical of what has been taught.
    \item Environmental Policy. Left: regulations needed to protect the environment. Right: the market can find its own solution to climate change.
    \item Gay Rights. Left: support gay marriage. Right: oppose gay marriage.
    \item Gun Rights. Left: favor laws such as background checking before buying a gun. Right: strongly support the Second Amendment.
    \item Health Care. Left: support universal healthcare and believe healthcare is a human right. Right: consider private healthcare to be more effective and it is not a human right.
    \item Immigration. Left: less restrictive legal immigration; support a moratorium on deporting certain undocumented immigrants; Right: More restrictive legal immigration; oppose a moratorium on deporting certain workers.
    \item Military. Left: decreasing spending. Right: increasing spending.
    \item Personal Responsibility. Left: laws are enacted to protect equality. Safety nets are provided for those in need. Right: government should hold accountable for personal responsibility. Favor fair competitions over safety nets.
    \item Regulation. Left: government regulations are necessary to protect consumers. Right: government regulations would prevent the free market.
    \item Social View. Left: based on community and social responsibility. Right: based on individualism and justice.
    \item Taxes. Left: higher rates for high-income earners (progressive taxation). Right: same tax rate regardless of income (flat tax).
    \item Voter ID. Left: oppose voter I.D. laws and believe there is no evidence of voter fraud. Right: Support vote I.D. laws to tackle the alleged voter fraud.
    \item Worker's/Business Rights. Left: support worker unions and protections. Right: 
\end{itemize}

\begin{table}[h!]
\begin{adjustbox}{width=1.2\textwidth,center}

    \centering
    \begin{tabular}{|c|c|c|c|c|c|c|c|} 
    \hline
    Political Ideology & Extreme-left & Left & Center-left & Center & Center-right & Right & Extreme-right \\
    \hline
    \multirow{3}{*}{Media Outlet} & Bossip & CNN & ABC News & The Hill & Daily Press & Whitehouse.gov\tablefootnote{Note that our tweets datasets are from 2020, when Whitehouse.gov is still considered to be right-leaning. This has changed on MBFC website after the U.S presidential election in 2020. } & Yes I'm Right \\

     & True Activist & MSNBC & BBC & Associated Press & The Australian & Fox News & Washington Times \\

     & Left Action & New Yorker  &  New York Times & Voice of America & India Today & Rebel News & CNS News \\
    \hline
    \end{tabular}

\end{adjustbox}

\caption{Examples of media outlets from each political ideology.}
\label{tab:media-example-leaning}
\end{table}

The factuality is determined by a score ranging from 0 - 10 inclusively, where low scores imply strong factuality. The standard used for factual reporting is shown below. Some instances on media of different factuality are shown in the table \ref{tab:media-example-factuality}.

% \fix{Maybe added the overlap between factuality and political ideology?}

\begin{itemize}
    \item Very high (score: 0): the media source is always factual and provides credible information. It makes correct on incorrect information immediately and has never failed a fact check. 
    \item High (score: 1 - 2): the media source is almost always factual.   
    \item Mostly factual (score: 3- 4): the media source is usually factual.   
    \item Mixed (score: 5 - 6): the media does not always use proper sourcing or provide biased/mixed factual sources.
    \item Low (score: 7 - 9): the media rarely uses credible sources and does not provide trustworthy and reliable information. 
    \item Very low (score: 10): the media rarely uses credible sources and does not provide trustworthy and reliable information at all times. 
\end{itemize}

\begin{table}[h!]
\begin{adjustbox}{width=1.2\textwidth,center}

    \centering
    \begin{tabular}{|c|c|c|c|c|c|c|} 
    \hline
    Factuality & Very High & High & Mostly Factual & Mixed & Low & Very Low \\
    \hline
    \multirow{3}{*}{Media Outlet} & Associated Press & ABC News & Whitehouse.gov & CNN & Turning Point USA & Breaking911  \\

    & Reuters & BBC & The Hill & MSNBC & The Truth Voice & We Love Trump \\

    & Pew Research & New York Times & Buzzfeed News & Fox News &  Alpha News & Time Times of America \\
    \hline
    \end{tabular}

\end{adjustbox}

\caption{Examples of media outlets from each factuality category.}
\label{tab:media-example-factuality}
\end{table}

Credibility is a unique measurement provided by MBFC. It is calculated mathematically based on various factors such as political ideology, media traffic, factuality, whether the media source contains conspiracy or questionable content with some additional subjective input. There are three levels of credibility:  

\begin{itemize}
    \item High: Any media source with high or very high factuality. Any media source rated mostly factual with a high or medium amount of traffic.
    \item Mixed: Any media source rated mostly factual with low traffic and left- or right-leaning. Any media source with mixed factuality and high or medium traffic. Any media source with mixed factuality and low traffic but only failed few fact checks.
    \item Low: Any media source with mixed factuality and low traffic but only failed many fact checks. Any media with low or very low factuality. Any media rated as questionable or conspiracy.
\end{itemize}

However, credibility ratings are missing for most media outlets, and hence it will not become the primary metric in the later analysis but rather serve as a reference only. Some examples of media of different credibility are shown in table \ref{tab:media-example-credibility}.

\begin{table}[h!]
\begin{adjustbox}{width=0.6\textwidth,center}

    \centering
    \begin{tabular}{|c|c|c|c|c|c|c|} 
    \hline
    Credibility & High & Mixed & Low  \\
    \hline
    \multirow{3}{*}{Media Outlet} & BBC & CNN & We Love Trump  \\

    & Reuters & The Australian & Alpha News  \\

    & Pew Research & MSNBC & Breaking911 \\
    \hline
    \end{tabular}

\end{adjustbox}

\caption{Examples of media outlets from each credibility category.}
\label{tab:media-example-credibility}
\end{table}

\subsection{Media Twitter Handle Collection}
\label{sec:media-twitter}

\begin{figure*}
  \centering
    \includegraphics[width=0.35\linewidth]{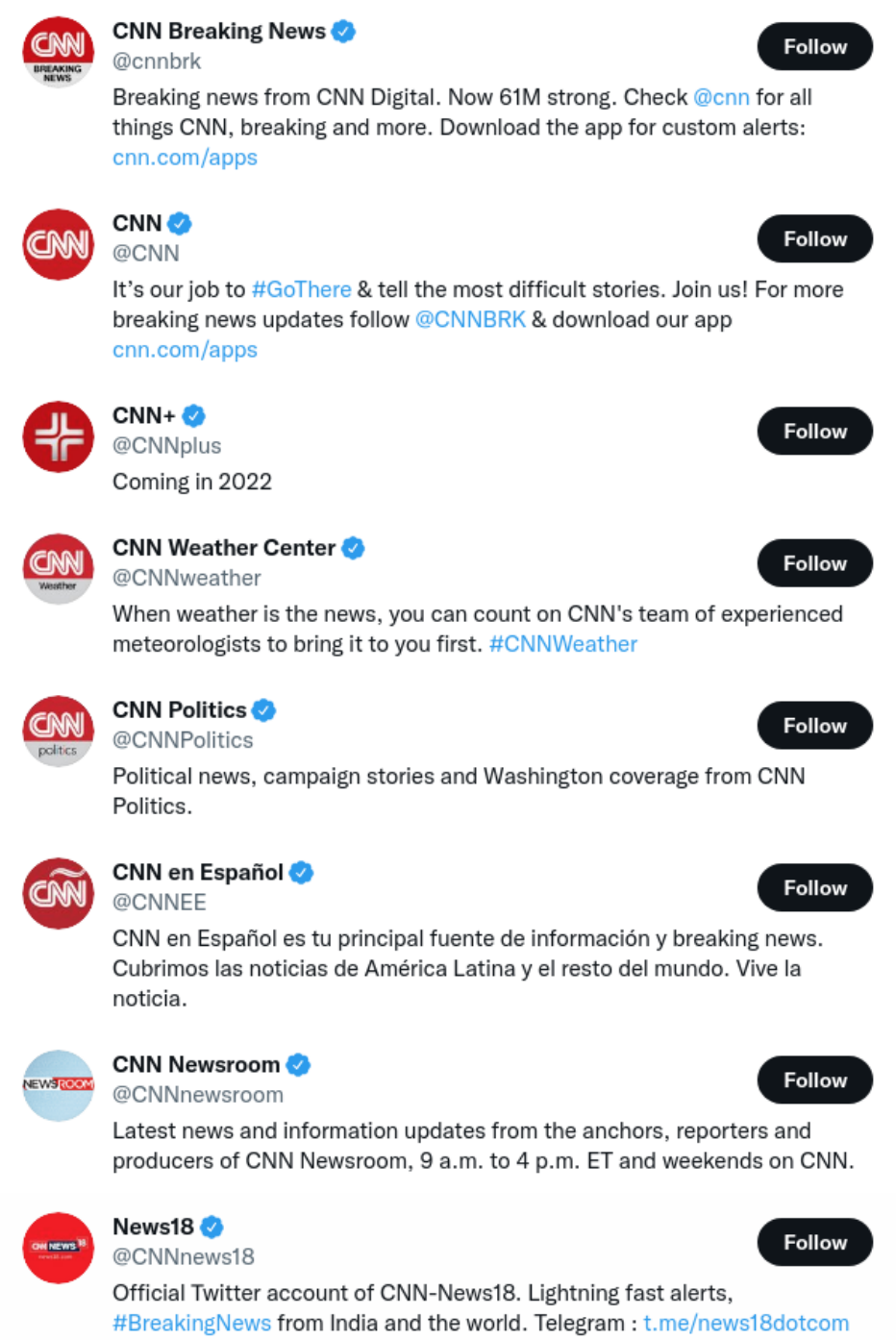}
    \includegraphics[width=0.35\linewidth]{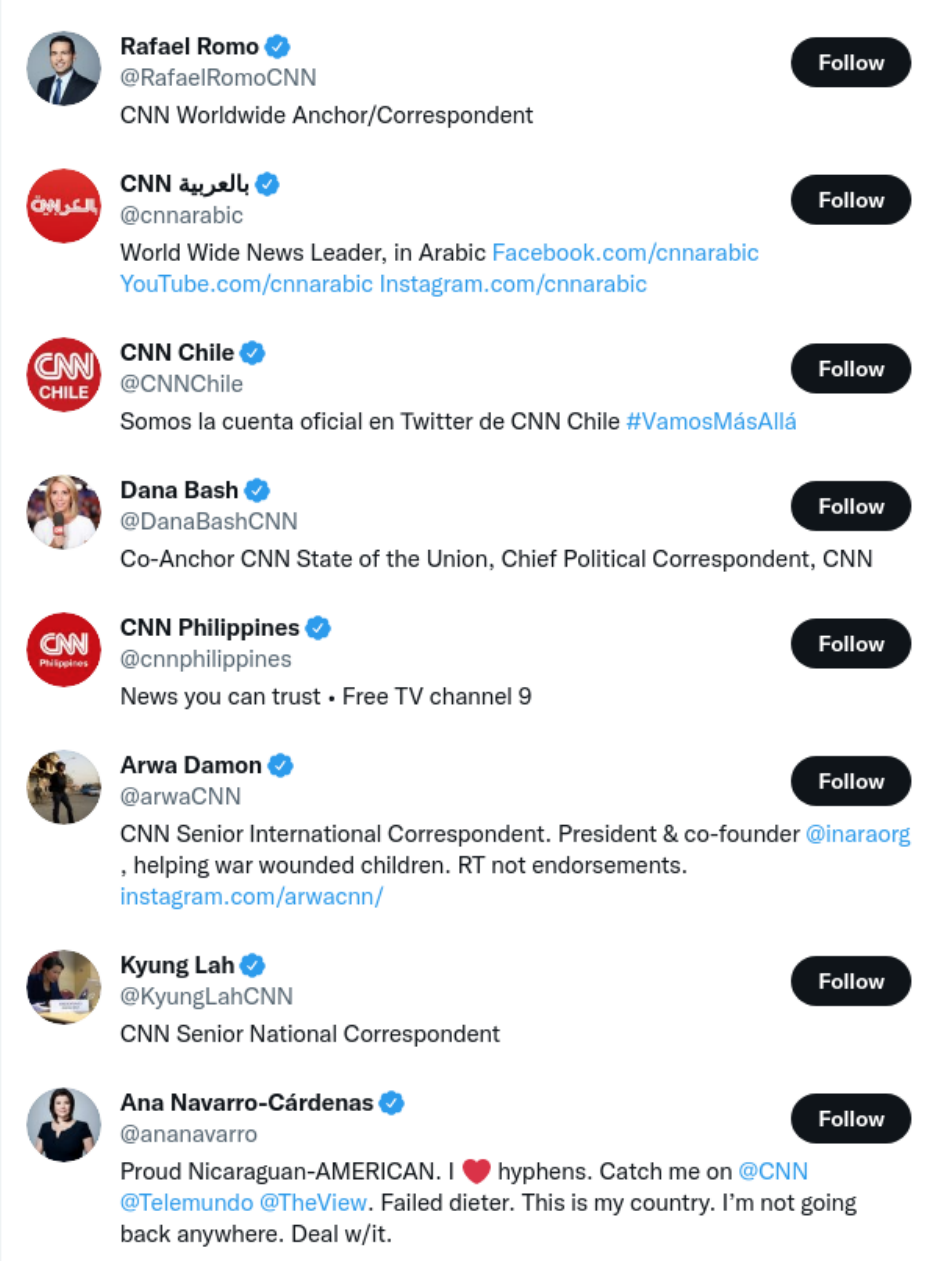}
  \caption{Twitter searching results returned by ``CNN''. The left figure are all related CNN media accounts while the right figure also contains several anchors from CNN. Searching results are subjective to changes due to personalization.}
  \label{fig:cnn-media-anochr}
\end{figure*}

There are \num{1724} media outlets crawled in total from the MBFC website. We then constructed the media Twitter handle dataset based on it. Each media outlet is likely to have multiple Twitter handles. For instance, as shown in figure \ref{fig:cnn-media-anochr}, various Twitter accounts are returned when searching ``CNN", which differs by their designated scopes, languages, and locations. Meanwhile, we also consider anchors of each media relevant such as \textit{Rafael Romo}, \textit{Dana Bash}, \textit{Kyung Lah}, \textit{Ana Navarro-Cárdenas} for CNN, as shown on the right in figure \ref{fig:cnn-media-anochr}.

Due to the existence of various Twitter accounts with very different names for each media outlet, traditional name-matching approaches are very likely to miss some Twitter handles and hence cannot sufficiently capture the relevant accounts. To tackle the problem, we use the URL-matching method instead. In detail, we query the Twitter Search API from \verb|tweepy| with the name of each media outlet. For each of the returned Twitter account, we extract all the URLs contained in its profile. Then we perform a redirection on the URL to resolve the shortened links (known as URL enrichment). We compare the URL provided by MBFC with the redirected URLs and consider the Twitter account relevant once they match. For each media Twitter handle, we set its location to be the same as the location of the query media from the MBFC dataset. Although the contents of some media Twitter accounts are designated for a specific country, we consider them to be located in the same country as the query media with specially tailored topics. For instance, \textit{BBC News Africa} is considered to be a British media but with topics on Africa. URL redirection is, in general time-consuming, and for the sake of saving time, we only consider the top 10 accounts returned by Twitter. In the following parts of this thesis, we by default refer to the media outlet Twitter accounts as media outlets or media.  

\begin{figure*}[!htbp]
  \centering
    \includegraphics[width=1.0\linewidth]{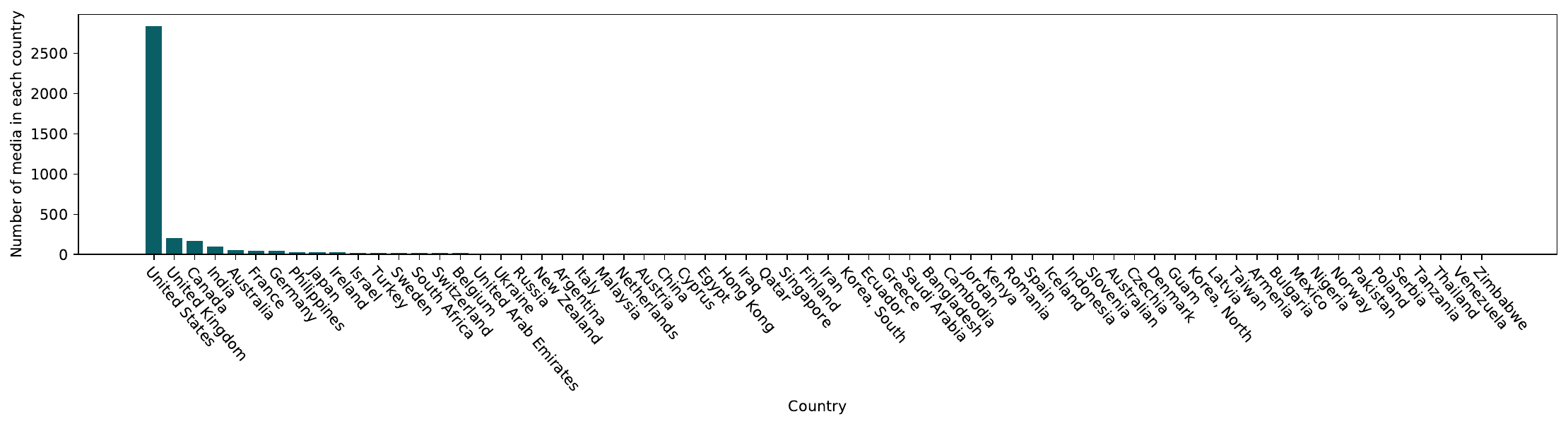}
    \includegraphics[width=0.8\linewidth]{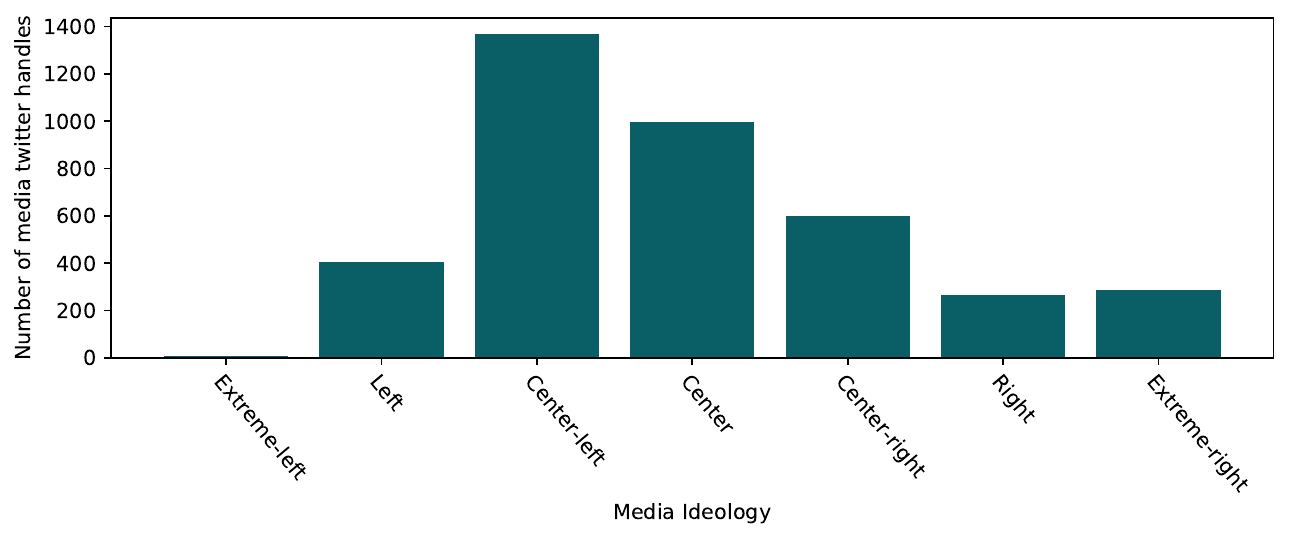}
  \caption{The number of media Twitter handles in each country (top) and ideology (bottom).}
  \label{fig:media-cnt-country-ideo}
\end{figure*}

The distributions of media by country and ideology are shown in figure \ref{fig:media-cnt-country-ideo}. There are \num{3927} Twitter handles found in total. There is an imbalance in terms of the amount of media in each country. There  is a dominant amount of media in the United States compared to others. Moreover, the long tail of the bar plot indicates that the majority of the countries only have one media.

The distributions of media factuality and credibility are shown below. There are \num{3754} media Twitter handles with factuality information while there are \num{1448} with credibility available. Most of the media sources fall into high factuality and credibility. Some media are rated as low factuality and credibility, most of which are right-biased media (center-right, right, and extreme-right).

\begin{figure*}[!htbp]
  \centering
    \includegraphics[width=1.0\linewidth]{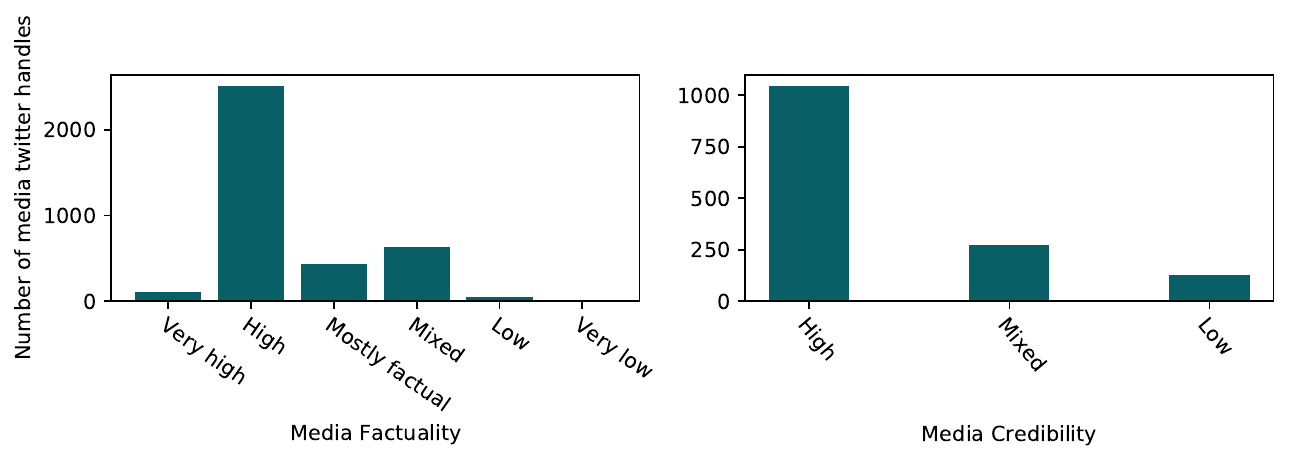}
  \caption{The number of media Twitter handles in each category of factuality (left) and credibility (right) provided by MBFC.}
  \label{fig:media-cnt-country-fact-cred}
\end{figure*}

% It is worth noting that in the following work, \textit{media outlet} and \textit{media Twitter accounts} will be used interchangeably. 

\section{Supplementary Data}
\label{sec:supp-data}

We briefly mention some supplementary datasets used in the project. Compared to the previous datasets, these additional datasets are mainly used to validate our methods or assist with visualization.

In addition to the media dataset above, we also utilize a second media dataset from \cite{Fischer2020AuditingLN} at the initial stage of the project. This dataset contains information on the locations, domains, types (local or international) of media. However, this dataset is mainly limited to United States media with less coverage than the MBFC dataset, which cannot help us analyze global level Twitter activities. We refer to the MBFC media dataset and the media Twitter handle dataset as the default datasets in the remaining chapters when there is no explicit mention.

Although our work aims at analyzing the interactions from a global perspective, it is unfeasible to perform case studies on every possible country. Hence we limit our scope to those countries with a relatively large population and active Twitter users based on the demographics obtained from \cite{simwiki} and \cite{statista}. At the United States level, the presidential elections in 2020 also played a role in validating our approach. The statistics on votes are obtained from \cite{cnnpol}.

\chapter{Interaction Matrix}
\label{cha:interaction-intro}

This chapter goes through the most crucial concept in this project: interactions and how they are extracted and constructed from tweets. Section \ref{sec:inter-define} introduces the definition of interaction and interaction matrix. Following the definition, Section \ref{sec:inter-construct} compares different approaches to build interactions and presents the final method adopted. Section \ref{sec:inter-rep} introduces three representations of interaction matrices used in later analysis.

% \fix{Move this chapter to a section under other chapters?}

\section{Definition}
\label{sec:inter-define}

\textbf{Interaction} can be viewed as a metric to quantify Twitter activities between two entities such as grassroots users, organizations, and media outlets. The interactions are associated with the Twitter activities and target accounts present in each tweet. Specifically, we consider retweets, quotes, replies, mentions, and URL sharing as our targeted Twitter activities. We exclude hashtags due to their potential usage as spamming and hijacking, and we believe very few users interact with media outlets through hashtags. Due to the intrinsicness of our dataset, the interactions here by default refer to COVID-19-related interactions.

% We limit our attention on the Twitter accounts to be the media outlet Twitter accounts mentioned in Chapter \ref{cha:data}.
% \fix{Added the original definition on interactions first - and show it's over counting}

The \textbf{interaction matrix} provides an overall measurement of interactions between two sets of entities (the set can contain only one entity). Let $I$ be the interaction matrix, then $I_{i, j}$ indicates the volume of interactions from a set of entities $i$ and another set of entities $j$. This thesis focuses on interactions between users and media outlets, namely $i$ is a single user and $j$ is a single media outlet. Each row of the interaction matrix ($I_{i, :}$) represents the interactions with all media outlets from a specific user $i$, named as \textbf{media consumption vector}. We name this interaction matrix as \textbf{user-to-media interaction matrix}.

\section{Construction}
\label{sec:inter-construct}

% Interaction, in this case, can be viewed as a metric to quantify Twitter activities. For each tweet, the interactions are determined by the unique number of target Twitter objects (users, media and organizations, etc.), ranging from 0 to 1. For example, if a tweet mentioned one media while replying to a second media, then the interaction received by each media from the poster will be 0.5.

% The \textbf{interaction matrix} provides an overall measurement of interactions between two sets of entities. Let $I$ be the interaction matrix, then $I_{i, j}$ indicates the volume of interactions from a set of entities $i$ and another set of entities $j$. In this thesis, we focus on interactions between users and media outlets, namely $i$ is a single user and $j$ is a single media outlet. And each row of the interaction matrix ($I_{i, :}$) represents the interactions with all media outlets from a specific user $i$, named as \textbf{media consumption vector}.

% namely $i$ is one single user or a set of users from the same state or country and $j$ is a single media outlet or a set of media outlets grouped by some attribute such as their locations or political leaning.

With previously mentioned Twitter activities, it is essential to determine the actual quantifying measurement based on them. We introduce three different quantification schemes below.

\begin{enumerate}
    \item The most straightforward approach is to set the number of interactions to be the same as the number of media Twitter accounts occurring in a tweet from all Twitter activities. For instance, if a tweet quotes a message from \textit{CNN} while mentioning \textit{the BBC Breaking News} in its text, the number of interactions will be 2.
    \item The number of unique countries determines the volume of interactions for a single tweet. For instance, if a tweet quotes a message from \textit{CNN} while mentioning \textit{The New York Times} in its text, the number of interactions will be 1 for both media outlets since they are both from the United States.
    \item The number of interactions for a single tweet is always 1, and there is a weight associated with each media outlet occurring in the tweet. The weight is determined by the number of times its corresponding Twitter handles show up in different Twitter activities. For instance, if a tweet quotes a message from \textit{CNN} while mentioning \textit{BBC Breaking News} and \textit{The New York Times} in its text, the number of interactions on \textit{CNN}, \textit{BBC Breaking News}, and \textit{The New York Times} will be 1/3, respectively.
\end{enumerate}

Our initial analysis began with the first approach. However, this approach is likely to overestimate the actual Twitter activities. First, if a single media Twitter handle shows up multiple times within a tweet, it will receive too many interactions. Second, spamming through mentioning and retweeting will increase the interactions.

To resolve this issue, we consider the latter two approaches. We regard the second scheme can be as a country-oriented method. Although it can mitigate the problems encountered in the initial method, it introduces new issues at the same time. When multiple media Twitter accounts from the same country occur in a single tweet, this method cannot effectively distinguish them since it assigns the same values to their received interactions. 

These reasons motivate us to accept the third approach as our final interactions quantifying measurement. Compared to previous methods, the last method has several advantages. First, it is media-targeted and weighs active media higher, which aligns with our research purpose. Second, it retains country-level information from media outlets. Last, it resolves potential spamming by distributing weights effectively.

\section{Representation}
\label{sec:inter-rep}

% Interactions are directional. According to our definition, each interaction is associated with one \textbf{sender} (the user of the tweet) and one or more \textbf{receivers} (other Twitter accounts being interacted with). By grouping similar senders and receivers according to some criterion, we can construct interaction matrices, where each row represents the interactions of a group of senders towards some receivers, named \textbf{media consumption vector}. 

% By aggregating users and media outlets with similar attributes together, we can have different representations for the user-to-media interaction matrix. In particular, we consider the locations of users and media outlets and the political ideology of media outlets in this thesis. The following subsections go through the user-to-media interaction matrix aggregated in three different ways, sorted by the granularity of information they provide.

% There are some criteria that can be used to group the media consumption vectors from Twitter users with similar attributes together.

The user-to-media interaction matrix could be represented in different forms. The most naive representation is to use the raw interaction matrix directly, that is, each cell $I_{i, j}$ represents the number of interactions from a single user towards a specific media outlet. This is also the most informative representation  as it demonstrates how the interactions change across users. However, due to the large user base in our dataset, this interaction matrix is usually huge and hard to work with directly. In the later chapters, we introduce some user cutoff techniques to reduce the size of the matrix. To avoid confusion on different representations, we name this representation as \textbf{individual-to-media} representation.

In addition, users and media outlets can also be grouped by some attributes. In particular, we consider the locations of users and media outlets and the political ideologies of media outlets in this thesis, which leads us to the following representations.

\textbf{GPE-to-GPE}. Geo-political entities (GPE) refer to countries, states, cities, etc. We have a GPE-to-GPE representation for the user-to-media interaction matrix by aggregating users and media by their locations. By using the same notation before, $I_{i, j}$ represents the volume of interactions from all the users in one location towards all the media outlets within another location, where locations are usually at the same level.

As a high-level overview of user-to-media interactions, GPE-to-GPE representation attempts to provide a highly abstract way of quantifying Twitter activities. As shown in our later analysis, this representation serves as a teaser for the interactions and cannot provide detailed information due to its lack of granular information. Each cell on the interaction matrix only conveys the number of interactions at a state-to-state or country-to-country level but does not distinguish users and media within the same locations.

\textbf{GPE-to-media}. Similarly, the GPE-to-Media representation groups users from the same locations together. What is different from the previous representation is that it will have each media outlet as its columns now, i.e., $I_{i, j}$ represents the number of interactions from users of the same location towards a single media outlet. We sum up all the interactions received by media outlets associated with more than one Twitter account. 

This representation allows us to investigate the number of interactions each media outlet receives in terms of user locations. But consequently, the dimension of the interaction matrix, especially the number of columns, will grow much larger compared to the previous case.

\section{Summary}

This chapter introduces the most fundamental concept used throughout the later analysis: interactions and the user-to-media interaction matrix with different representations. As mentioned at the beginning of the chapter, we can extend the interaction matrix to other types with more datasets. For instance, we could have a user-to-user interaction matrix to quantify the interactions between followers and followees. We could also construct a user-to-government interaction matrix to study the activities between users and government Twitter accounts. We leave these studies as future work.

The following two chapters will divide into the analysis of the user-to-media interaction matrix based on the three representations on the United States level and global level.  %\fix{This section may be redundant.}
\chapter{United States Interactions Analysis}
\label{cha:us-inter}

This chapter studies two representations of the user-to-media interaction matrix on the United States level. Before diving into the analysis, section \ref{sec:us-usercut} presents additional post-processing methods on users to further reduce potential spamming behaviors. Section \ref{sec:sts-analysis} presents state-to-state representation, where users and media are all located in the United States. It also attempts to link the overall political ideologies of states with the information flowing between them. Section \ref{sec:stm-analysis} disaggragates media from states representation back into individual representation and shows the media consumption patterns at the state level. Meanwhile, it will construct the validity of the media consumption vector by correlating online interactions with offline votes, which also provides convincing evidence for our later work at the global level. Section \ref{sec:us-dis} will summarize this chapter and motivate the work on a global level.

\section{User Cutoff}
\label{sec:us-usercut}

\begin{figure*}[!htbp]
  \centering
    \includegraphics[width=1.0\linewidth]{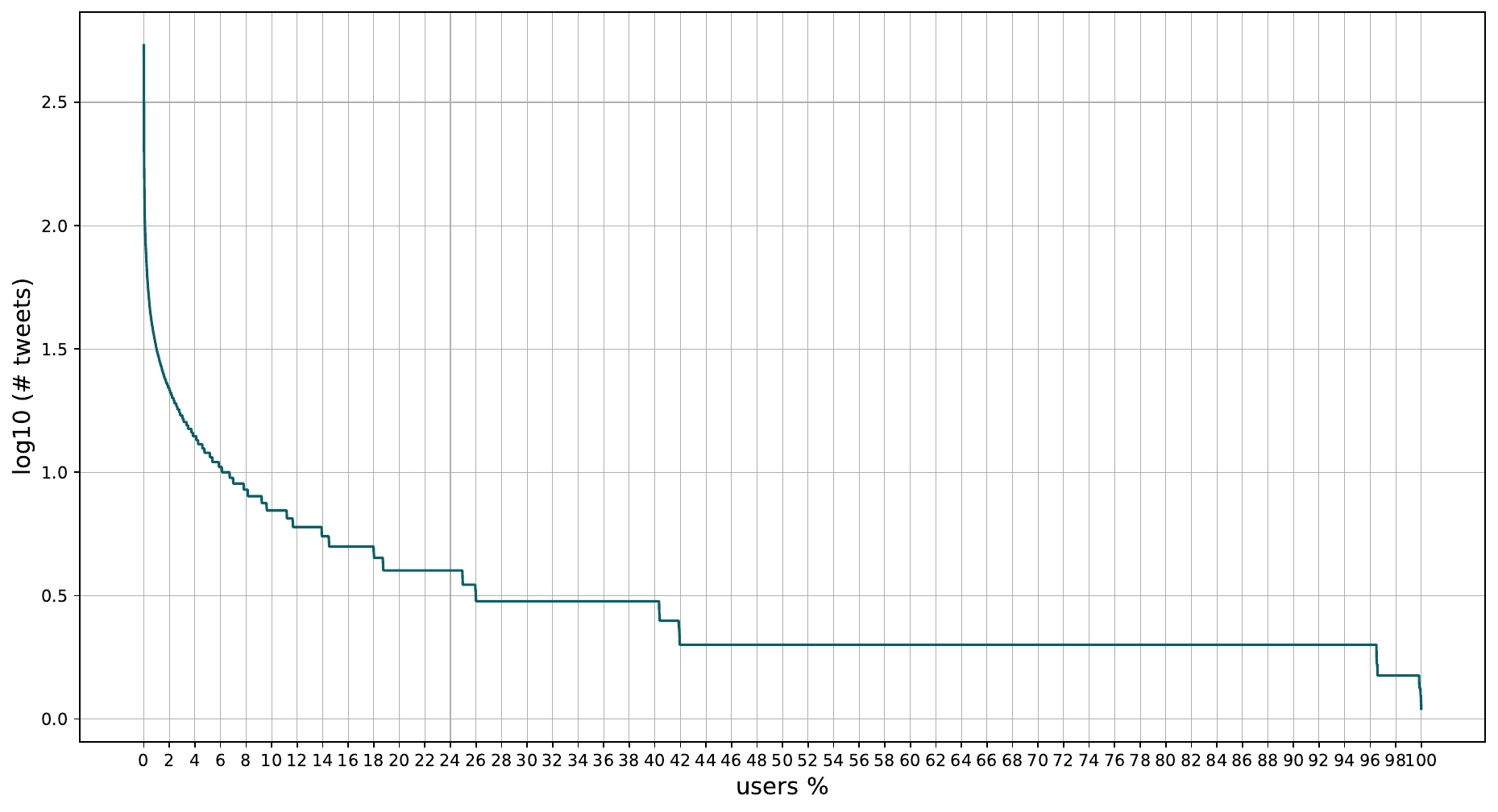}
  \caption{United States User percentiles and their corresponding interactions in log10.}
  \label{fig:us-user-cutoff}
\end{figure*}

Removing hashtags and having weighted interactions can resolve potential spamming within a single tweet effectively. However, if a user repetitively posts tweets with spamming behaviors on specific media outlets (e.g., mentioning a media outlet), the accumulated interactions can still affect our analysis. To tackle this issue, we perform a user cutoff. Figure \ref{fig:us-user-cutoff} shows the change of interactions of each user along with the user percentile. In general, we expect the number of interactions to change smoothly along with the user percentile. However, the figure has indicated a sharp change for the top 2\% users compared to the rest, which implies potential user spamming. We decide to remove users from the top 2\% and proceed with the rest.

\section{State-to-State Representation}
\label{sec:sts-analysis}

As mentioned in chapter \ref{cha:interaction-intro}, state-to-state representation aims to provide an overview of the interactions between United States-located users with their local media. However, the lack of state-level information from MBFC has made it difficult to aggregate United States media outlets by their states. We additionally make use of the media dataset (\cite{Fischer2020AuditingLN}) mentioned in chapter \ref{cha:data} section \ref{sec:supp-data} to tag media outlets with state information. However, the lack of coverage discussed earlier will reduce the number of media outlets a lot. We consider the following results as a teaser for the actual interactions.

\begin{figure*}[!htbp]
  \centering
    \includegraphics[width=1.0\linewidth]{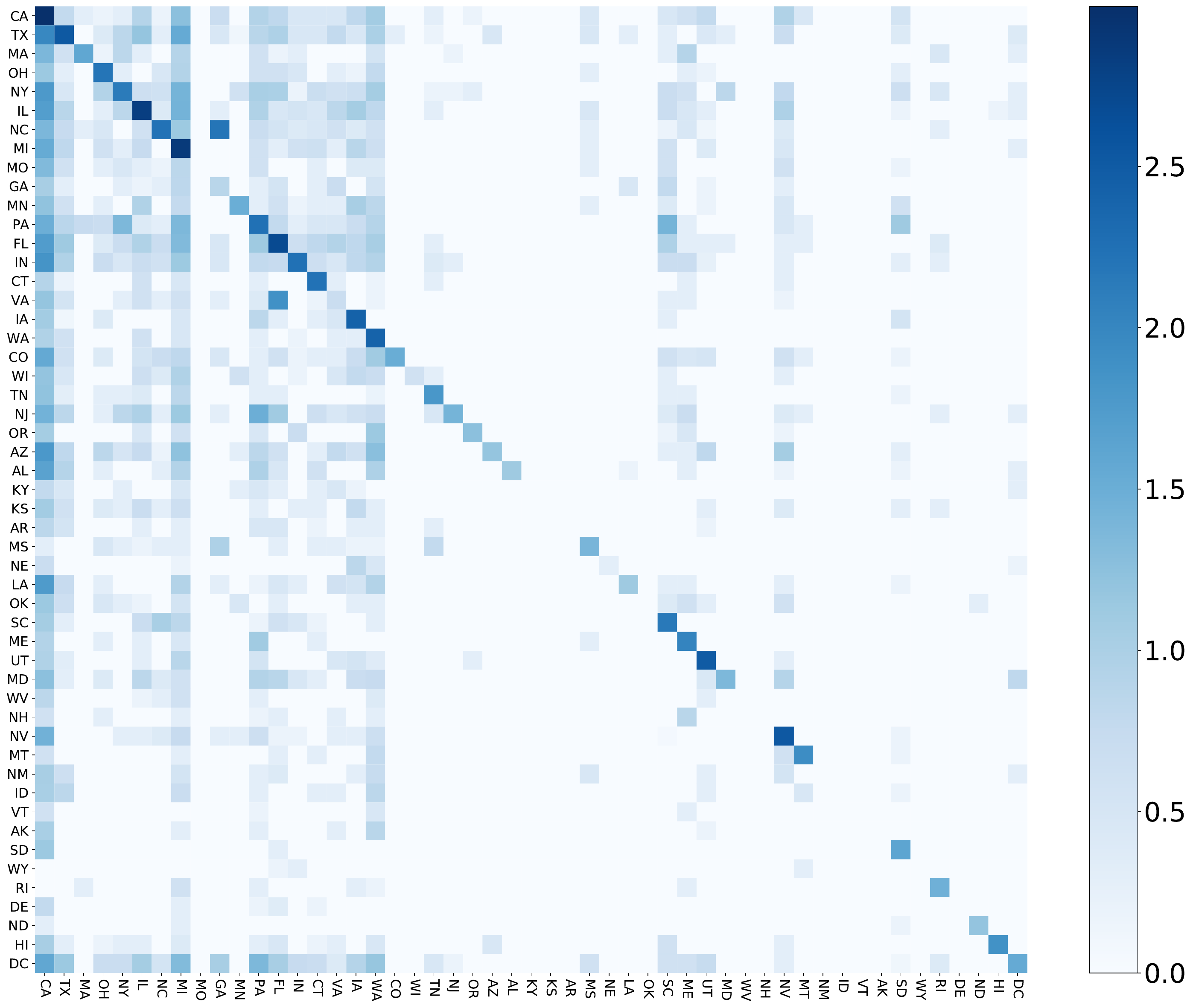}
  \caption{State-to-state representation of the interaction matrix, where each cell represents log10 of the interactions. Each row represents the volume of interactions from users of a state towards media of different states.}
  \label{fig:state-to-state}
\end{figure*}

Figure \ref{fig:state-to-state} displays a heatmap visualization on the state-to-state representations of the user-to-media interaction matrix, where we take log10 over the interactions on each cell for better visualization and comparison. The shallow colored cells in the middle of the heatmap are likely to be caused by the lack of media coverage in the corresponding states. Note that we do not explicitly distinguish the District of Columbia (DC) from other states and treat them the same in the subsequent context.

It is clear that the diagonal is darker than the rest cells, indicating that users interact more with media outlets within the same state. It makes sense due to the geographical proximity. There are also columns with a darker color, implying media from the corresponding states receiving lots of interactions from users across the United States such as California (CA), Texas (TX), Michigan (MI), Pennsylvania (PA), and Massachusetts (MA) due to both its population and media amount. This also provides a partial answer to RQ1.

\subsection{Information Flow and Political Ideology}

We consider the interactions with media as an exchange of information across users and media outlets. To better understand how the information is exchanged across different states, we further analyze the information flow from a spatial perspective. There are states where users actively consume media outlets from other states and states where media outlets supply lots of information to users in other states. The consumption over information can be quantified through the row sum of the interaction matrix while the supply can be calculated by the column sum. Information flow cares more about the consumption-supply relation across states and hence we remove the self-interactions from the interaction matrix by setting the values on the diagonal to 0.

We define the information flow ratio as the logarithm of the volume of interactions consumption from users over the logarithm of interactions supplied by media. If the ratio is larger than 0, users consume more information than the amount of information supplied by media. Otherwise, the media supply more information.

We assume there is a pattern associated with the ratio and the political ideology of each state. For example, most Republican states may have a positive ratio while most Democratic states may have a negative ratio. We focus our attention on the three presidential elections in 2012, 2016, and 2020 to select Democratic, Republican states. We also select swing states based on the alignment of the voting between the two elections.

\begin{figure*}[!htbp]
  \centering
    \includegraphics[width=0.75\linewidth]{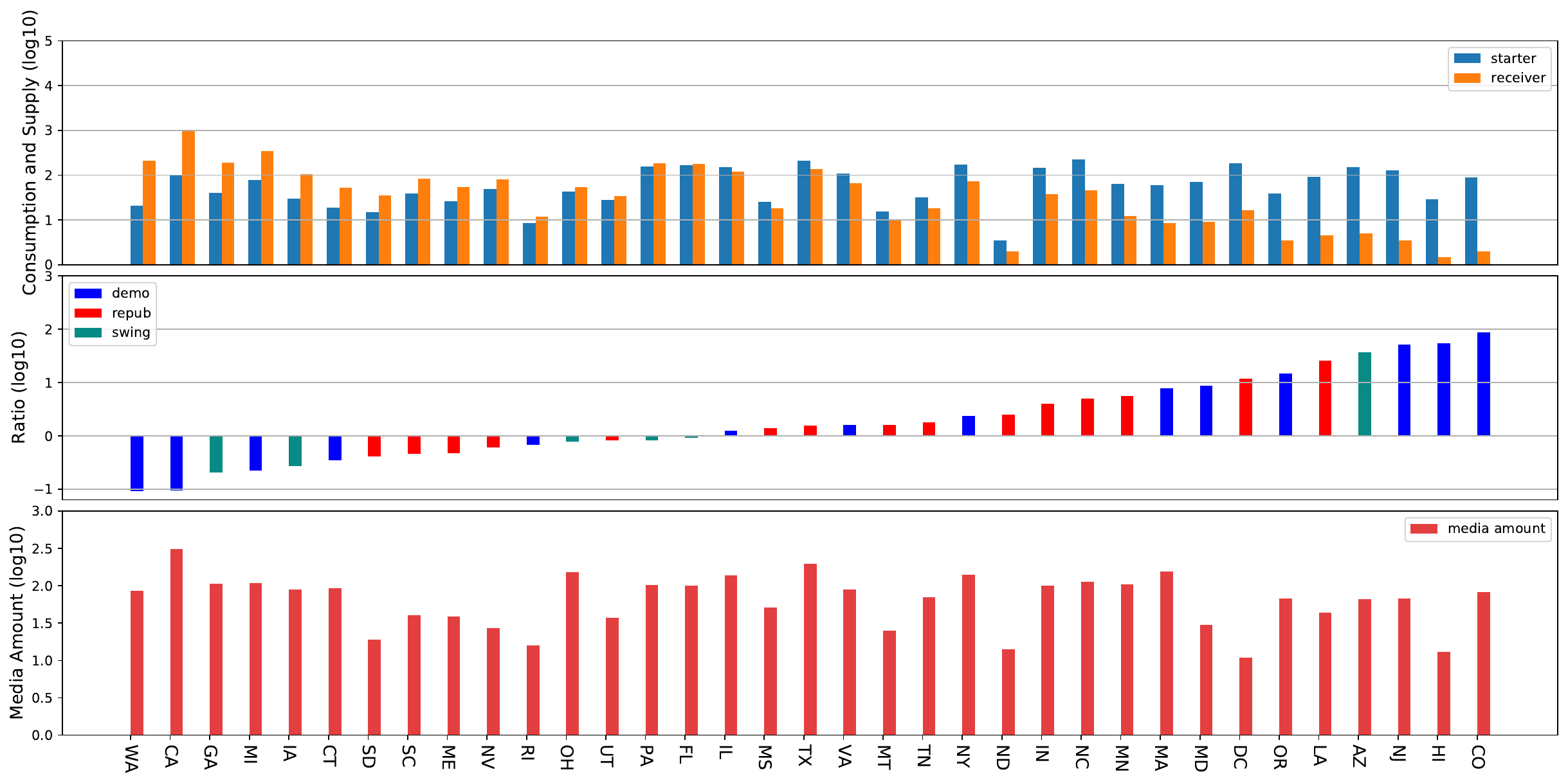}
  \caption{Information flow based on 2012 and 2020 presidential elections.}
  \label{fig:info-flow-US-1}
\end{figure*}

\begin{figure*}[!htbp]
  \centering
    \includegraphics[width=0.75\linewidth]{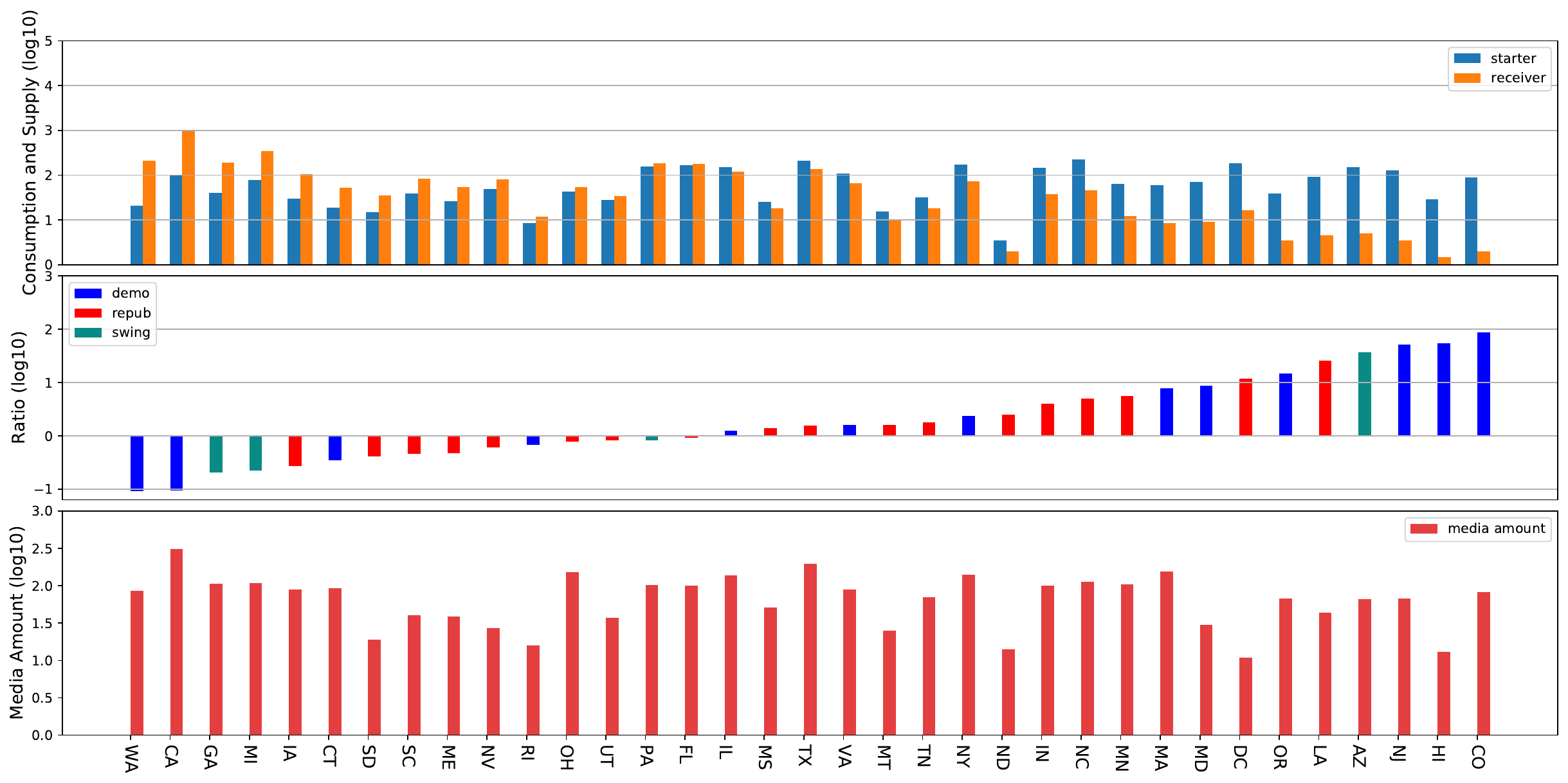}
  \caption{Information flow based on 2016 and 2020 presidential elections.}
  \label{fig:info-flow-US-2}
\end{figure*}

Figure \ref{fig:info-flow-US-1} and \ref{fig:info-flow-US-2} show the number of interactions consumed by users and supplied by media in each state (top) and the corresponding ratio (middle) along with the amount of media in each present state (bottom). The figures have been re-ordered by the ratio from small to large. To assist with visualization, bars representing Democratic and Republican states have colored blue and red respectively. Swing states have been colored green. Not all states are present in the figures above due to the lack of media coverage in some states.

Unfortunately, different from our assumption, for the 35 present states we did not observe any meaningful pattern associated with their the political leanings. There are both Democratic and Republican states with negative and positive ratios. Moreover, it is also clear that these ratios do not correlate with the amount of media in each state. This has forced us to re-assess our approach, which has revealed the following issues:

\begin{enumerate}
    \item The most straightforward reason, again, is due to the poor coverage from the state-level media dataset, which does not sufficiently support us to find meaningful results.
    \item Our definition of information flow along with its calculations need to be further refined. The information flow is defined in an asymmetric manner, i.e. the consumption is defined as user behaviors while the supplies are defined regarding media behaviors. However, there are other potential information flows that are not captured in this way such as (a) information flowing from media back to users; (b) interactions between users from two states; and (c) interactions between media from two states. The incompleteness in the definition has made the results not representative and prone to failure.
\end{enumerate}

However, our existing datasets do not support us to consider all potential information flow and hence we must move onto a more granular level of analysis.

\section{State-to-Media Representation}
\label{sec:stm-analysis}

The previous highly aggregated representation attempts to quantify the information flow and investigate potential patterns of it, but with no luck, it fails with existing data. It has also motivated us to conduct studies on media outlets directly, that is, the state-to-media representation on the user-to-media interaction matrix.

\begin{landscape}
\begin{figure}%[!htbp]
  \centering
    \makebox[0pt]{\includegraphics[width=1.75\textwidth]{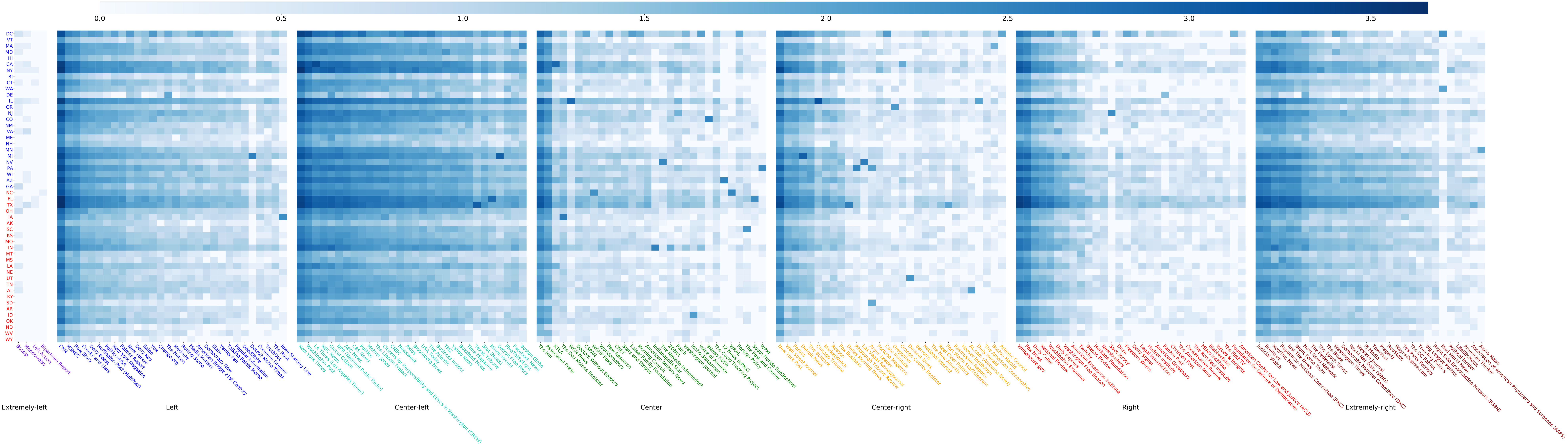}}
  \caption{State-to-media outlet interaction matrix in log10 scale. Top 30 media outlets of each leaning based on their interactions received have been displayed except extreme-left leaning. Starting from leftmost: extreme-left, left, center-left, center, center-right, right, extreme-right. States have been colored by their ideology in blue and red.}
  \label{fig:st-to-med}
\end{figure}\end{landscape}

Due to the high dimensionality of the state-to-media representation, we only take the top 30 media outlets from  each leaning based on the number of interactions they receive for visualization in figure \ref{fig:st-to-med}. Note that there are only 4 media outlets of extreme-left leaning due to its small media base. Similarly, we take log10 transformation of the data for better visualization.

Meanwhile, each row has been re-ordered according to the voting results in 2020. Rows in top represent states with a higher ratio of votes towards the Democratic Party such as the District of Columbia, Vermont (VT), Massachusetts (MA), etc. Rows in the bottom represent states with a higher ratio of votes towards the Republican Party (in other words, lower votes towards the Democratic Party) such as Wyoming (WY), West Virginia (WV), North Dakota (ND). Rows in the middle represent states with a close amount of votes towards both parties. The state names have been colored in blue and red to be distinguishable.

% \newpage
\newgeometry{hmargin=3cm,vmargin=5cm}
\thispagestyle{lscape}
\pagestyle{headings}

\restoregeometry
\pagestyle{headings}

The figure has revealed some interesting findings: 

\begin{enumerate}
    \item Media outlets of left and center-left leaning receive more interactions than others in general. This phenomenon can be attributed to (a) there being more left-biased media outlets in our dataset; (2) the majority of the underlying Twitter population have been identified as Democrats, as pointed out by \cite{doi:10.1177/2053168017720008, pewsize, pewdiff}.
    \item Meanwhile, unlike extreme-left media outlets, media outlets of extreme-right leaning also receive a lot of interactions, which indicates the far-right media contents on Twitter attract more audience compared to center-right and right media. This also reveals some potential user consumption habits for right-leaning users, as pointed out by \cite{Freelon2020FalseEO}.
    \item Another difference between left- and right-biased media outlets lies in their received interaction patterns. For interactions received by media of center-left- and left-leaning, its log-transformed values generally indicate a smooth decreasing tendency. On the contrary, the interactions received by right-biased media outlets have indicated a less smooth change. The very top media outlets received more interactions compared to the rest.
    \item The volume of Interactions is correlated with the population of each state, which is within our expectation States like California, New York, Texas, Florida have a large population base which implies potential more Twitter users. On the contrary, states like Delaware and Wyoming have much smaller population sizes, and hence the color is much shallower.
    % \item Purely based on the heatmap visualization, it seems there are correlations between 
    \item Furthermore, we can also find some scattered dark cells, which are likely to be the interactions from users in a state with local media. For instance, users from Michigan (MI) with Detroit Metro Times; users from Texas (TX) with Texas Tribune; users from Colorado (CO) with 9news KUSA; users from Oregon (OR) with Oregonian; users from New Jersey (NJ) with Save Jersey.
\end{enumerate}

\subsection{From Online Consumption to Offline Votes}
\label{sec:online-offline}

We did not answer in the previous section whether users in left- or right-leaning states consume more content from left- or right-biased media outlets, respectively. More formally, is there a correlation between the political ideology of states and consumption of users on media of different leanings within each state? 

We believe users' online behaviors can reflect offline political preference. Specially, we choose the United States presidential election in 2020 as this offline signal due to its proximity to the period covered by our dataset. We only focus on the votes towards either Democratic Party or Republican Party and ignore the rest parties, such as Libertarian Party, Green Party, etc, due to they received a non-significant amount of votes compared to the two major parties.

% hence it is important to verify the assumption by validating the media consumption vectors. Hence we assume there exist some correlation between their interactions with offline signals. 

% The first step with the media consumption vector is to construct their validity. We believe users' online behaviors, in this case interactions with media outlets on Twitter, can reflect the offline political preference. Hence we assume there exist some correlation between their interactions with offline signals. 

Linking media consumption vectors with offline votes is modelled as a regression task. We use coefficient of determination (\cite{wikir2, skr2}), also known as $R^2$, as evaluation metric, which is defined as $R^2(Y, \tilde{Y} ) = 1 - \frac{ \sum_{i=0}^n (y_i - \tilde{y}_i)^2 }{ \sum_{i=0}^n (y_i - \bar{y}_i)^2 }$, where $y_i$ and $\tilde{y}_i$ are the real values and predicated for sample $i$ and $\bar{y}_i$ is the sample mean. $R^2$ has a maximum value of 1 and it can be interpreted as the proportion of the variance that can be explained by the regression model.

The training samples to the regressor, in this case, is the state-level media consumption vectors and hence the data size is the same as the number of states. The target values are set to be the percentage of votes towards the Democratic Party.  We train the regressor through 5-fold cross-validation to avoid overfitting.

% We further process it by summing interactions with media of similar leaning together, producing left, center and right leanings. We leave the number of media chosen from each leaning as a hyperparameter.

Our initial attempts began with a simple ordinary least squares linear regression model (\cite{Gray2002IntroductionTL, sklr}) based on 70 media from each leaning. We tested a set of regularisation parameters to avoid overfitting, ranging from $10^{-5}$ to $10$. The highest average $R^2$ obtained on the testing data through 5-fold cross-validation is less than 0.3. The score did not increase when the number of media selected changed. We think linear regression is not expressive enough to capture the general relations between media consumption and votes and hence we decide to choose a more complicated regressor.

Specifically, we tested the following three ensemble methods, along with media outlets from three different sets of media: United States media outlets including both nation-wise and state-wise media, international media outlets, state-wise local media outlets in the United States. The results have been displayed in table \ref{tab:r2-regression}.

\begin{itemize}
    \item Gradient Boosting regressor (\cite{Friedman2002StochasticGB, Friedman2001GreedyFA, skgrad}) starts with a dummy regressor such as constant values or mean values over training samples, it fits a set of weaker learners like decision tree regressor on training samples and gradients obtained based on a model from the previous step to construct an estimator iteratively.
    \item Ada Boost regressor (\cite{Freund1995ADG}) is a meta-regressor that fits a regressor on the input dataset and then fits multiple copies of the regressors on the same dataset with adjusted weights on samples according to the errors at each prediction. Specifically, we choose the algorithm from \cite{Drucker1997ImprovingRU} provided by Scikit-learn (\cite{skada}) that uses decision tree (\cite{Breiman1983ClassificationAR}) as a base estimator.
    \item Random Forest regressor (\cite{Breiman2004RandomF, skrand}) is a meta-regressor that fits a set of decision trees (\cite{Breiman1983ClassificationAR}) on different sub-samples of the original dataset and uses average predictions from each tree to improve the final predictive performance.
\end{itemize}

\begin{table*}[!htbp]
%\begin{adjustbox}{width=\textwidth,center}
    \centering
    \begin{tabular}{|c||c||c||c|} 
    \hline
     & Gradient Boosting & Ada Boost & Random Forest  \\
    \hline
    U.S. Media - 20 & 0.6146 & 0.6106 & 0.5800 \\
    \hline 
    U.S. Media - 30 & 0.5392 & 0.6305 & 0.6026  \\
    \hline 
    U.S. Media - 50 & 0.4954 & 0.6679 & 0.5860  \\
    \hline 
    U.S. Media - 70 & 0.5371 & 0.6584 & 0.5746  \\
    \hline
    U.S. Media - 100 & 0.6484 & 0.6623 & 0.6318  \\
    \hline
    U.S. Media - 120 & \textbf{0.7098} & 0.6813 & 0.6658  \\
    \hline
    \hline
    International Media - 20 & 0.5881 & 0.6606 & 0.5858 \\
    \hline
    International Media - 30 & 0.6177 & 0.6245 & 0.5943 \\
    \hline
    International Media - 50 & 0.5938 & 0.6660 & 0.5976 \\
    \hline
    International Media - 70 & 0.6488 & 0.6648 & 0.6396 \\
    \hline
    International Media - 100 & 0.6795 & 0.6717 & 0.6275 \\
    \hline
    International Media - 120 & 0.6641 & 0.6465 & 0.6259 \\
    \hline
    \hline
    Local U.S. Media - 5 & 0.2631 & 0.3150 & 0.3318   \\
    \hline
    Local U.S. Media - 10 & 0.2383 & 0.3068 & 0.3162 \\
    \hline
    Local U.S. Media - 20 & 0.1406 & 0.3443 & 0.3110  \\
    \hline
    \end{tabular}
%\end{adjustbox}
\caption{$R^2$ on votes predictions by using different regressors and media sets. U.S. Media represents all national media outlets and state-level local media in the United States. International media represents all the media outlets across the world. Local U.S. media represents only state-level local media in the United States.}
\label{tab:r2-regression}
\end{table*}

\begin{figure*}[!htbp]
  \centering
    \includegraphics[width=0.75\linewidth]{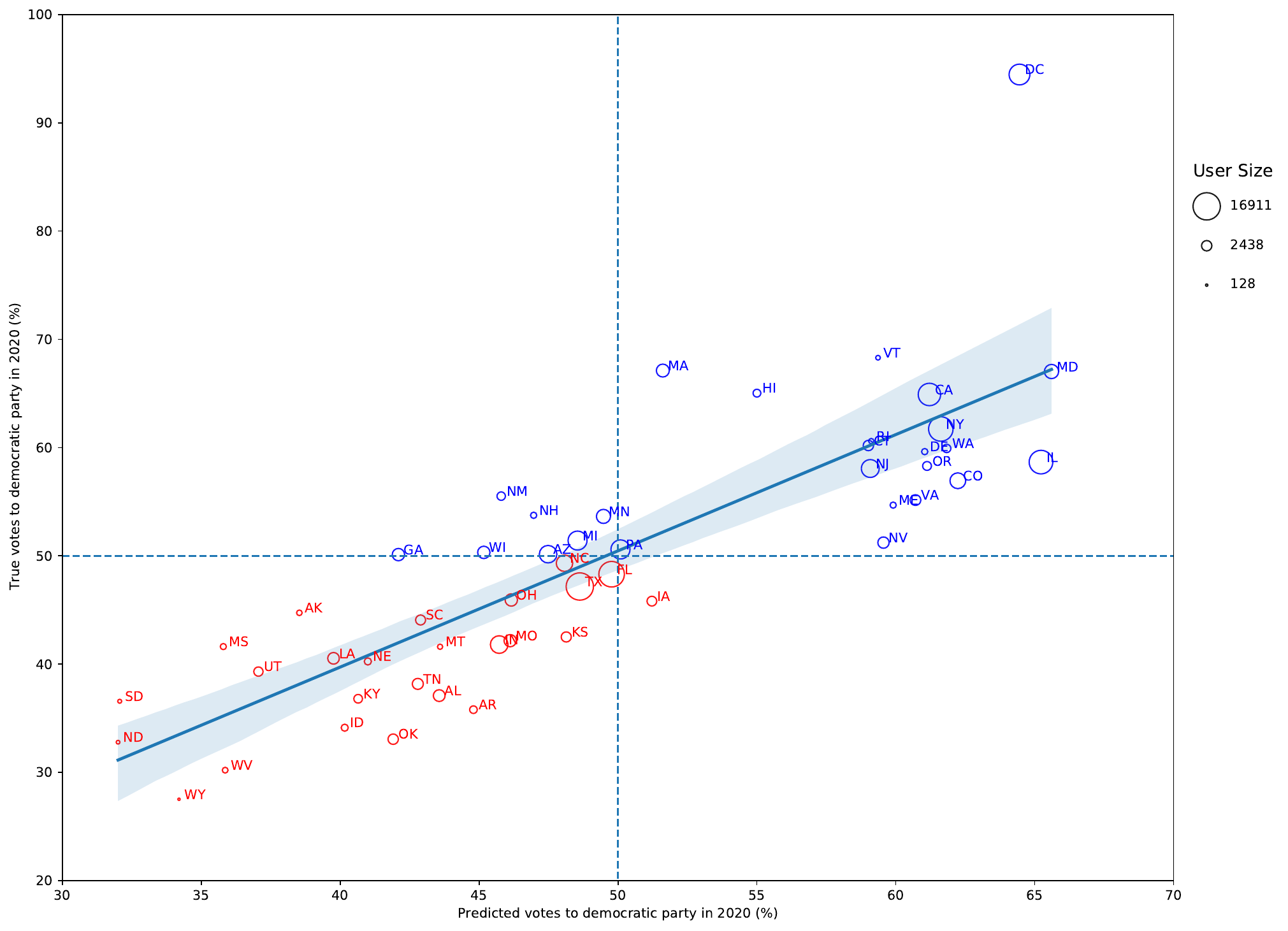}
  \caption{Correlation plot on predicted votes and real votes with a 95\% confidence interval. States are colored by the 2020 United States presidential election results. Each state is represented by a bubble proportional to its user size.}
  \label{fig:corr-plot}
\end{figure*}

The highest $R^2$ obtained through the three regressors and different sets of media outlets is 0.7098 through Gradient Boosting Regressor. Figure \ref{fig:corr-plot} visualizes the correlations based on the predicated votes produced by the regressor and the actual votes with a 95\% confidence interval. The horizontal and vertical lines indicate 50\% votes towards the Democratic Party. Values higher than the threshold (higher than the horizontal line or on the right on the vertical line) imply more actual or predicated votes towards it. The size of the bubble represents the number of users from each state.

The final $R^2$ score and the regression plot have shown the correlation between online media consumption on Twitter and offline voting behaviors. The results have verified our assumption and further constructed the validity of the media consumption vectors from the interaction matrix. It has provided a solid foundation for our later work and answered RQ2.

\section{Summary}
\label{sec:us-dis}

This chapter offers a first glance at the actual interaction matrix constructed from the 5-day Twitter dataset. Starting with the most high-level representation of the United States user-to-media interaction matrix, it demonstrates that people tend to interact with media outlets on Twitter that share geographical proximity to them. Meanwhile, this chapter also attempts to capture the information flow between states but fails due to the coarse approach and a lack of datasets. Meanwhile, in this chapter, we have also linked the online user activities and the offline political preference by demonstrating the high correlation between predicted votes from online Twitter media consumption vectors with offline 2020 Presidential election results. We believe this has also validated our definition and construction on Twitter interactions, making it possible to generalize onto research where political preference has been involved. The prediction could also be used in scenarios where there is a lack of offline surveys.

% User-to-media interaction matrix on United States level is not studied in this chapter. We instead analyze the global interaction matrix in the next chapter rather than separate a particular country. The motivation is that studying the global interaction matrix as a whole is likely to reveal underlying relations across countries that are lost when zooming onto a specific country.

The individual-to-media representation on the United States level is not studied in this chapter. We instead analyze the representation of the global user-to-media interaction matrix in the next chapter. The motivation is that analyzing the representation from a global perspective is more likely to reveal underlying relations across countries, which would have been lost when zooming onto a specific country.

\chapter{Global Interactions Analysis}
\label{cha:global-inter}

This chapter shifts attention from the United States level interactions to global level interactions. We conduct clustering on the user-to-media interaction matrix to depict some potential user groups found through the 5-day dataset. Meanwhile, the clustering results on media consumption of specific country pairs have revealed the non-alignment of user behaviors on cross-country interactions.

% and user behaviors on cross-country consumption. It also demonstrates the underlying difference between political spectrum of United States and others by showing that the consumption patterns are 

The chapter is structured as follows. The first three sections are echos to the work done on the United States level. Section \ref{sec:glo-usercut} reviews the cutoff on user interactions that have been done in the previous chapter. Then, section \ref{sec:ctc-analysis} shows the interaction patterns in country-to-country representation. As a mirror work to the state-to-media representation from the last chapter, section \ref{sec:ctc-analysis} shows the media consumption patterns from users of different countries through country-to-media representation. Section \ref{sec:utm-analysis} provides very detailed analyses on individual-to-media representation, including identification of user groups from a global perspective and users' media consumption behaviors across countries.

% Meanwhile, it will also go through the abortive study on linking interactions with the demographics of a country.

\section{User Cutoff}
\label{sec:glo-usercut}

\begin{figure*}[!htbp]
  \centering
    \includegraphics[width=1.0\linewidth]{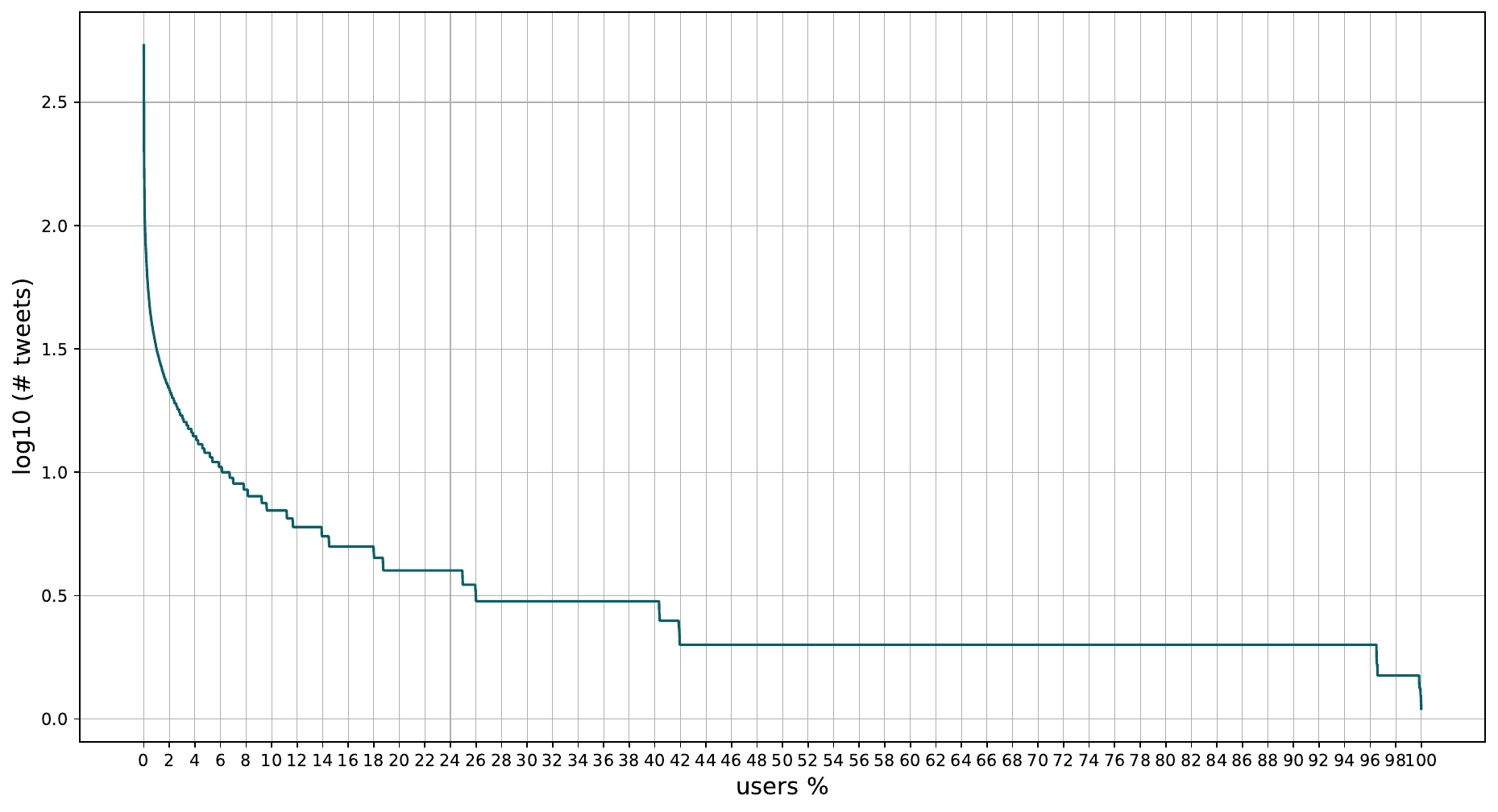}
  \caption{Global User percentiles to the corresponding interactions in the log10 scale.}
  \label{fig:glo-cutoff}
\end{figure*}

Due to the reason mentioned in section \ref{sec:us-usercut} of chapter \ref{cha:us-inter}, we perform the cutoff on the users across all the available countries once again. Figure \ref{fig:glo-cutoff} displays the change of interactions in the log10 scale along with the percentile of users. Similar to the previous case, the change on the curve for the top 2\% users is more rapid compared to the rest and hence we remove the top 2\% users from the overall user base.

\section{Country-to-Country Representation}
\label{sec:ctc-analysis}

\begin{figure*}[!htbp]
  \centering
    \includegraphics[width=1.0\linewidth]{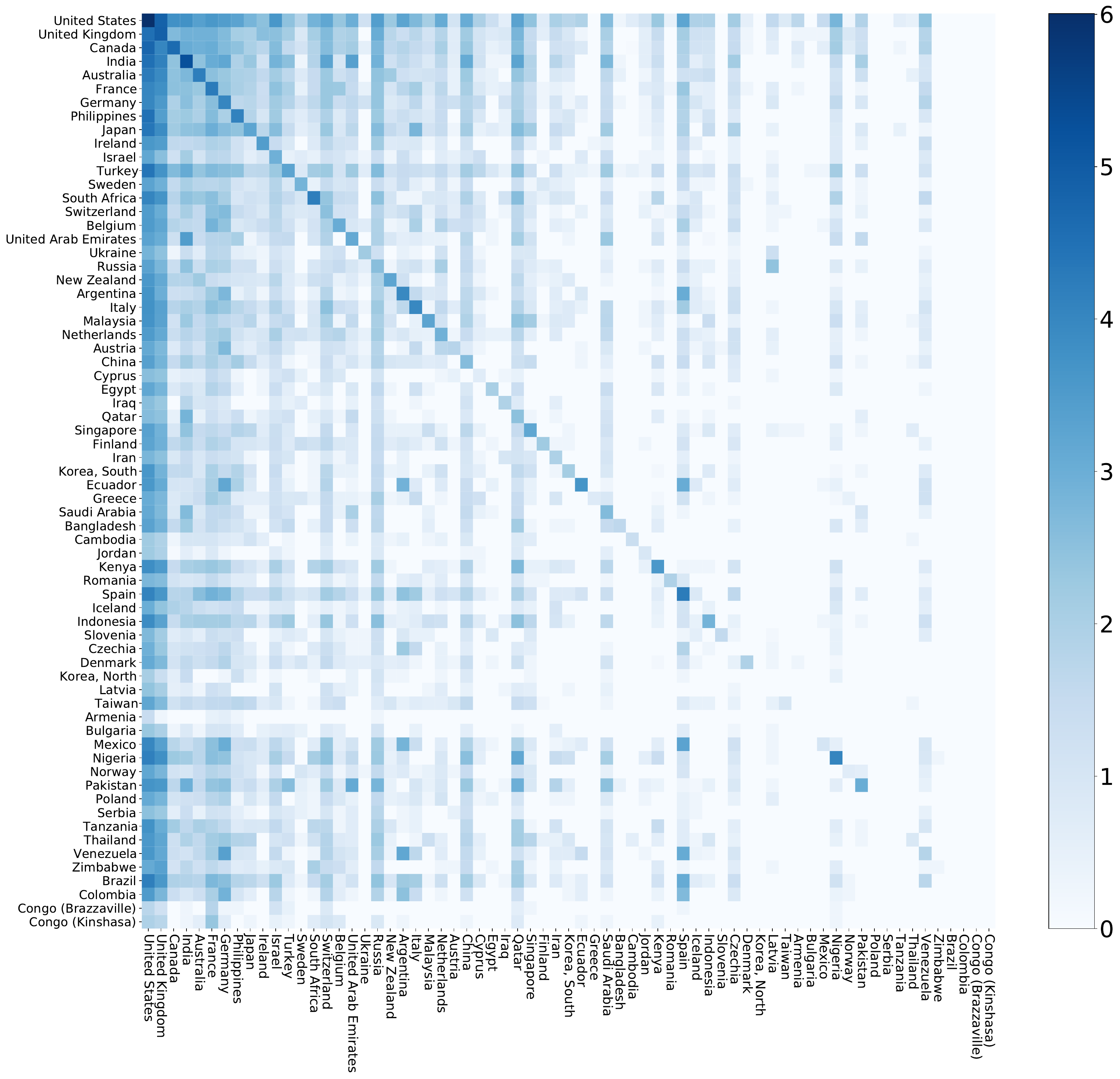}
  \caption{Country-to-country interaction matrix, where each cell is in the log10 scale. Each row represents the volume of interactions from users of a country towards media of different countries.}
  \label{fig:country-to-country}
\end{figure*}

The heatmap visualization of country-to-country representation in the log10 scale is displayed in figure \ref{fig:country-to-country}. We only select a subset of countries based on the data from chapter \ref{cha:data} section \ref{sec:supp-data} to construct the interaction matrix due to not having enough interactions from users in the remaining countries. Nevertheless, there are still cells with very shallow colors in the heatmap due to the low interactions volume.

Similar to the previous chapter, we can observe that most interactions happen from users to media within their own countries due to their closeness geographically. The attention received by media outlets in each country is affected by the number of media outlets in the corresponding countries, which explains why the leftmost five columns have darker colors than the rest.

% \subsection{Online Consumption and Offline Demographics}

% \begin{figure*}[!htbp]
%   \centering
%     \includegraphics[width=1.0\linewidth]{figs/information-flow/global_sender_receiver_media.pdf}
%   \caption{Information flow on countries.}
%   \label{fig:state-to-state}
% \end{figure*}

% This work is conducted in parallel to the study on information flow of United States level. There is no consistent voting data to use when it comes to global perspective. Instead of predicting the offline political preference, the other assumption we believe is that the online media consumption in a country can reflect the demographics of the country to some extent. For instance, the consumption on Spanish media outlets in United States may reflect there is a significant amount of Spanish population. 

\section{Country-to-Media Representation}
\label{sec:ctm-analysis}%\fix{This section may have to be dropped.}

\begin{landscape}

\begin{figure}%[!htbp]
  \centering
    \makebox[0pt]{\includegraphics[width=1.75\textwidth]{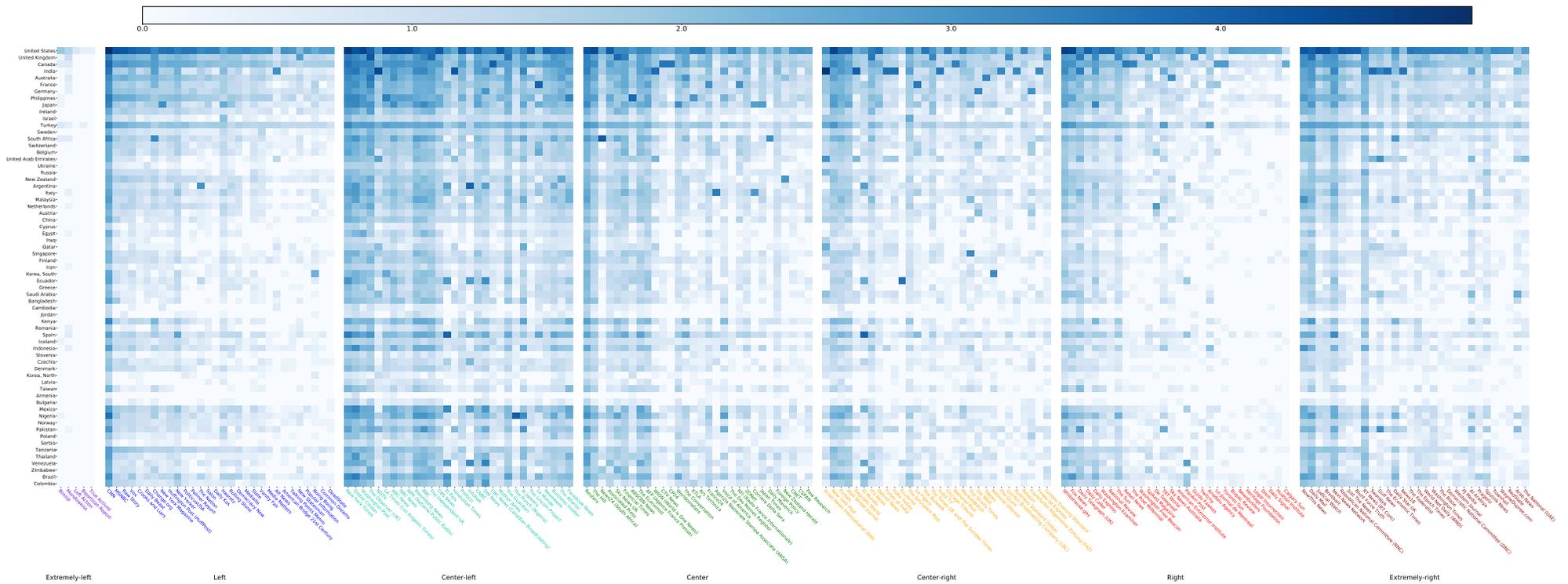}}
  \caption{Country-to-media outlet interaction matrix in log10 scale. Top 30 media outlets of each leaning based on their interactions received have been displayed except extreme-left leaning. Starting from leftmost: extreme-left, Left, center-left, Center, center-right, Right, extreme-right. States have been colored by their ideologies in blue and red.}
  \label{fig:ct-to-med}
\end{figure}\end{landscape}

The interactions aggregated from each country to media outlets have been shown in figure \ref{fig:ct-to-med}. Similarly, the top 30 media outlets based on their interactions received have been chosen for display except extreme-left leaning. 

% Top 15 countries where users have most interactions towards media outlets have shown below.

% \newpage
\newgeometry{hmargin=3cm,vmargin=5cm}
\thispagestyle{lscape}
\pagestyle{headings}

\restoregeometry
\pagestyle{headings}

There are some general findings spotted from the figure. First, left-biased media outlets seem to be more popular than the right-biased and hence receive more interactions across the world. Second, consumption on left-biased media outlets is imbalanced, mostly towards center-left media outlets. On the other hand, right-biased media outlets receive more balanced interactions across three leanings. We can attribute these to the distribution of media outlets within each ideology. Third, the interactions between users from a specific country and local media or media in their languages are also reflected in the figure. For instance, Spain/Mexico - EL Pais; Nigeria - The Punch (Nigeria); Italy - Agenzia Nazionale Stampa Associata (ANSA); Ecuador - Ecuavisa; Singapore - The Straits Times, etc. These two sections, together with previous analysis on the United States level, answer RQ1.

\section{Individual-to-Media Representation}
\label{sec:utm-analysis}

This section presents the analysis conducted on individual-to-media representation of the global user-to-media interaction matrix. Specifically, we aim to find potential user groups among selected countries and study the cross-country media consumption behaviors. 

\begin{figure*}[!htbp]
  \centering
    \includegraphics[width=1.0\linewidth]{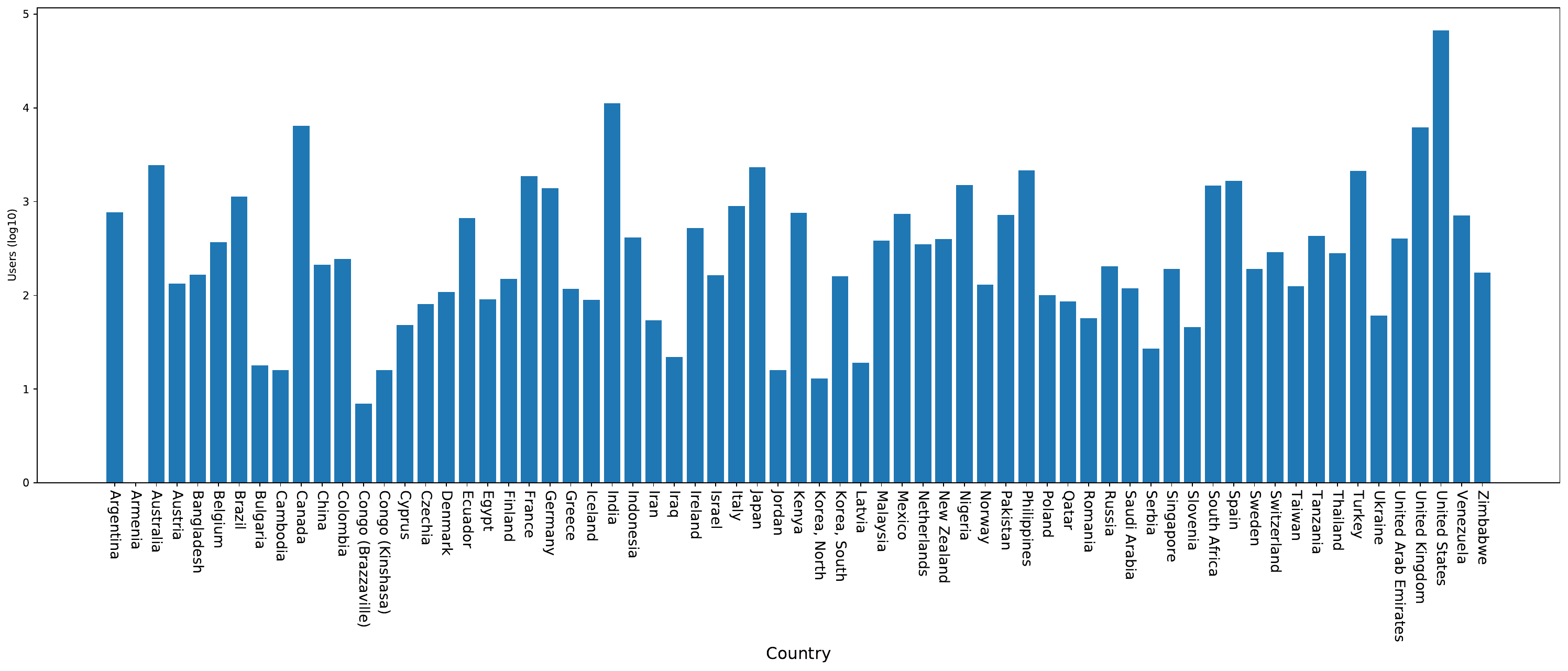}
  \caption{The number of users in each country after cutoff by a threshold of 5 interactions.}
  \label{fig:glo-country-user}
\end{figure*}

To ensure each user has enough media consumption, we decide to filter less active users from each country. Two approaches can be used for the filtering: (1) taking a fixed amount of users (e.g., top 300) from each country; (b) taking users with at least some amount of interactions. The first approach suffers from tie-breaking issues, namely users with the same amount of interactions may be identified as both active and inactive users. We take the second approach with a threshold of 5 interactions for user selection instead. Figure \ref{fig:glo-country-user} displays the number of users in each country in the log10 scale.

\subsection{Background}

The discovery of user groups within and across countries relies on KMeans clustering. In particular, this subsection provides a brief background on KMeans clustering and several evaluation metrics.

\textbf{KMeans Clustering} aims to partition the input dataset into predefined K clusters by minimizing with-cluster variance. It starts with randomly initialized centroids and performs re-assignment (assigns data points to nearest clusters) and re-computing (compute the new cluster centroid) in an iterative way to obtain the final clusters (\cite{Lloyd1982LeastSQ}). The initialization in KMeans could determine the speed of convergence to local-minima and hence a good initialization could reduce a significant amount of computation time. In particular, \textbf{KMeans++} (\cite{Arthur2007kmeansTA}) initializes the centroids via weighted probability distribution based on the distance between selected centroids and remaining data.

In our later task, clustering results are evaluated from two major perspectives: (1) \textbf{quantitative measurement}; (2) the \textbf{usefulness} of the resulting clusters as mentioned by \cite{Luxburg2012ClusteringSO}. 

For quantitative measurement, we make use of some of the following metrics:

\begin{enumerate}[a.]
    \item Elbow method is based on the distortion of the clusters, which is defined as the sum of squared distances between each data point to its closest centroid. Distortion usually decreases with the number of clusters.
    \item Silhouette Coefficient is defined as $ s = \frac{x-y}{max(x, y)} $, where $x$ is the average distance between a data point and all other points within the same cluster and $y$ is the average distance between a data point and all other points in the next nearest cluster. Mean Silhouette Coefficient measures the average scores over all the samples. Silhouette Coefficient for a single data point is bounded between -1 to 1 with higher values indicating better partitions.
    \item Davies-Bouldin Index measures the averaged similarity between each cluster and its most similar cluster. Mathematically, $DB = \frac{1}{K} \sum_{i=1}^{K} max_{i \neq j} R_{i, j}$, where $R_{i, j} = \frac{s_i + s_j}{d_{i,j}}$. $s_i$ is known as the cluster diameter and defined as the average intra-cluster distance. $d_{i, j}$ is the distance between two centroids. Lower scores implies better partitions.
    % \item Calinski-Harabasz Index measures the ratio of total inter-cluster dispersion and total intra-cluster dispersion, where dispersion is defined as sum of squared distance. Higher values imply better clusters.
\end{enumerate}

To evaluate the ``usefulness'' of the clusters, we consider: 

\begin{enumerate}[a.]
    \item Stability of the clustering algorithm. It is explored by applying the clustering algorithm repeatedly to the same data under different initialization. The algorithm is said to be stable if the produced user groups are robust under different initialization.
    \item Interpretability of the user groups. As our final goal is to discover the potential user groups, the interpretability of the groups and their media consumption need to be taken into account.
\end{enumerate}

We emphasize that human inspections on clustering results play a critical role when evaluating the clustering results. The quantitative measurement here serves merely as a heuristic and sometimes could provide insufficient and even misleading results.

\subsection{Global-level User Clustering}

\begin{figure*}[!htbp]
%   \centering
    \includegraphics[width=0.5\linewidth, height=0.175\textheight]{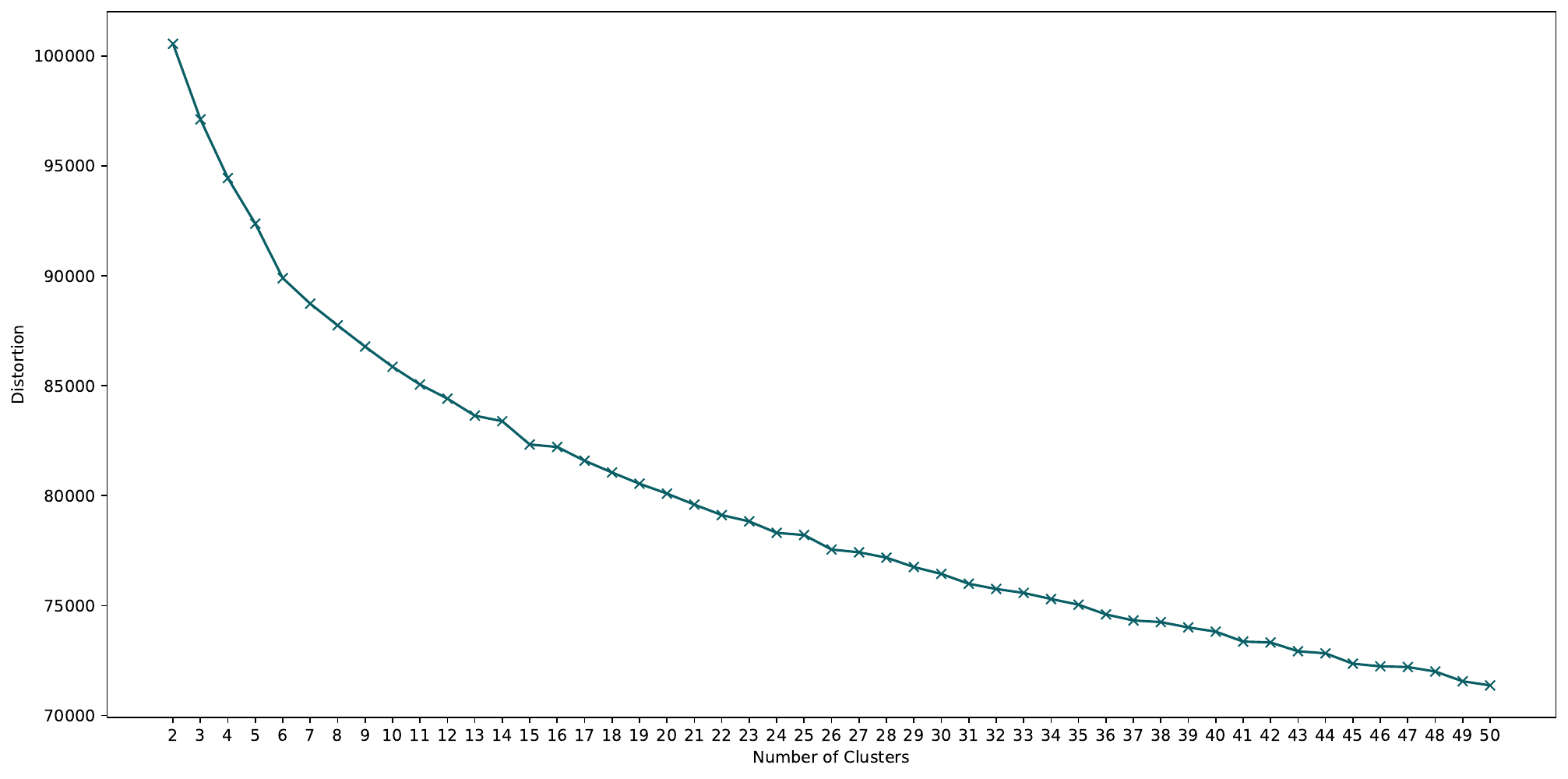}
    \includegraphics[width=0.5\linewidth, height=0.175\textheight]{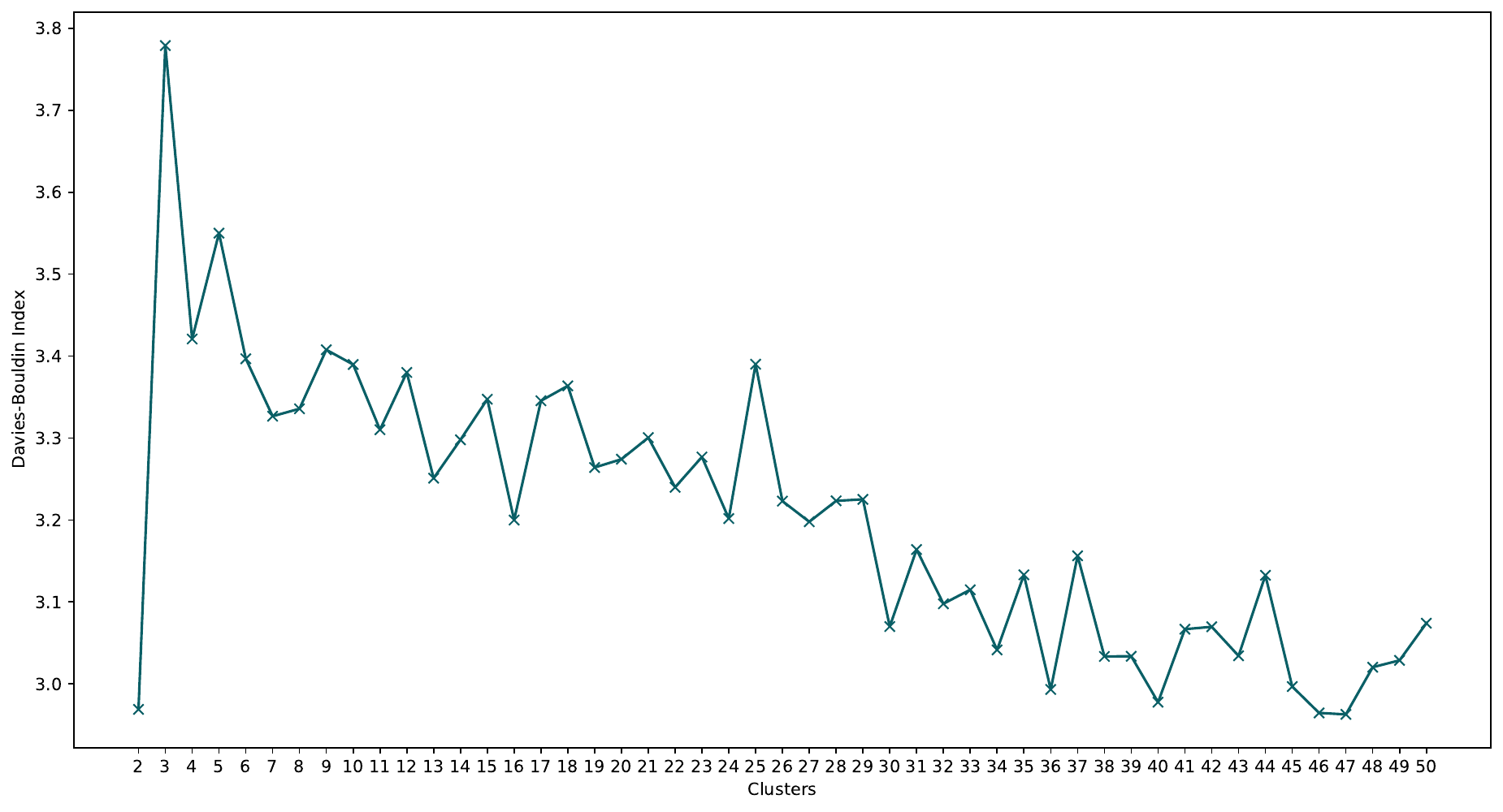}
    \includegraphics[width=0.5\linewidth, height=0.175\textheight]{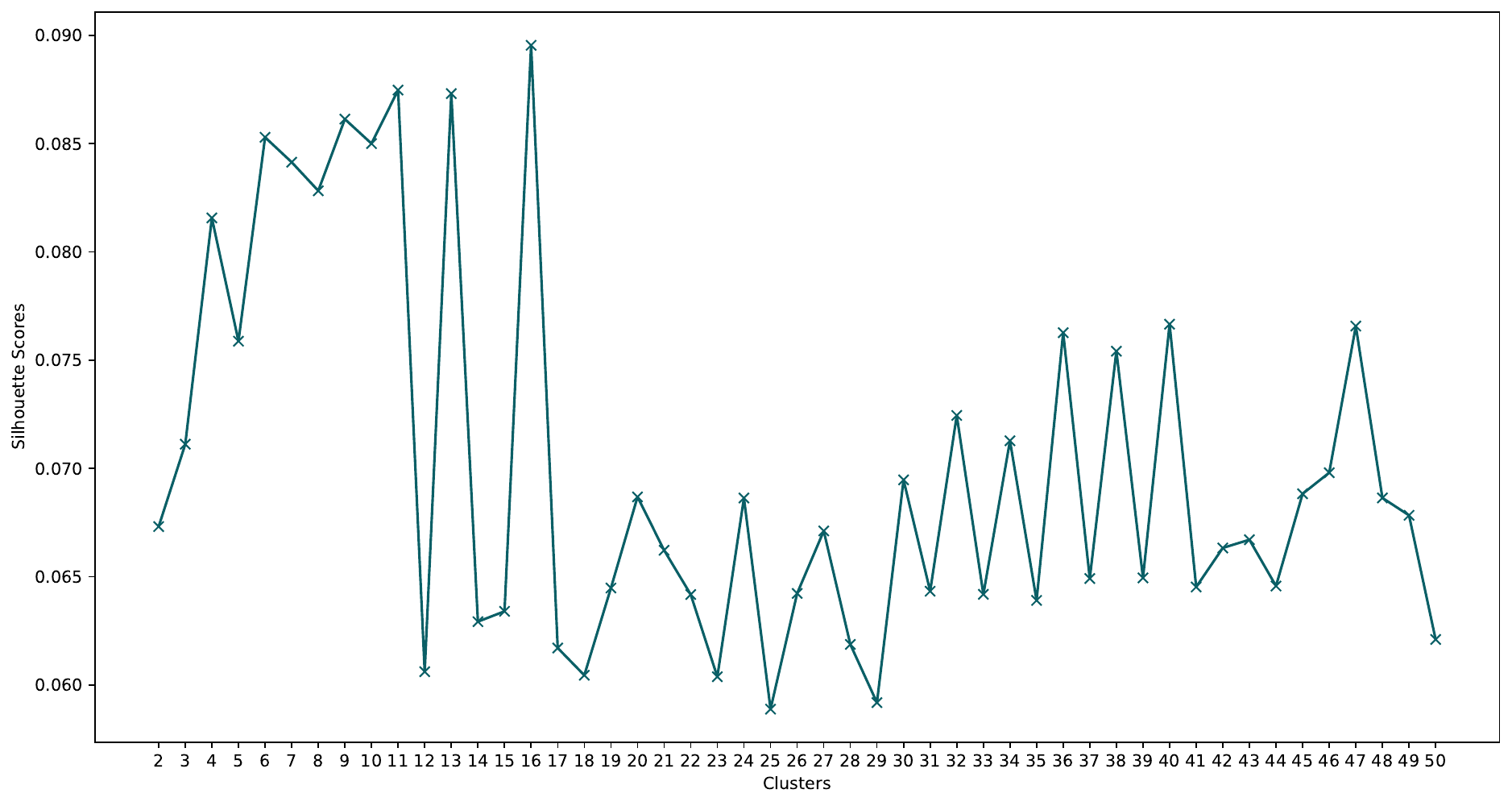}
  \caption{Evaluation of KMeans clustering on user-to-media interaction matrix. Top left: distortion; Top right: Davies-Bouldin scores; Bottom left: average Silhouette Coefficient scores.}
  \label{fig:kmeans-metrics}
\end{figure*}

To make the later visualization easier, the original seven ideologies on media have been grouped into three categories based on their bias as an additional preprocessing on global-level clustering. Figure \ref{fig:kmeans-metrics} displays the quantitative  evaluation of KMeans clustering on the global user-to-media interaction matrix.% A decreasing tendency can be observed on the change of distortion and Davies-Bouldin scores. 

However, The average Silhouette Coefficient scores keep oscillating when the number of clusters changes. Although it demonstrates a very weak increasing tendency when increasing the number of clusters, the maximum value achieved is less than 0.1. One possible reason can be attributed to the high dimensionality of the input data, which makes the average Silhouette Coefficient scores unable to provide meaningful guidance on evaluation on the clustering results. We ignore this metric in the future and only refer to the distortion and Davies-Bouldin scores for cluster selection. 

Although distortion and Davies-Bouldin scores show a declining propensity over the number of clusters, the overall slope gradually reduces. It implies that increasing the number of clusters beyond a certain number will not bring potential gain over the heuristic values compared to the previous range. We eventually select the number of clusters as 25.

\begin{figure*}[!htbp]
  \centering
  \advance\leftskip-3.6cm
%   \vspace{0pt}
%   \captionsetup{width=1.4\linewidth}
  \includegraphics[width=1.4\linewidth, height=70mm]{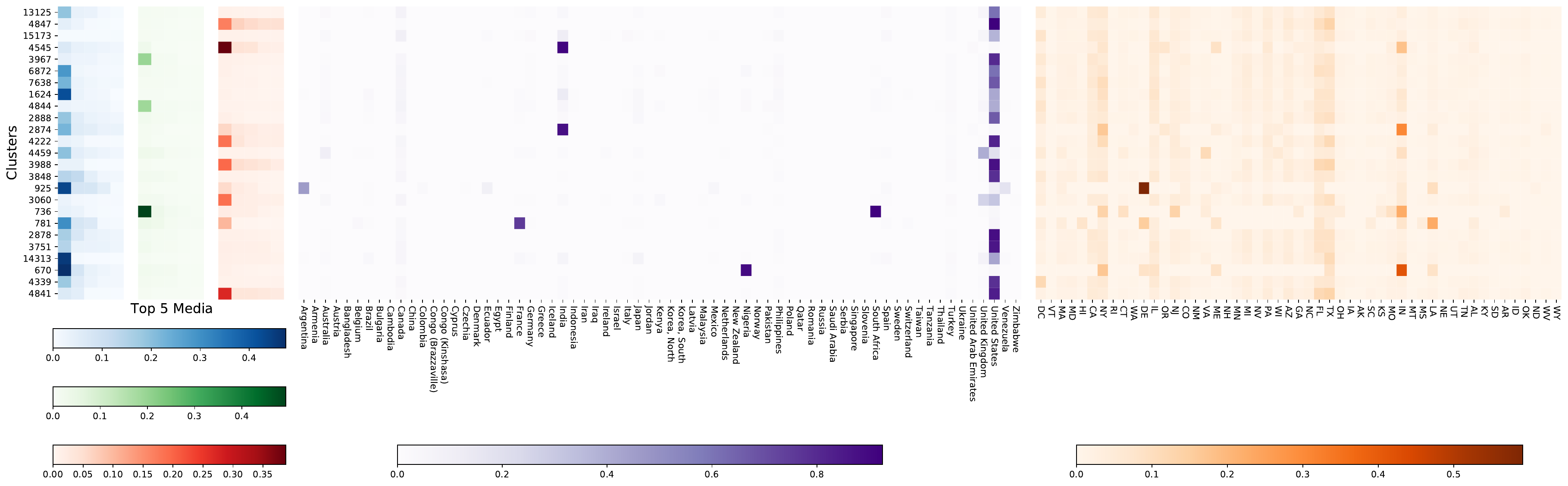}
  \caption{Global user-media consumption clustering results visualization. Each row in the figure represents related information on a cluster. The left sub-figure visualizes the distribution of the ratio of interactions with the top 5 media from each merged ideology. Note that users within each cluster interact with different media sets and hence  cells across clusters are not comparable. The size of each cluster is shown as Y-axis tick labels. The middle sub-figure presents the composition of each cluster by the distribution of the ratio of users' nationalities with it. The right sub-figure is a supplementary visualization of the state composition of the United States users from each cluster.}
  \label{fig:glo-vis}
\end{figure*}

In order to ensure the existence of user groups are robust rather than coincidental due to randomness, We apply KMeans clustering multiple times on the interaction matrix. We distinguish different user groups by the distributions of nationalities of their containing users. And We consider a user group robust if there is a group with close size and similar nationality distribution showing up most of the time for different runs. Figure \ref{fig:glo-vis} visualizes one of the clustering results, and the remaining visualizations based on different initialization are available in Appendix \ref{app:glo-clusters}. For simplicity, we use \textit{X-cluster} to represent clusters that are dominated by users from \textit{country X} or \textit{area X}. Based on the heatmap in figure \ref{fig:glo-vis}, some of the robust clusters are \textbf{India} clusters (the 3rd row and the 11th row), \textbf{South Africa} cluster (the 18th row), \textbf{Nigeria} cluster (the 23rd row), \textbf{South America} cluster (the 16th row), \textbf{Continental Europe} cluster (the 19th row)

% Each row in the figure represents related information on a cluster. The left sub-figure visualizes the distribution of the ratio of interactions with top 5 media from each merged ideology. Note that each cluster interacts with different sets of media and hence different cells across clusters are not comparable. The size of each cluster is shown as Y-axis tick labels. The middle sub-figure presents the composition of each cluster by the ratio of its consisting users' nationality. The right sub-figure is a further visualization on the state composition of the United States users from each cluster.

It is also clear that users from the United States constitute a significant portion of the overall population and hence most clusters are United States-dominated. Such clusters do not provide anything interesting for finding global user groups as they do not convey additional information about other countries. We hence limit our attention to clusters consisting of users mainly from countries outside the United States and conduct 4 case studies here. % For simplicity, we use \textit{X-cluster} to represent clusters which are dominated by users from \textit{country X} or \textit{area X}. 

% Each cluster is visualized as follows: (a) radar plot showing the percentage of interactions with media of different leaning. (b) bar plots showing the percentage of interactions received by media of different factuality and credibility; (c) three treemaps showing top 5 media based their received interactions from each leaning along with the ratio; (d) an additional treemap displaying the country composition along with the number of users from each country and corresponding ratio.

\begin{figure*}[!htbp]
  \centering
  \advance\leftskip-3.6cm
  \vspace{0pt}
    \includegraphics[width=1.4\linewidth]{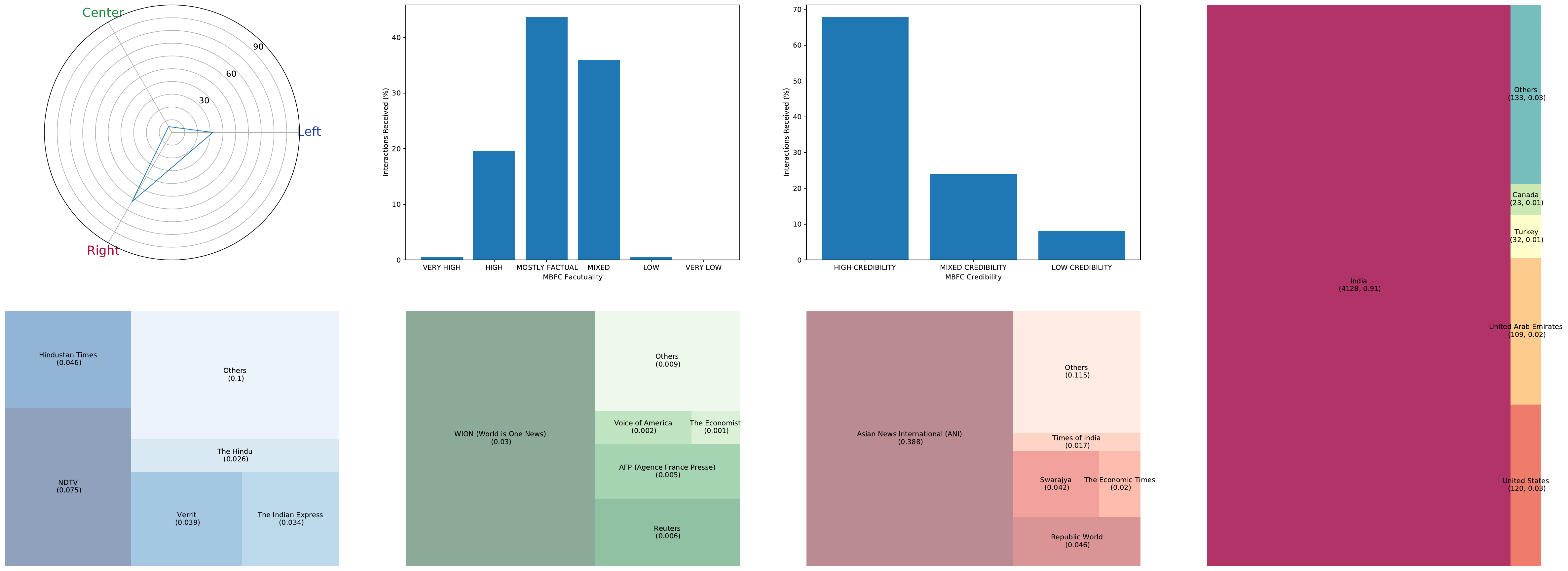}
     \par\bigskip
    \includegraphics[width=1.4\linewidth]{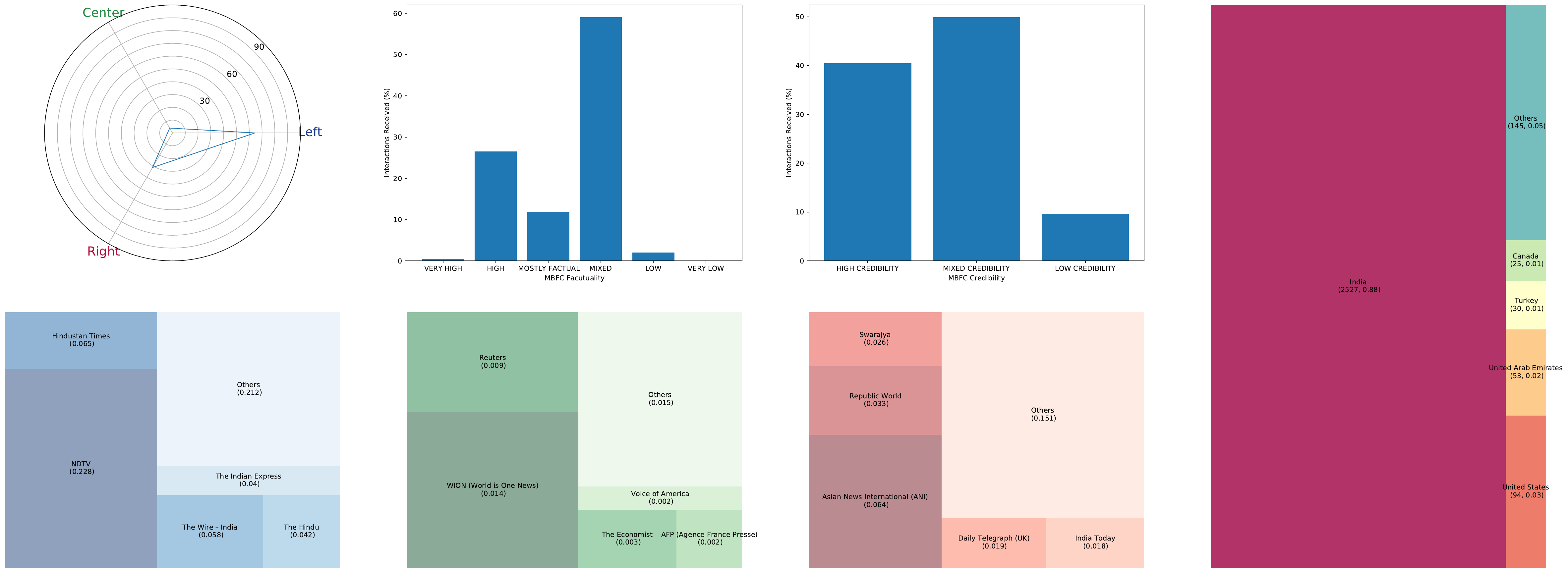}
  \caption{India cluster visualization. The radar plot shows the percentage of interactions with media of different learning. Bar plots display the rate of interactions received by media from different factuality and credibility. Three treemaps on the bottom present the top 5 media based on their received interactions from each leaning along with the ratio; Treemap on the rightmost shows the country composition and the number of users from each country and corresponding ratio.}
  \label{fig:in-cluster}
\end{figure*}

\textbf{India} clusters compositions are shown in figure \ref{fig:in-cluster}. Roughly 90\% of users from the two clusters are from India and consume a significant portion of biased media outlets. The two clusters together demonstrate the polarized media consumption with opposing ideologies in India. Approximately 60\% of interactions are towards right-biased media outlets in the upper cluster (e.g., \textit{Asian News International (ANI)}), while roughly left-biased media receive 60\% interactions in the bottom cluster (e.g., \textit{NDTV}). Meanwhile, there also lies a difference in the factuality and credibility of the media outlets consumed. Users from the upper cluster tend to interact more with mostly factual and highly credible media outlets, while users from the other cluster interact more with media outlets of mixed factuality and credibility. 

\begin{figure*}[!htbp]
  \centering
   \advance\leftskip-2.4cm
  \vspace{0pt}
   \includegraphics[width=1.4\linewidth]{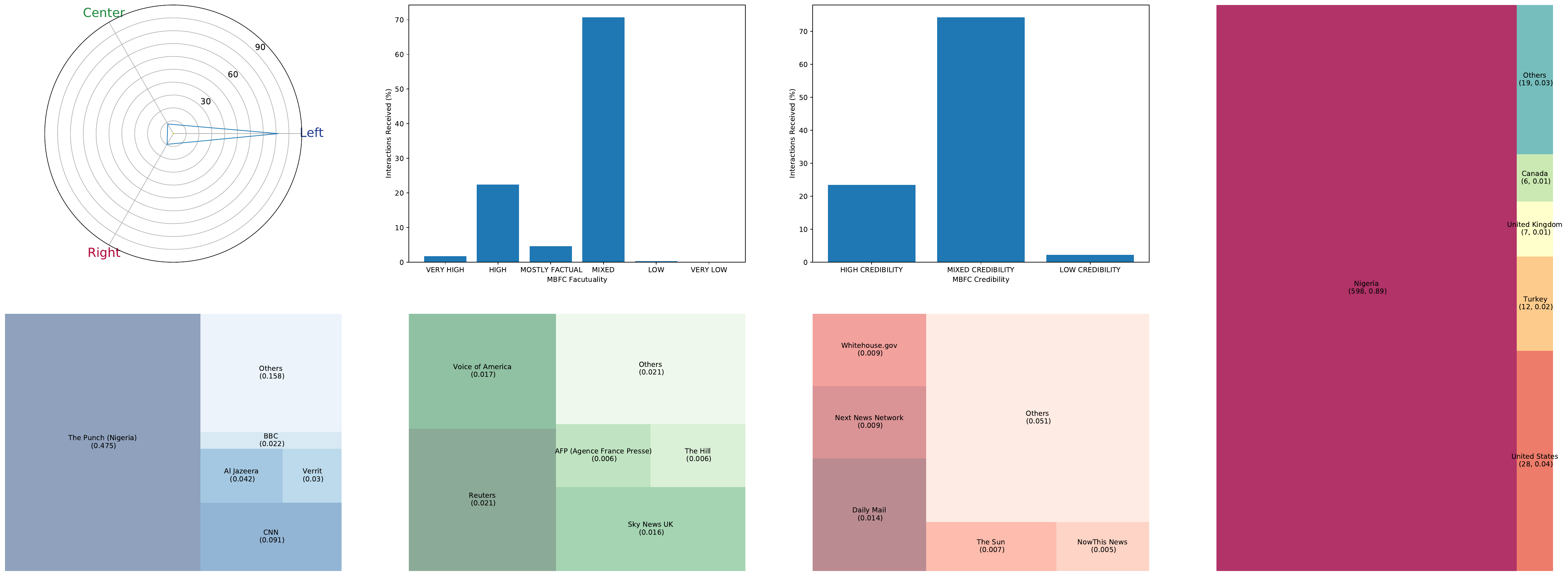}
    \includegraphics[width=1.4\linewidth]{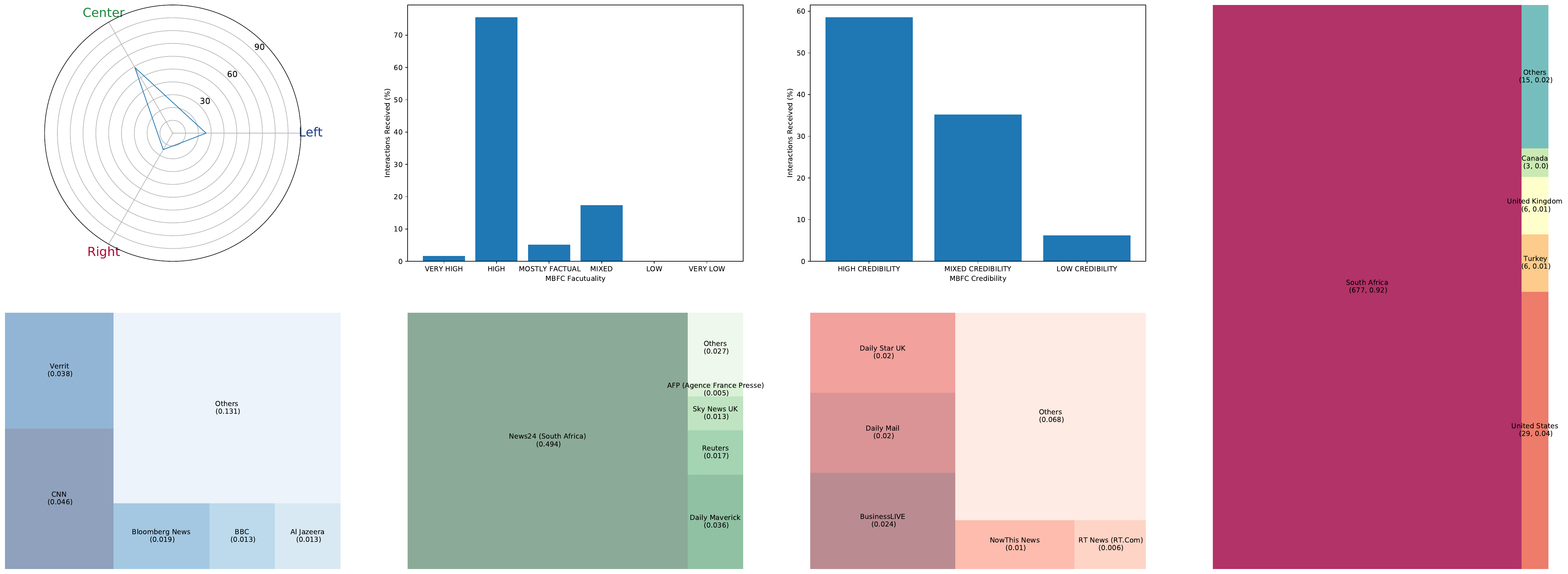}
    %  \par\bigskip
  \caption{Nigeria (top) and South Africa (bottom) cluster visualization.}
  \label{fig:safr-cluster}
\end{figure*}

\textbf{Nigeria} cluster and \textbf{South Africa} cluster can be viewed as some instances from \textbf{Africa} clusters, as shown in figure \ref{fig:safr-cluster}. Users from the Nigeria cluster exhibit highly polarized consumption towards left-biased media outlets (e.g. \textit{The Punch (Nigeria)}), which received about 80\% of the overall interactions. Meanwhile, the media diet of Nigerien users mainly consists of media from other countries such as \textit{CNN}, \textit{BBC}, \textit{Daily Mail}, which could be attributed to a lack of media coverage in our dataset. About 70\% of the interactions are towards media outlets of mixed factuality and credibility.

Users from South Africa, on the other hand, manifest a much less biased media consumption. Approximately 60\% of interactions are towards least-biased media outlets such as \textit{News24 (South Africa)} and \textit{Daily Maverick}. At the same time, the media diet of South African users is a mixture of local media and foreign media, and it consists of a large proportion of highly factual and credible media.

\begin{figure*}[!htbp]
  \centering
   \advance\leftskip-3.6cm
  \vspace{0pt}
    \includegraphics[width=1.4\linewidth]{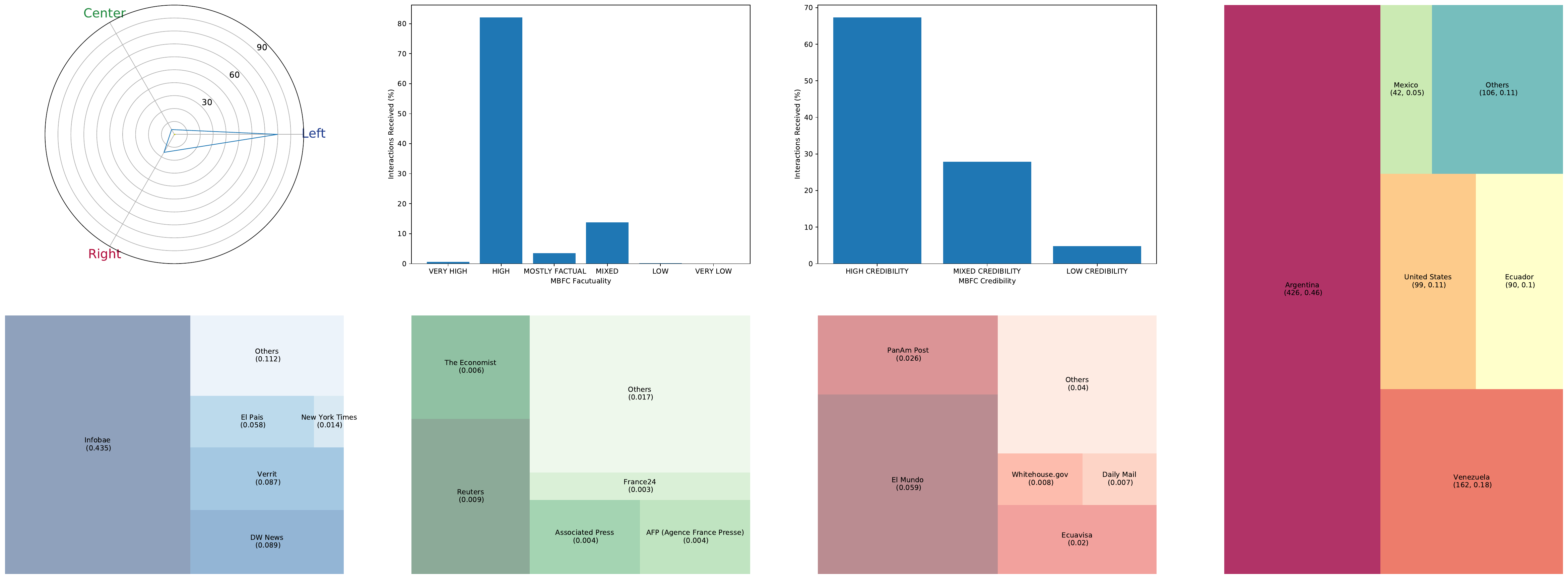}
  \caption{South America cluster visualization.}
  \label{fig:samer-cluster}
\end{figure*}

\textbf{South America} cluster, displayed in figure \ref{fig:samer-cluster}, mainly consists of users from South American countries such as Argentina, Venezuela, Ecuador, and Mexico.  Media consumed within this cluster are primarily left-biased, and most of them exhibit high credibility and factuality. These media are either local media due to geographical proximity or foreign media from countries with the same language. These include \textit{Infobae} in Argentina, \textit{EI Mundo} in Spain, \textit{france24} in France with Spanish available, \textit{PanAm Post} in the United States with Spanish available.

\begin{figure*}[!htbp]
  \centering
   \advance\leftskip-2.4cm
  \vspace{0pt}
    \includegraphics[width=1.4\linewidth]{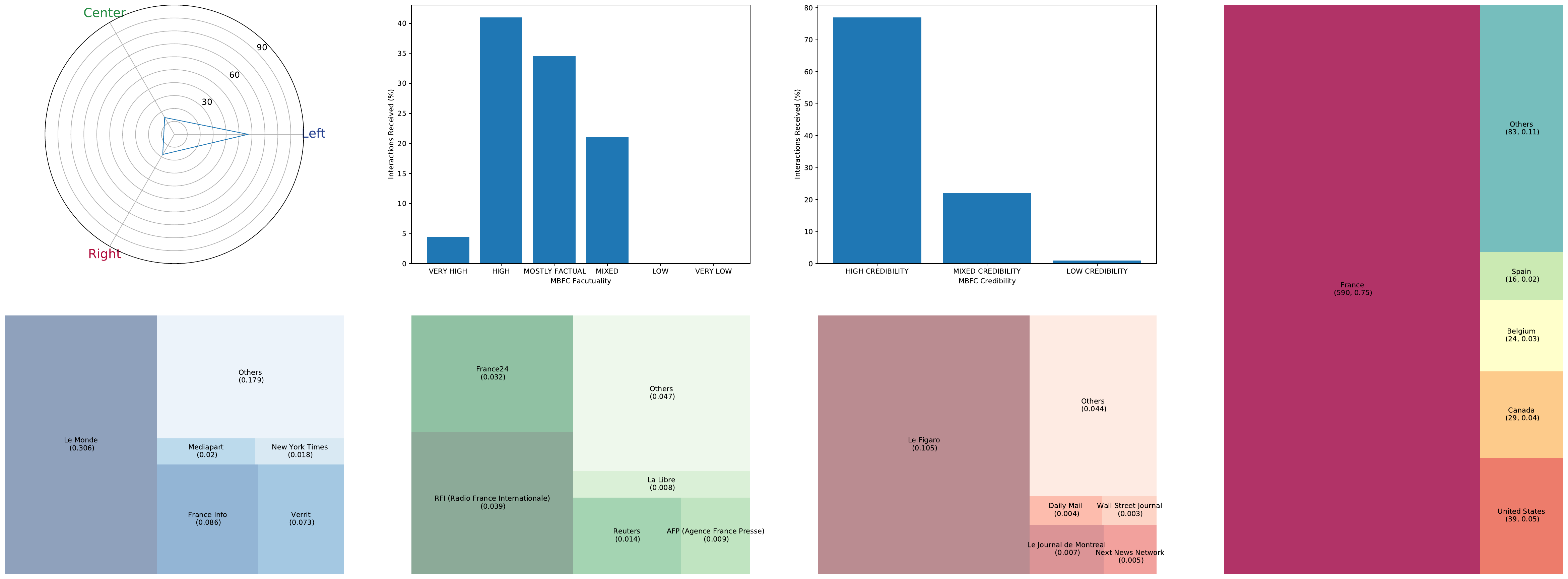}
  \caption{Continental Europe cluster visualization.}
  \label{fig:eu-cluster}
\end{figure*}

\textbf{Continental Europe} cluster is present in figure \ref{fig:eu-cluster}, which consists of users mainly from France and several other countries in mainland Europe such as Belgium and Spain. Users within these interact with right-biased media. Media of right-leaning and center-leaning receive a similar amount of interactions. Meanwhile, media within this cluster exhibits a diverse level of factuality, ranging from mixed-level to very high. Like the previous cluster, media here are mainly local media with commonly used languages or geographically proximate such \textit{as Le Monde}, \textit{France Info}, and \textit{Le Figaro}.

% The clustering on global user-to-media interaction matrix has revealed the existence of some user groups. However, one potential weakness is that we do not have control over the clusters in this case. In other words, each cluster consists of mainly users from a single country but there are still a non-negligible amount of users from other countries at the same time. Meanwhile there are many clusters with same dominant countries such as United States and United Kingdom. Although the reason could be the uneven distributed population from each country, such impurity and repetition has made it difficult to conduct detailed analysis on every cluster. To complement the global level analysis, we further conduct cross-country analysis on some pre-selected country pairs in the next section.

% However, due to the impurity of the cluster nationality composition, it is hard to conduct further study on it from a country-level perspective.

The clustering results on the global user-to-media interaction matrix have revealed the existence of some user groups and answered RQ3. Meanwhile, we could also notice some common membership across each cluster: users from the United States, the United Kingdom, and Canada also show up in most of the clusters. We ascribe this to the uneven population across countries and demographic reasons. We believe user groups not only exist globally but also within each country. To complement the global-level analysis, we further conduct cross-country analysis on some selected country pairs in the following part.

%%%%%%%%%%%%%%%%%%%%%%%%%%%%%%%%%%%%%%%%%%%%%%%%%%%%%%%%%%%%%%%%%%%%%%%%%%%%%%%%%%%%%%%%%%%%%%%%%%%%%%%%%%%%%%%%%%%%%%%%%%%%%%%%%%%%%%%%%%%%%%%%%%%%%%%%%%%%%%%%%%%%%%%%%%%%%%%%%%%%%%%%%%%%%%%%%%%%%%%%%%%%%%%%%%%%%%%%%%%%%%%%%%%%%%%%%%%%%%%%%%%%%%%%%%%%%%%%%%%%%%%%%%%%%%%%%%%%%%%%%%%%%%%%%%%%%%%%%%%%%%%%%%%%%%%%%%%%%%%%%%%%%%%%%%%%%%%%%%%%%%%%%%%%%%%%%%%%%%%%%%%%%%%%%%%%%%%%%%%%%%%%%%%%%%%%%%%%%%%%%%%%%%%%%%%%%%%%%%%%%%%%%%%%%%%%%%%%%%%%%%%%%%%%%%%%%%%%%%%%%%%%%%%%%%%%%%%%%%%%%%%%%%%%%%%%%%%%%%%%%%%%%%%%%%%%%%%%%%%%%%%%%%%%%%%%%%%%%%%%%%%%%%%%%%%%%%%%%%%%%%%%%%%%%%%%%%%%%%%%

\subsection{Cross-Country Analysis}

Cross-country analysis, in this case, refers to analyzing how users in country A consume media from country A and another country B. What is different from the previous section is now we can control the targeting countries. We make use of two sets of media consumption in this analysis. One is made of interactions with media in country A while the other is made of interactions with media in country B. The media interaction vectors are all aggregated into 7-dimensional vectors, consisting of the ratio of interactions with media of 7 different ideologies.

Finding the group that a user within a country belongs to provides a basis for later analysis. In this case, users within the same group shall share a similar media consumption pattern. There are two methods explored that could help with creating and identifying user groups:

\begin{enumerate}
    \item Predominant consumption: assignment of a user to a group is determined by one's predominant consumption of media of a particular leaning. For instance, if 80\% of the user consumption is on media of center-left-leaning, then we identify this user from a \textit{center-left} group.
    \item Clustering-based: clustering is applied on users with their media consumption. The resulting clusters are considered as underlying user groups.
\end{enumerate}

We decide to use a clustering-based approach to discover potential groups. Although the predominant consumption approach enjoys fast computation and could be easily adapted onto different country pairs, one potential drawback is we lose the information on a user if he has roughly equally consumption towards media of two different leanings. For example, if 45\% consumption of a user is on left-leaning media while 40\% of consumption is on center-right-leaning media, then the user will be put into the \textit{left} group. We believe it is not rare for users to consume media of more than one leaning. Assigning users to their groups based only on media of one particular leaning will prevent us from finding groups where users consume media of multiple leanings. 

% Specifically, we use KMeans clustering twice on user-media consumption based on the country pair using the two sets of the media consumption described earlier. The selection of the number of clusters, in this case, is based on similar metrics used in the previous section and limited to a small number in order to ensure the interpretability and conciseness of user groups. After obtaining the user groups, our next task would be to explore the users' ``movement'' across groups when consuming media from different countries. To avoid confusion in the following parts, we use group notations $A_i$ or $B_i$ to represent user groups $i$ constructed from their consumption of media from country $A$ or $B$. 

% Let $P_{i}$ be the probability that the user comes from a group $A_i$. Let $P_{i, j}$ be the probability that given the users fall in the group $B_j$ given that they are from a group $A_i$. Let $P_{j}^{r}$ be the probability of a random user falling in the group $B_j$. Inspired by \cite{Freelon2020FalseEO}, we further define the risk ratio as  $ r_{i, j} = P_{i, j} / P_{j}^{r} $ which indicates that, compared to a random user, how likely it is for a user from group $A_i$ to fall into group $B_j$. Risk ratio can be used to demonstrate echo chamber effects of users. When users from $A_i$ and $B_j$ have similar media consumption, higher $r_{i,j}$ indicates users from $A_i$ are ``stickier'' than a random user in the sense that they keep consuming media of the same leanings across countries.

Specifically, we use KMeans clustering twice on user-media consumption based on the country pair using the two sets of media consumption described earlier. The selection of the number of clusters, in this case, is based on similar metrics used in the previous section and limited to a small number in order to ensure the interpretability and conciseness of user groups. After obtaining the user groups, our next task would be to explore the users' ``movement'' across groups when consuming media from different countries. To avoid confusion in the following parts, we use $SG_i$ (source groups $i$) to represent user group $i$ based on local media consumption and $TG_i$ (target groups $i$) to represent user group $i$ based on foreign media consumption.

Let $P_{i}$ be the probability that the user comes from a group $SG_i$. Let $P_{i, j}$ be the probability that given the users fall in the group $TG_j$ given that they are from a group $SG_i$. Let $P_{j}^{r}$ be the probability of a random user falling in the group $TG_j$. Inspired by \cite{Freelon2020FalseEO}, we further define the risk ratio as  $ r_{i, j} = P_{i, j} / P_{j}^{r} $ which indicates that, compared to a random user, how likely it is for a user from group $SG_i$ to fall into group $TG_j$. Risk ratio can be used to demonstrate echo chamber effects of users. When users from $SG_i$ and $TG_j$ have similar media consumption, higher $r_{i,j}$ indicates users from $SG_i$ are ``stickier'' than a random user in the sense that they keep consuming media of the same leanings across countries.

We choose a few country pairs and conduct some case studies. We limit countries to be those with a relatively large set of media outlets such as the United Kingdom, India, Australia. The case studies will not go through every detail but rather pick up some highlights and unusual behaviors. They will also provide answers to RQ4.

% The visualization on each country pair consists of 6 sub-figures. The sub-figures on the rightmost represent the user groups produced by KMeans clustering based on interactions with local and foreign media outlets respectively. We label the user groups based on local and foreign media outlets as \textbf{SG} and \textbf{TG}, representing source groups and target groups. The total amount of users has been left in the title of the top right sub-figure as a reference. The values under the labels on Y-axis are the ratio of users falling into each group. The whole sub-figure presents the mean ratio of interactions towards media outlets of each leaning within each group, that is the centroid of each cluster. The leftmost two sub-figures along with the top middle sub-figure present the concepts mentioned earlier. Specifically, the top left sub-figure visualizes the transition probability for a user from one group to the other; the bottom left sub-figure provides information on the number of users falling into each group; the top middle figure displays the risk ratio across user groups. The bottom middle sub-figure visualizes the number of media outlets in each ideology within each country as a reference. 

\begin{figure*}[!htbp]
  \centering
  \advance\leftskip-1.5cm
  \vspace{0pt}
  \includegraphics[width=1.15\linewidth,,height=90mm]{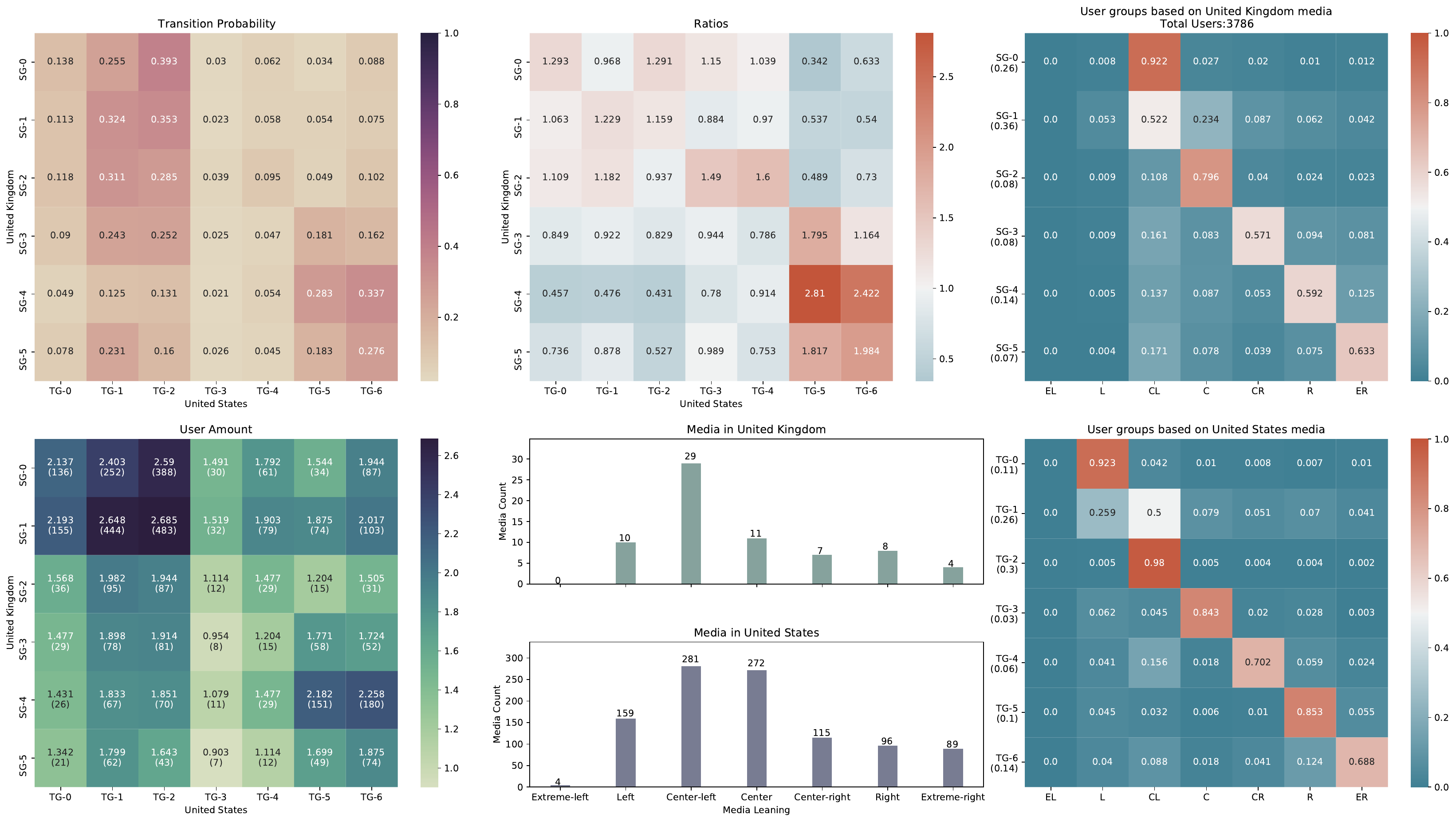}
  \caption{United Kingdom - United States interactions and user groups. The visualization consists of 6 sub-figures. The sub-figures on the rightmost represent the user groups produced by KMeans clustering based on interactions with local media outlets (the United Kingdom) and foreign media outlets (the United States). We label the user groups based on local and foreign media outlets as \textbf{SG} and \textbf{TG}, representing source groups and target groups. The total amount of users has been left in the title of the top right sub-figure as a reference. The values under the labels on Y-axis are the ratio of users falling into each group. The whole sub-figure presents the mean proportion of interactions towards media outlets of each leaning within each group, that is, the centroid of each cluster. The leftmost two sub-figures along with the top middle sub-figure present the concepts mentioned earlier. In detail, the top left sub-figure visualizes the transition probability for a user from one group to the other. The bottom left sub-figure provides information on the number of users falling into each group. The top middle figure displays the risk ratio across user groups, where cells with a ratio higher/less than 1 are colored red/blue. The bottom middle sub-figure visualizes the number of media outlets in each ideology within each country as a reference. }
  \label{fig:uk-us}
\end{figure*}

\textbf{United Kingdom - United States} interactions along with the underlying user groups are displayed in figure \ref{fig:uk-us}. The heatmaps on user groups demonstrate the homophily of user behaviors empirically. There are 6 SGs, which mainly consume British media outlets of political ideology ranging from center-left to extreme-right-leaning. Similarly, there are 7 TGs identified using American media outlets, where users consume media outlets of political ideology ranging from left-to extreme-right-leaning. The cluster sizes are highly imbalanced for both SGs and TGs due to the unbalanced distributions of media outlets among different leanings.

The risk ratio matrix has sub-blocks that can be identified through color. One sub-block is located in the top left, excluding a few cells in shallow blue. The other one is located in the bottom right. Based on the user groups visualization, the two blocks have demonstrated the echo chambers on user consumption across the United Kingdom and the United States. British users consuming center-left- and center- British media outlets tend to consume more American media outlets of left-, center-left-and center-leaning. And British users who consume right-biased and extreme-right-biased British media outlets are also much more likely to keep consuming American media outlets of the same ideology. In fact, for users falling into groups SG-4 and SG-5, the risk ratio is much higher when comparing the chance they stay in TG-5 and TG-6 compared to a random user., which implies that British right-media-consuming users are more loyal compared to the rest.

\begin{figure*}[!htbp]
  \centering
  \advance\leftskip-0.7cm
  \vspace{0pt}
  \includegraphics[width=1.15\linewidth,height=90mm]{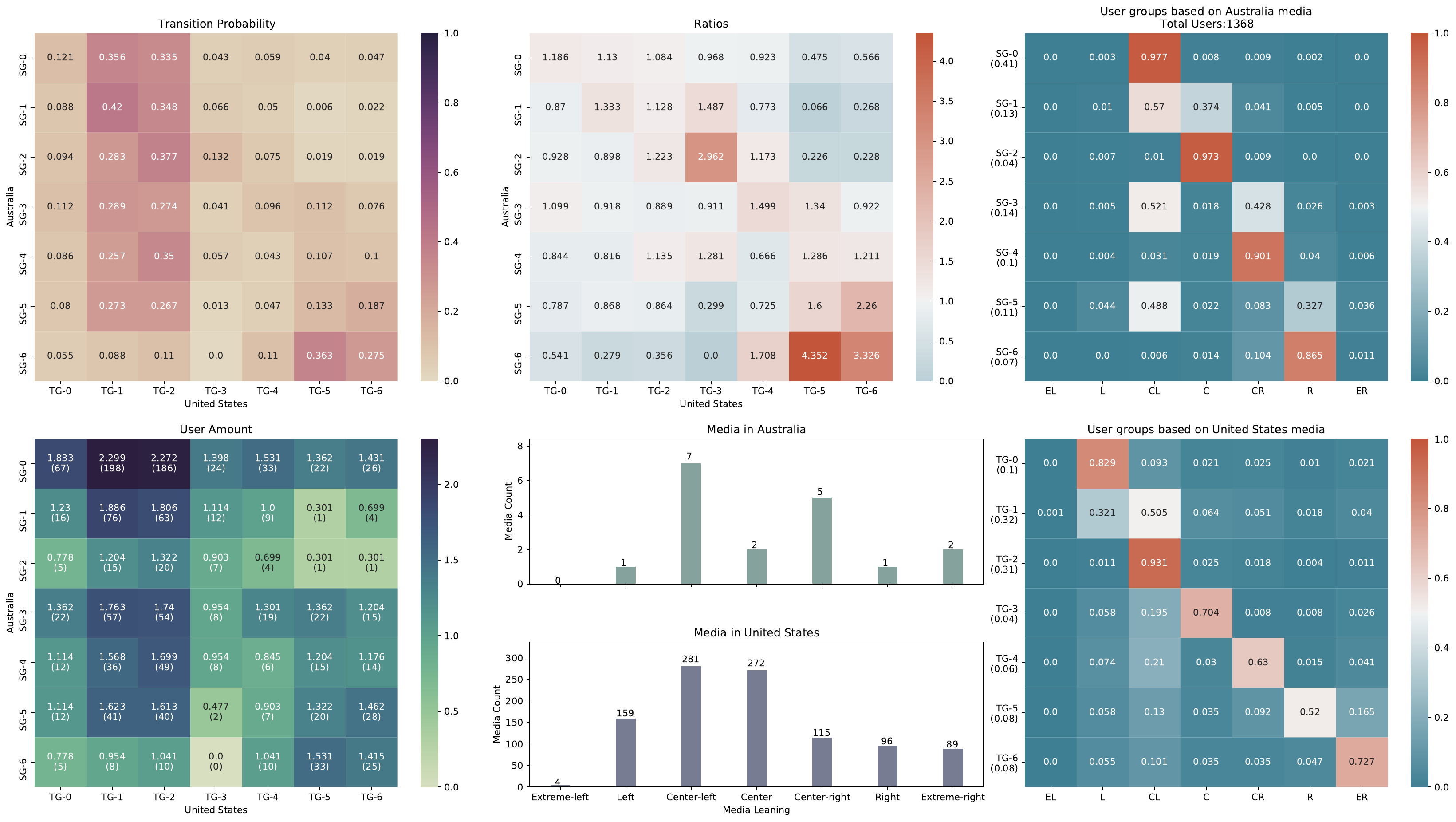}
  \caption{Australia - United States interactions and user groups.}
  \label{fig:au-us}
\end{figure*}

\textbf{Australia - United States} interactions along with the underlying user groups are displayed in the figure \ref{fig:au-us}. Compared to the previous case, one major difference is that users from several SGs tend to all consume center-left media outlets. Some of them are even interacting media outlets of opposing ideology like those from SG-5.

\begin{figure*}[!htbp]
  \centering
  \advance\leftskip-1.5cm
  \vspace{0pt}
  \includegraphics[width=1.15\linewidth]{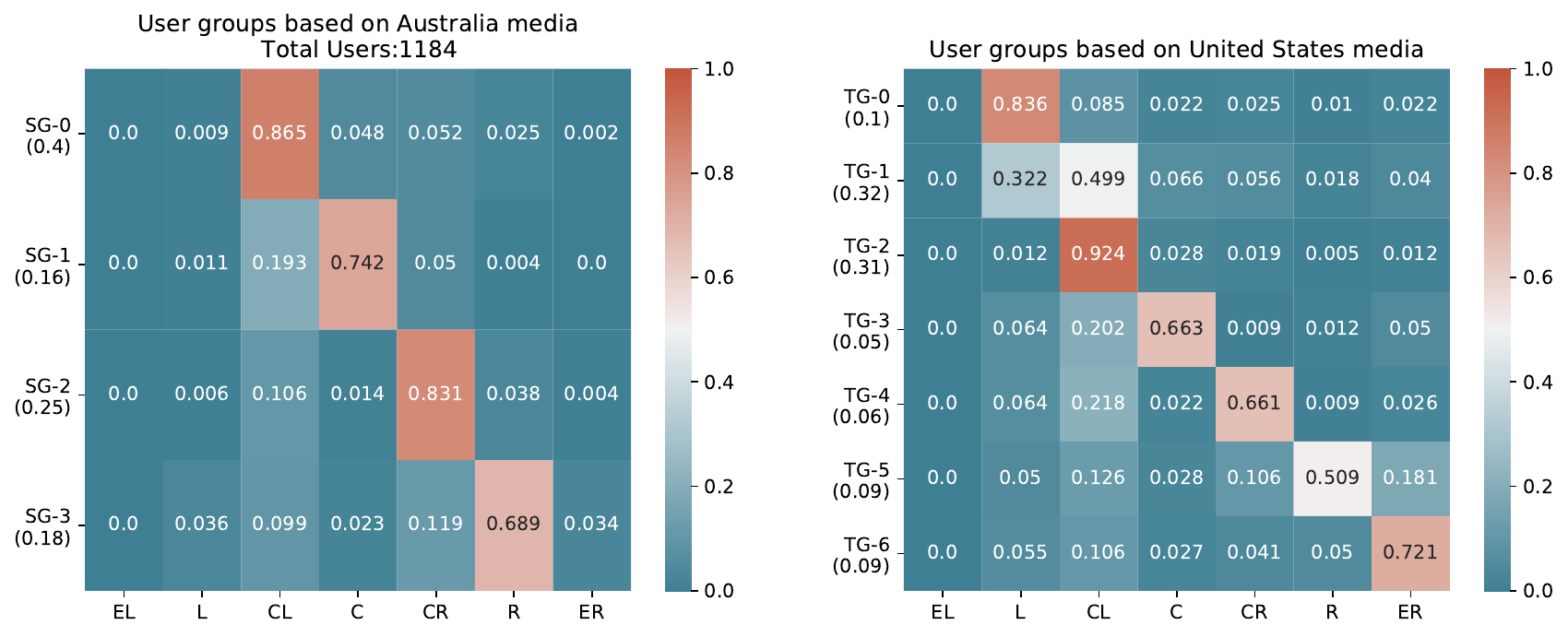}
  \caption{Australia - United States user groups constructed after removing ABC News Australia.}
  \label{fig:au-abc-removed}
\end{figure*}

The cause can be attributed to the fact that some center-left media outlets are too popular. In fact, to investigate the abnormal behaviors, we compare the media outlets being interacted with in SG-1, SG-3, and SG-5. It turns out ABC News Australia is always the one receiving the most interactions compared to other center-left-biased media. The interactions from Users among different groups with ABC News Australia are not due to their political preference but instead the popularity of the media outlet. The tweets from ABC News Australia are likely to be spread among the Australian Twitter user network, and hence users have significant exposure to it. We verify this by removing ABC News Australia from the Australian media set and applying KMeans again. The new user groups constructed are shown in \ref{fig:au-abc-removed}. As it shows, the media consumption among SGs is now more concentrated towards one leaning.

Similarly, the risk ratio matrix also demonstrates the echo chamber effect among SG-4, SG-5, SG-6, and TG-4 and TG-5. Users consuming right-biased content, regardless of the level of bias, are more likely to consume right-biased content from American media. Another obvious echo chamber happens between SG-2 and TG-3, where users are more likely to consume the least-biased media from both countries.

\begin{figure*}[!htbp]
  \centering
  \advance\leftskip-1.5cm
  \vspace{0pt}
  \includegraphics[width=1.15\linewidth,height=90mm]{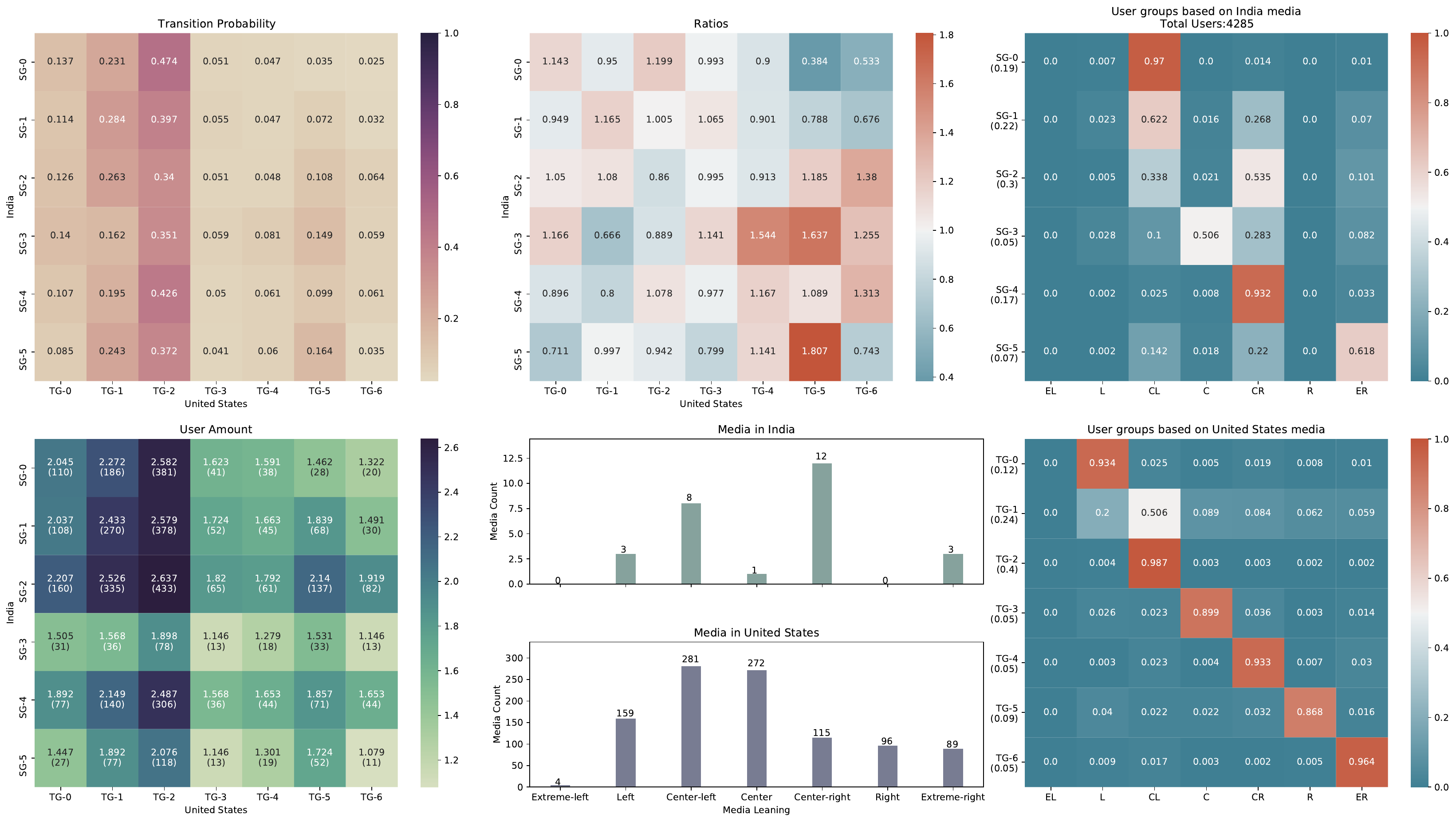}
  \caption{India - United States interactions and user groups.}
  \label{fig:in-us}
\end{figure*}

\textbf{India - United States} interactions along with the underlying user groups are present in figure \ref{fig:in-us}. Users from SG-1 and SG-2 are interacting with media outlets of opposing leaning. We again attribute the reason to the popularity of media similar to the previous case. The echo chamber effect still exists among groups where right-leaning content is the primary consumption. 

The risk ratio matrix still shows some echo chamber effects. For instance, users from SG-7 are still more likely to stay within TG-5. However, one unusual event happens between SG-7 and TG-6. Users from SG-6, where extreme-right Indian media are consumed, are much less likely to stay within TG-6, where extreme-right American media are consumed. This has contradicted with what happened in previous case studies, where users consuming extreme-right media in the United Kingdom and Australia demonstrate their loyalty when consuming media in a different country. 

We rule out the possibility of a clustering artifact by repetitively applying KMeans clustering with a slightly changed number of clusters and random seeds (shown in Appendix \ref{app:ind-us-extra}). And the ratio remains low among all the cases. We then speculate there may be one potential weakness of the current media bias labeling mechanism. There may exist differences between far-right leaning in different countries compared to the United States. However, verifying this speculation is difficult without a detailed comparison between the political spectrum, and we leave it as future work.

\section{Summary}
\label{sec:glo-sum}

This chapter presents the analysis of the global user-to-media interaction matrix. As an echo to the United States level analysis, it briefly goes through the two high-level representations: country-to-country and country-to-media. The observations show an overall pattern among the global level interactions.

Then the chapter presents a detailed analysis of the global user-to-media interaction matrix from two perspectives. The overall clustering on users across countries reveals the existence of underlying user groups outside the United States, such as India, France, Argentina, and South Africa. 

Last, the cross-country analysis provides complementary work by focusing on specific country pairs. The media consumption on the user groups illustrates empirical evidence on the homophily of Twitter users' online behaviors. Furthermore, the risk ratio presents the echo chamber effects when users consume media outlets across countries. Users consuming biased media content tend to be more likely to consume media of similar bias. Users consuming a dominant amount of right-biased media are more loyal in general.
\chapter{Conclusion}
\label{cha:conc}

This thesis pictures the media consumption on COVID-19 related topics by using a large-scale Twitter data. We use the interaction metric to quantify Twitter user activities with media outlets, which can be generalized onto different locations such as states or countries straightforwardly. With the help of the interaction matrices, we have discovered the media consumption patterns of users from different countries. To the best of our knowledge, we are the first to apply COVID-19 media consumption vectors to link online behaviors on Twitter and offline political preference. Moreover, unlike prior works that only focus on one specific country (the United States in most cases) regarding COVID-19 communications, we also present the first-of-its-kind study on COVID-19 media consumption across multiple countries simultaneously. We believe our findings in the thesis can shed light on understanding COVID-19 media consumption and user behaviors.

\section{Limitations}
\label{sec:limit}

We acknowledge our work takes the first step to study the media consumption over different countries during COVID-19, hence there are some limitations.

First, the 5-day dataset used in this thesis is not representative enough to support us to draw any conclusion from a long-term perspective but instead, serves as a snapshot of user media consumption during the whole pandemic.

Second, on the global level analysis, our work does not explore the user groups from a social science perspective but only draws data-driven observations. Such observations could be incomplete due to the potential bias introduced by the dataset such as a lack of media converge in some countries. We are uninformed whether there are counterparts on the user groups in the real world.

Third, the media bias rating provided by MBFC may not be accurate for some countries. In this thesis, we proceed with our work without questioning the accuracy of MBFC media bias. However, as mentioned in chapter \ref{cha:data}, MBFC rates media according to the American political perspective. Obviously, there is non-alignment between the United States politics and other countries, such as the definition of different ideologies and even the number of political parties. For instance, \cite{Cointet2021UncoveringTS} show the French media ecosystem has less polarization when compared to the media ecosystem of the United States. Meanwhile, as shown in the India - United States analysis, such inconsistency will likely produce some counter-intuitive behaviors and affect our understanding.

Last, the number of media outlets are mainly English-based and very imbalanced in the current dataset, making it hard to generalize onto countries whose language is not English such as countries from Europe and Asia.

\section{Future Work}
\label{sec:future}

Previous limitations have pointed out some future directions. First, we want to conduct similar studies using datasets from other periods so that the observations are more representative. Meanwhile, by combining the results obtained from multiple periods, we can further study the temporal media consumption patterns during the pandemic. Based on it, we can answer questions such as ``how does the media consumption towards media outlets of a specific ideology evolve?'', ``How does the polarization in online consumption change?'', ``Is there any change on the user groups such as occurrence and vanish of some particular user group within some countries?'' and ``Are there users who change their media consumption drastically?''.

Second, this thesis focus on quantitative measurement and analysis of interactions extracted from tweets. We would like to move from the analysis of numerical data to non-numerical data such as the texts from tweets and conduct research such as opinion mining and sentiment analysis. We believe the outcome from the study on text data can provide more explanations over user media consumption patterns.

Third, resolving the third limitation mentioned above requires a detailed understanding of the political spectrum over many countries and re-rating the media bias, which is expensive to complete. Hence, another possible direction would be studying methods that could help with alignment between political spectrums across different countries with existing data. This direction could be achieved by either inventing a new rating process that could consider the difference among various political spectrum or methods that can help describe the political spectrum of a country based on another country like the United States.

Last but not least, to generalize the analysis onto non-English speaking countries, future work can focus on obtaining user-media interactions through additional platforms. For instance, Google Trends\footnote{\url{https://trends.google.com/trends}} provides searching volumes of given keywords during a time window and identifiers for addressing languages. Previously, user groups were constructed directly from all users without distinguishing each other. It is also promising to compare user sub-groups such as Spanish-speaking users in the United States and users in Spain.

%%%%%%%%%%%%%%%%%%%%%%%%%%%%%%%%%%%%%%%%%%%%%%%%%%%%%%%%%%%%%%%%%%%%%%
% Here begins the end matter

\appendix

\chapter{COVID-19 Keywords and Crawler Settings}
\label{app:keywords}

There are 7 crawlers used to collect the tweets. The settings and assigned keywords are displayed below in JSON format. The \verb|language| attribute indicates the corresponding collected language of tweets, with empty list indicating all languages.

\subsubsection*{crawler 0}
\begin{itemize}
    \item keywords: ["coronavirus", "covid19", "covid", "covid–19", "COVID---19", "pandemic", "covd", "ncov", "corona", "corona virus", "sars-cov-2", "sarscov2", "koronavirus", "wuhancoronavirus", "wuhanvirus", "wuhan virus", "chinese virus", "chinesevirus", "china", "wuhanlockdown", "wuhan", "kungflu", "sinophobia", "n95", "world health organization", "cdc", "outbreak", "epidemic", "lockdown", "panic buying", "panicbuying", "socialdistance", "social distance", "socialdistancing", "social distancing"]
    \item languages: []
\end{itemize}

\subsubsection*{crawler 1}
\begin{itemize}
    \item keywords: ["wuhanlockdown", "wuhan", "kungflu", "sinophobia", "n95", "world health organization", "cdc", "outbreak", "epidemic"]
    \item languages: []
\end{itemize}

\subsubsection*{crawler 2}
\begin{itemize}
    \item keywords: ["lockdown", "panic buying", "panicbuying", "socialdistance", "social distance", "socialdistancing", "social distancing"]
    \item languages: []
\end{itemize}

\subsubsection*{crawler 3}
\begin{itemize}
    \item keywords: ["pandemic", "covd", "ncov"],
    \item languages: ["en", "es"]
\end{itemize}

\subsubsection*{crawler 4}
\begin{itemize}
    \item keywords: ["coronavirus"]
    \item languages: ["en"]
\end{itemize}

\subsubsection*{crawler 5}
\begin{itemize}
    \item keywords: ["covid"]
    \item languages: ["en"]
\end{itemize}

\subsubsection*{crawler 6}
\begin{itemize}
    \item keywords: ["covid–19", "COVID---19", "covid19"]
    \item languages: ["en"]
\end{itemize}

\newpage 
\chapter{Global Level Clustering Results}
\label{app:glo-clusters}

The following figures visualize KMeans clustering results on the global user-to-media interaction matrix with different initialization. We consider a group robust if a group with similar size and nationality distribution shows up multiple times among all runs.

\begin{figure*}[!htbp]
  \advance\leftskip-2.4cm
  \vspace{0pt}
  \includegraphics[width=1.4\linewidth, height=70mm]{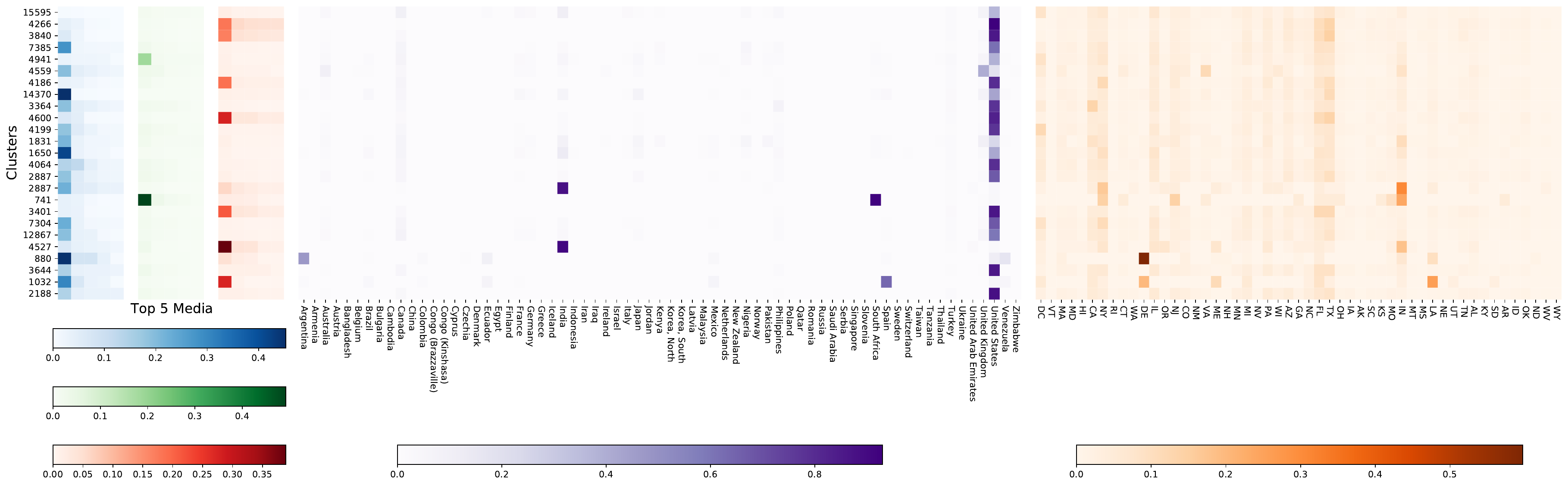}
  \caption{Global user-media consumption clustering results visualisation using random seed 26.}
%   \label{fig:glo-vis}
\end{figure*}

\begin{figure*}[!htbp]
  \advance\leftskip-3.6cm
  \vspace{0pt}
  \includegraphics[width=1.4\linewidth, height=70mm]{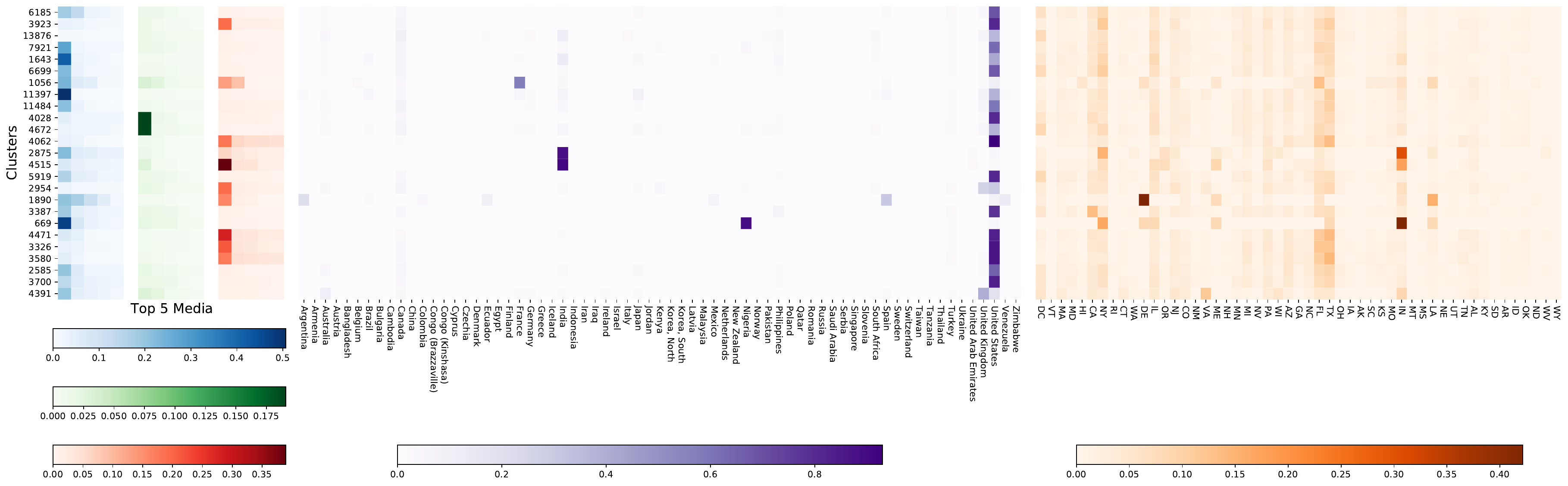}
  \caption{Global user-media consumption clustering results visualisation using random seed 32.}
%   \label{fig:glo-vis}
\end{figure*}

\begin{figure*}[!htbp]
  \advance\leftskip-3.6cm
  \vspace{0pt}
  \includegraphics[width=1.4\linewidth, height=70mm]{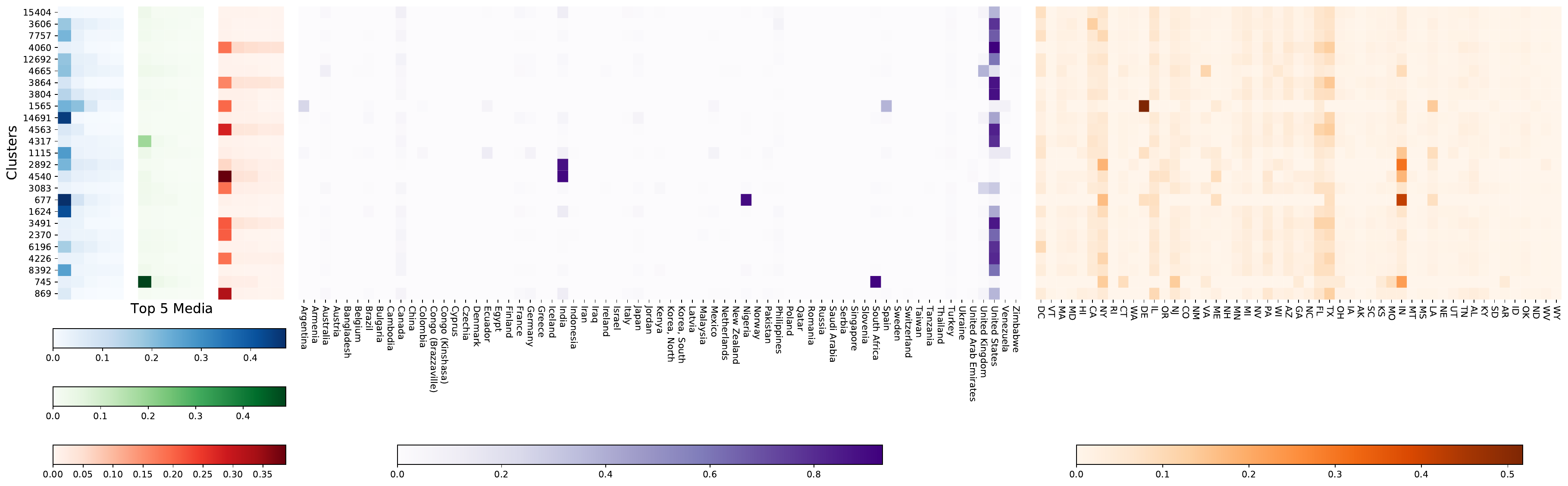}
  \caption{Global user-media consumption clustering results visualisation using random seed 38.}
%   \label{fig:glo-vis}
\end{figure*}

\begin{figure*}[!htbp]
  \advance\leftskip-2.4cm
  \vspace{0pt}
  \includegraphics[width=1.4\linewidth, height=70mm]{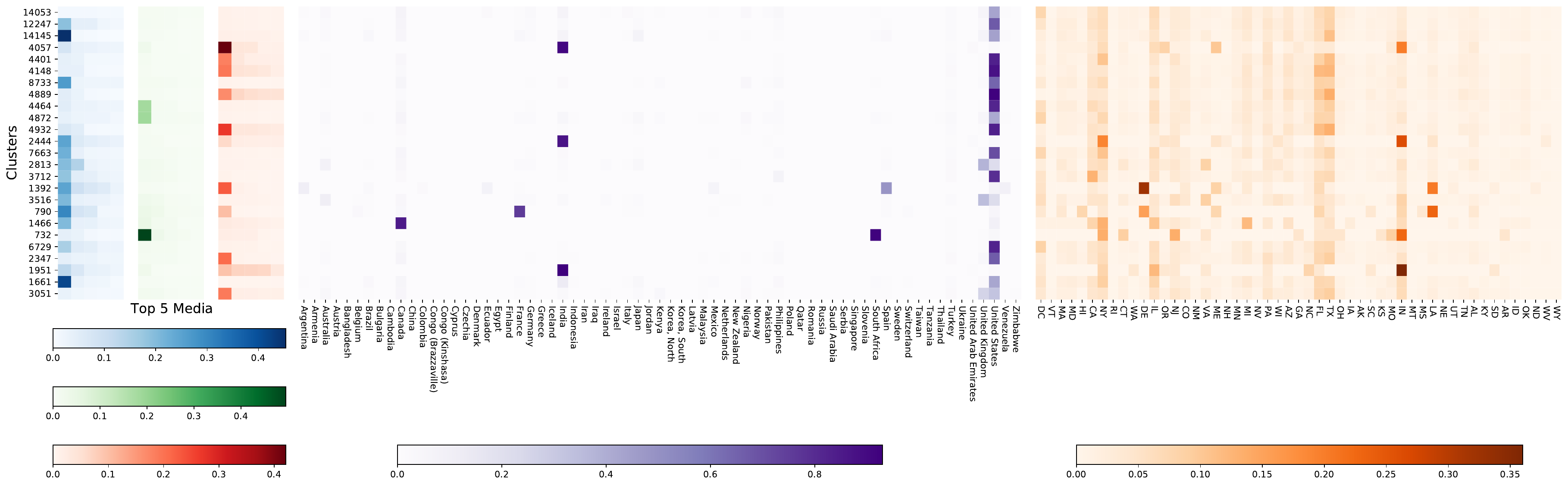}
  \caption{Global user-media consumption clustering results visualisation using random seed 123.}
%   \label{fig:glo-vis}
\end{figure*}

\begin{figure*}[!htbp]
  \advance\leftskip-2.4cm
  \vspace{0pt}
  \includegraphics[width=1.4\linewidth, height=70mm]{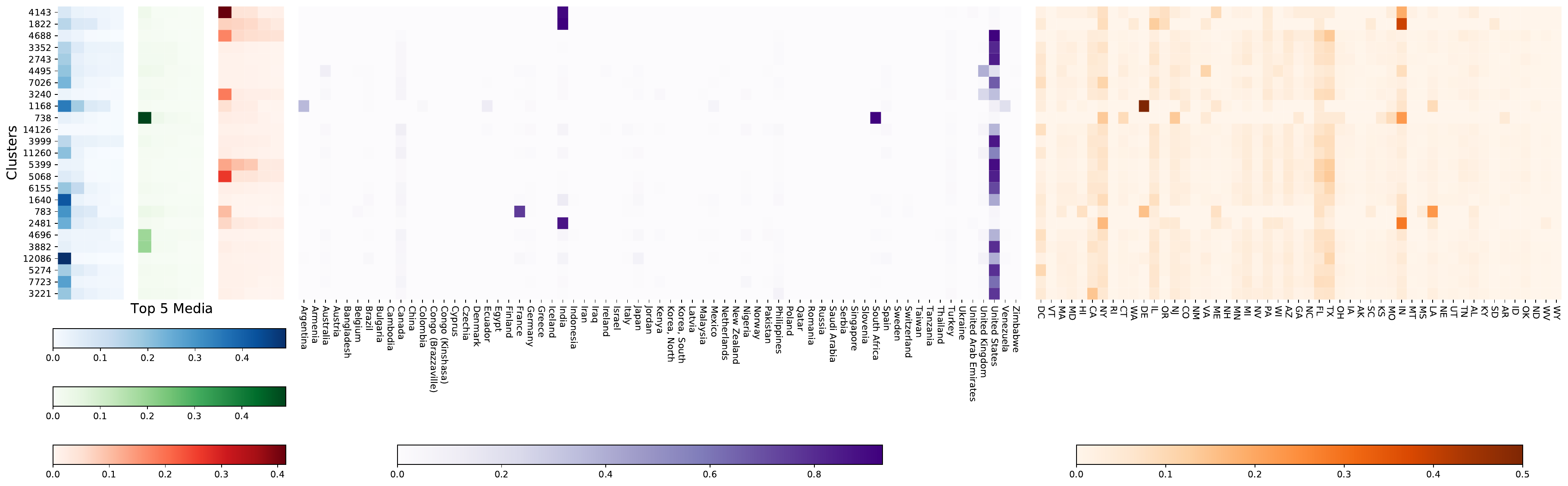}
  \caption{Global user-media consumption clustering results visualisation using random seed 123.}
%   \label{fig:glo-vis}
\end{figure*}

\begin{figure*}[!htbp]
  \advance\leftskip-3.6cm
  \vspace{0pt}
  \includegraphics[width=1.4\linewidth, height=70mm]{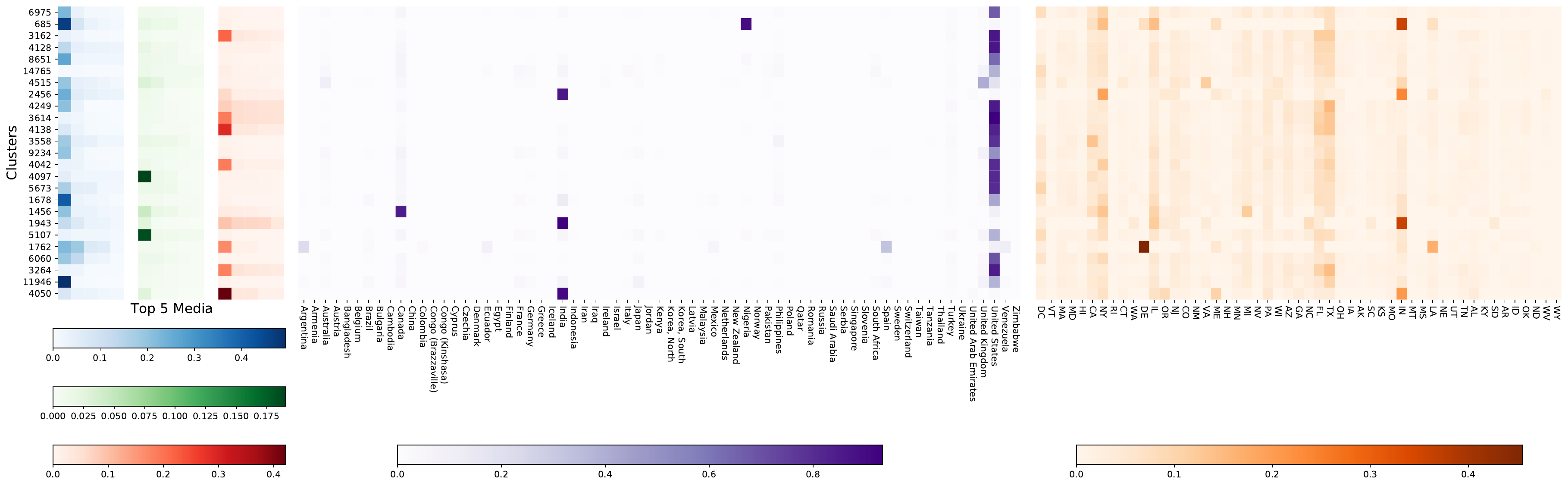}
  \caption{Global user-media consumption clustering results visualisation using random seed 123.}
%   \label{fig:glo-vis}
\end{figure*}

\newpage 
\chapter{India - United States Analysis}
\label{app:ind-us-extra}

The following are the user groups identified from Indian users with media consumption on India media and American media. We run KMeans clustering on the two sets of media consumption multiple times with different initialization and randomness. The risk ratio for users who previously consume extreme-right Indian media to keep consume extreme-right American media remains low in all cases.

\begin{figure*}[!htbp]
  \centering
  \advance\leftskip-1.5cm
  \vspace{0pt}
  \includegraphics[width=1.15\linewidth,height=90mm]{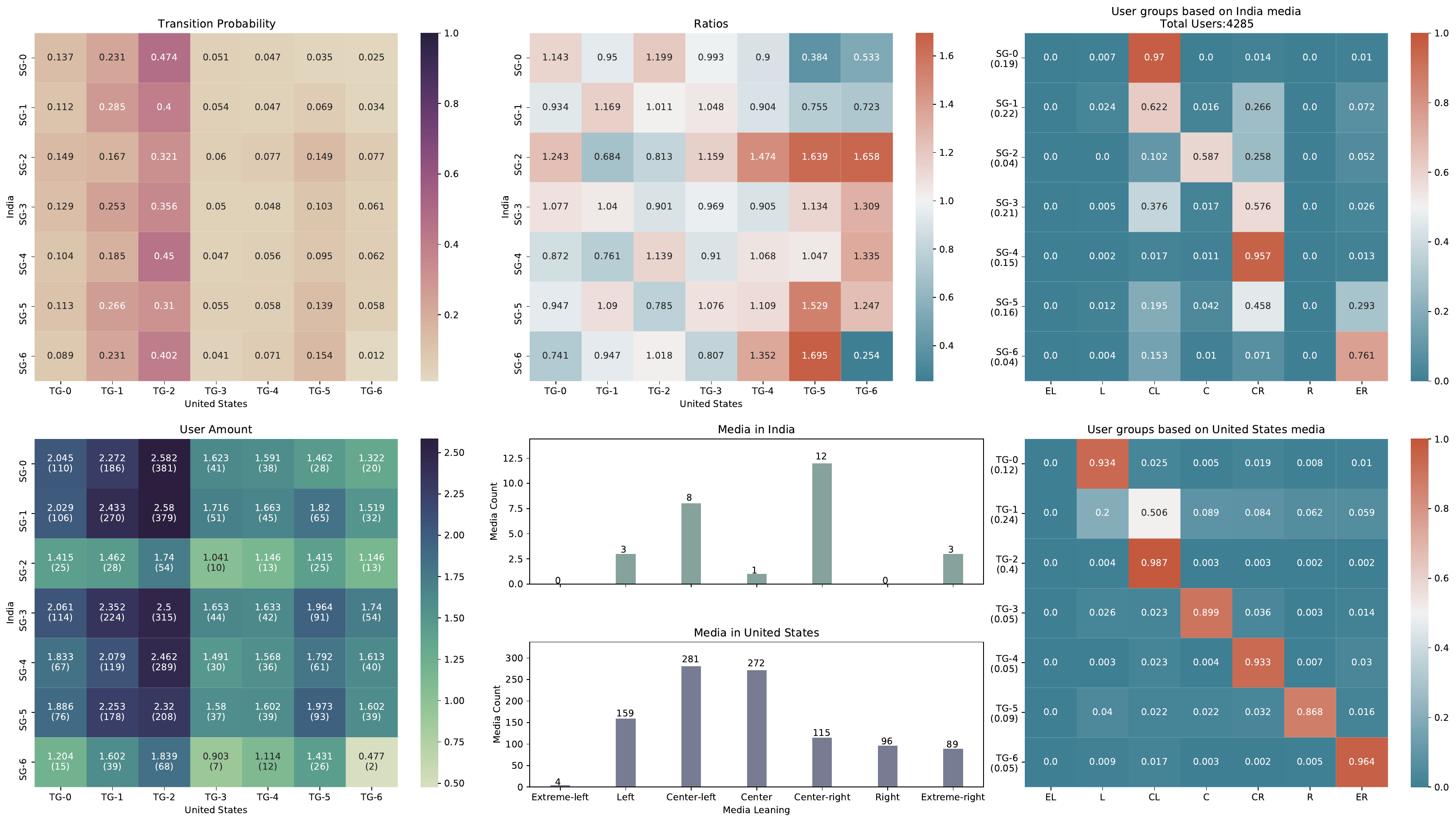}
  
  \caption{India - United States interactions and user groups. The number of cluster for identifying SGs is set to be 7.}
\end{figure*}

\begin{figure*}[!htbp]
  \centering
  \advance\leftskip-1.5cm
  \vspace{0pt}
  \includegraphics[width=1.15\linewidth,height=80mm]{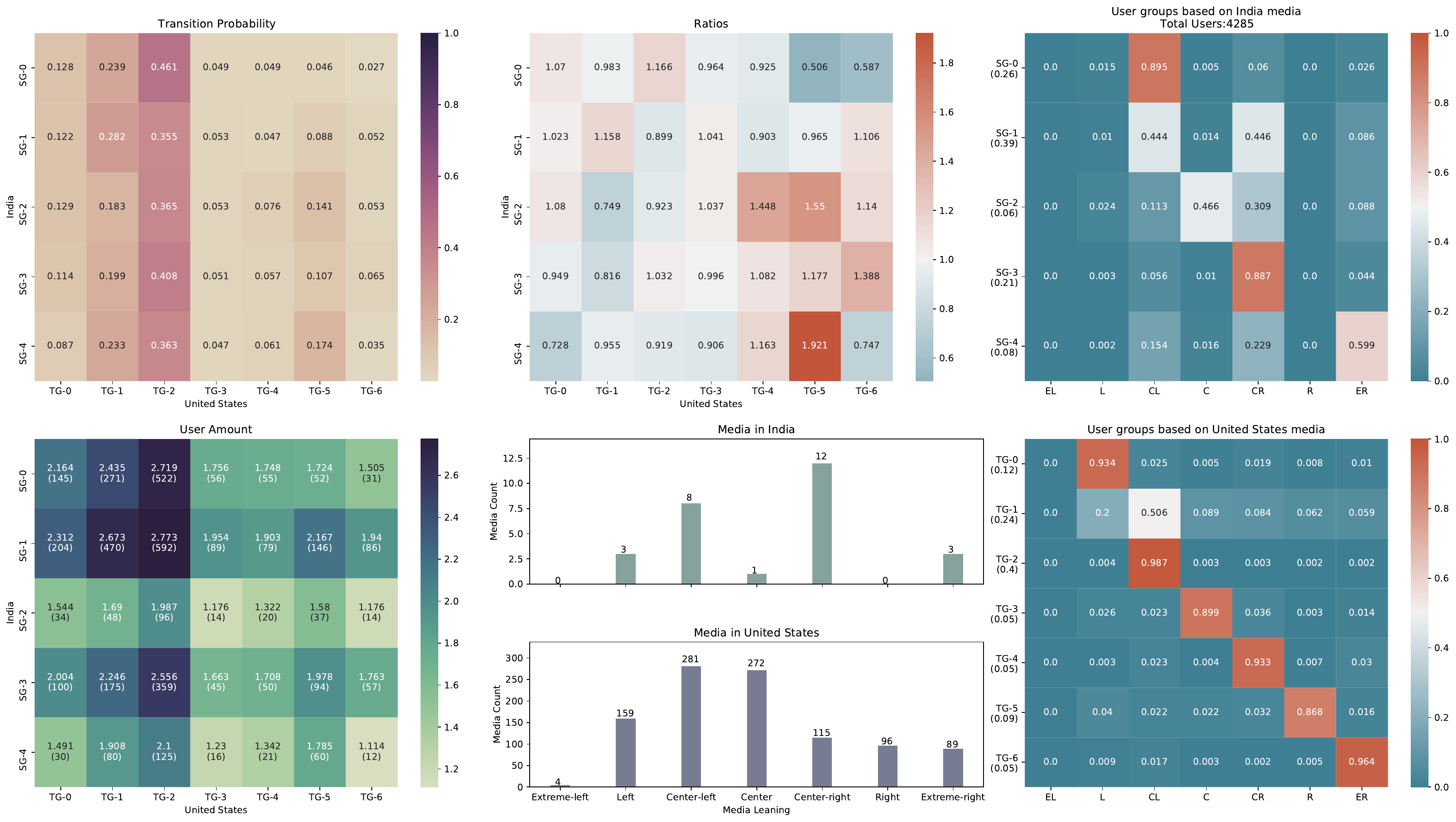}
  \caption{India - United States interactions and user groups. The number of cluster for identifying SGs is set to be 5.}
\end{figure*}

\begin{figure*}[!htbp]
  \centering
  \advance\leftskip-1.5cm
  \vspace{0pt}
  \includegraphics[width=1.15\linewidth,height=80mm]{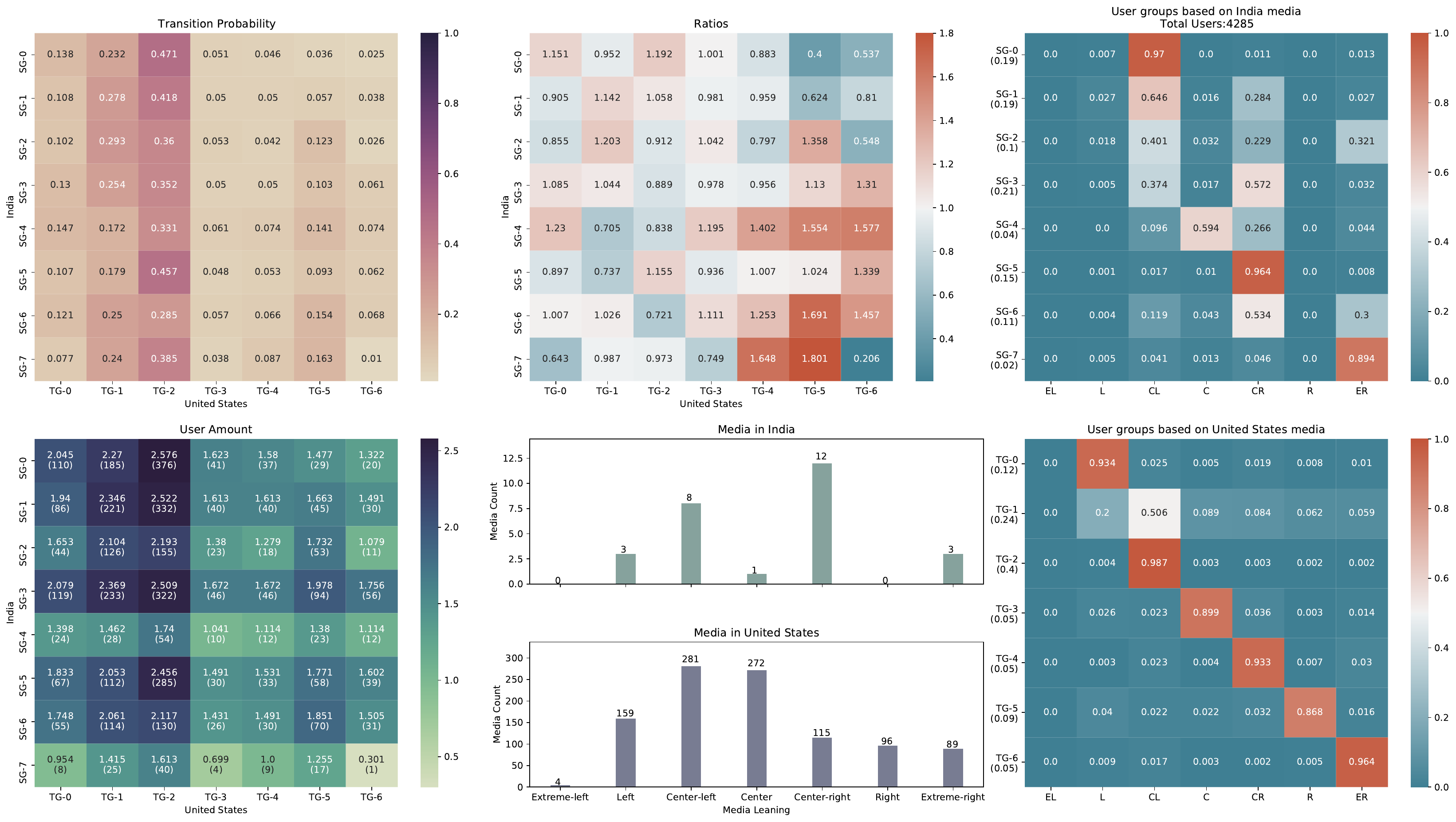}
  \caption{India - United States interactions and user groups. The number of cluster for identifying SGs is set to be 8.}
\end{figure*}

\backmatter

\bibliographystyle{anuthesis}
\bibliography{thesis}
\addcontentsline{toc}{chapter}{Bibliography}

\printindex

\end{document}